\documentclass[9pt,twocolumn,twoside]{pnas-new}

\templatetype{pnasresearcharticle} 

\usepackage{amsmath,amssymb}
\usepackage{bm}
\usepackage[round]{natbib}

\usepackage{amssymb}
\usepackage{mathrsfs}
\usepackage{graphicx}
\usepackage{subcaption}
\usepackage{pdfpages}
\usepackage{afterpage}
\usepackage{caption}
\usepackage{multirow}
\usepackage{rotating}
\usepackage[flushleft]{threeparttable}
\usepackage{multirow}
\usepackage{enumitem} 
\usepackage{algorithm2e}
\usepackage{svg}
\usepackage{stackengine}
\usepackage{float}
\RestyleAlgo{ruled}
\SetAlgorithmName{Procedure}{procedure}{List of Procedure}
\usepackage{tikz} 
\usetikzlibrary{matrix}

\usepackage{preamble/star}
\usepackage[group-separator={,},group-minimum-digits={3}]{siunitx}

\usepackage{lineno}       
\usepackage[colorlinks,linkcolor=blue,anchorcolor=blue,citecolor=blue,urlcolor=black]{hyperref}
\usepackage{cleveref}

\newcolumntype{P}[1]{>{\centering\arraybackslash}p{#1}}

\newcommand{\Rom}[1]{\text{\uppercase\expandafter{\romannumeral #1\relax}}}

\def\v{{\varepsilon}}
  
\def\$#1\${\begin{align*}#1\end{align*}}

\renewcommand{\max}{\mathop{\mathrm{max}}}

\usepackage{color}

\renewcommand{\P}{\operatorname{\mathbb{P}}}
\newcommand{\ex}{\mathrm{e}}

\title{The 2020 US Decennial Census is more private than you (might) think}

\author[a,1]{Buxin Su} 
\author[a,1,2]{Weijie J.\ Su} 
\author[b,1]{Chendi Wang}

\affil[a]{University of Pennsylvania}
\affil[b]{Xiamen University}

\leadauthor{Su, Su, and Wang}

\significancestatement{
The U.S.\ Decennial Census informs critical policy decisions, yet ensuring privacy while maintaining data accuracy is a challenge. The Census Bureau adopted differential privacy to protect the confidentiality of individual data in 2020. Our research demonstrates that the 2020 Census achieves significantly stronger privacy guarantees than officially reported, enabling reduced noise injection by 15.08\% to 24.82\% while maintaining nearly the same privacy guarantee at each geographical level. The enhanced accuracy of private census data would benefit applications like redistricting and socioeconomic research. This work addresses an open problem posed by the Census Bureau, advancing methodologies in data privacy crucial to societal decision-making.
}

\authorcontributions{The authors designed research, performed research, analyzed data, and wrote the paper.}
\authordeclaration{The authors declare no competing interest.}
\equalauthors{\textsuperscript{1}Authors are listed in alphabetical order.}
\correspondingauthor{\textsuperscript{2}To whom correspondence should be addressed. E-mail: suw@wharton.upenn.edu.}


\begin{abstract}
The U.S.\ Decennial Census serves as the foundation for many high-profile policy decision-making processes, including federal funding allocation and redistricting. In 2020, the Census Bureau adopted differential privacy to protect the confidentiality of individual responses through a disclosure avoidance system that injects noise into census data tabulations. The Bureau subsequently posed an open question: Could stronger privacy guarantees be obtained for the 2020 U.S.\ Census compared to their published guarantees, or equivalently, had the privacy budgets been fully utilized?

In this paper, we address this question affirmatively by demonstrating that the 2020 U.S.\ Census provides significantly stronger privacy protections than {the officially published guarantees} suggest at each of the eight geographical levels, from the national level down to the block level. This finding is enabled by our precise tracking of privacy losses using $f$-differential privacy, applied to the composition of private queries across these geographical levels. Our analysis reveals that the Census Bureau introduced unnecessarily high levels of noise to meet the specified privacy guarantees for the 2020 Census. Consequently, we show that noise variances could be reduced by 15.08\% to 24.82\% while maintaining nearly the same level of privacy protection for each geographical level, thereby improving the accuracy of privatized census statistics. We empirically demonstrate that reducing noise injection into census statistics mitigates distortion caused by privacy constraints in downstream applications of private census data, illustrated through a study examining the relationship between earnings and education.
\end{abstract}

\dates{This manuscript was compiled on \today}
\doi{\url{www.pnas.org/doi/pdf/10.1073/pnas.2500337122}}

\begin{document}

\maketitle
\thispagestyle{firststyle}
\ifthenelse{\boolean{shortarticle}}{\ifthenelse{\boolean{singlecolumn}}{\abscontentformatted}{\abscontent}}{}

\section{Introduction}
The U.S.\ Census Bureau conducts a decennial national census, with the most recent one held in 2020. The census provides critical information about population distribution, economic indicators, and demographic trends, significantly influencing the nation's political and economic decisions with consequential and lasting effects. Specifically, the census impacts resource allocation, including federal funding distributions \citep{hotchkiss2017uses}, redistricting \citep{funding2023census,Kenny2023Comment,cohen2021census}, labor markets \citep{autor2003rise,bureau2023guidance}, and the apportionment of congressional representation \citep{eckman2021apportionment,bureau2021approtionment}.

\begin{figure*}%
\centering
\subfloat[Composition of the 2020 U.S.\ Decennial Census. \label{fig:composition-chart}]{\includegraphics[ width=0.57\textwidth]{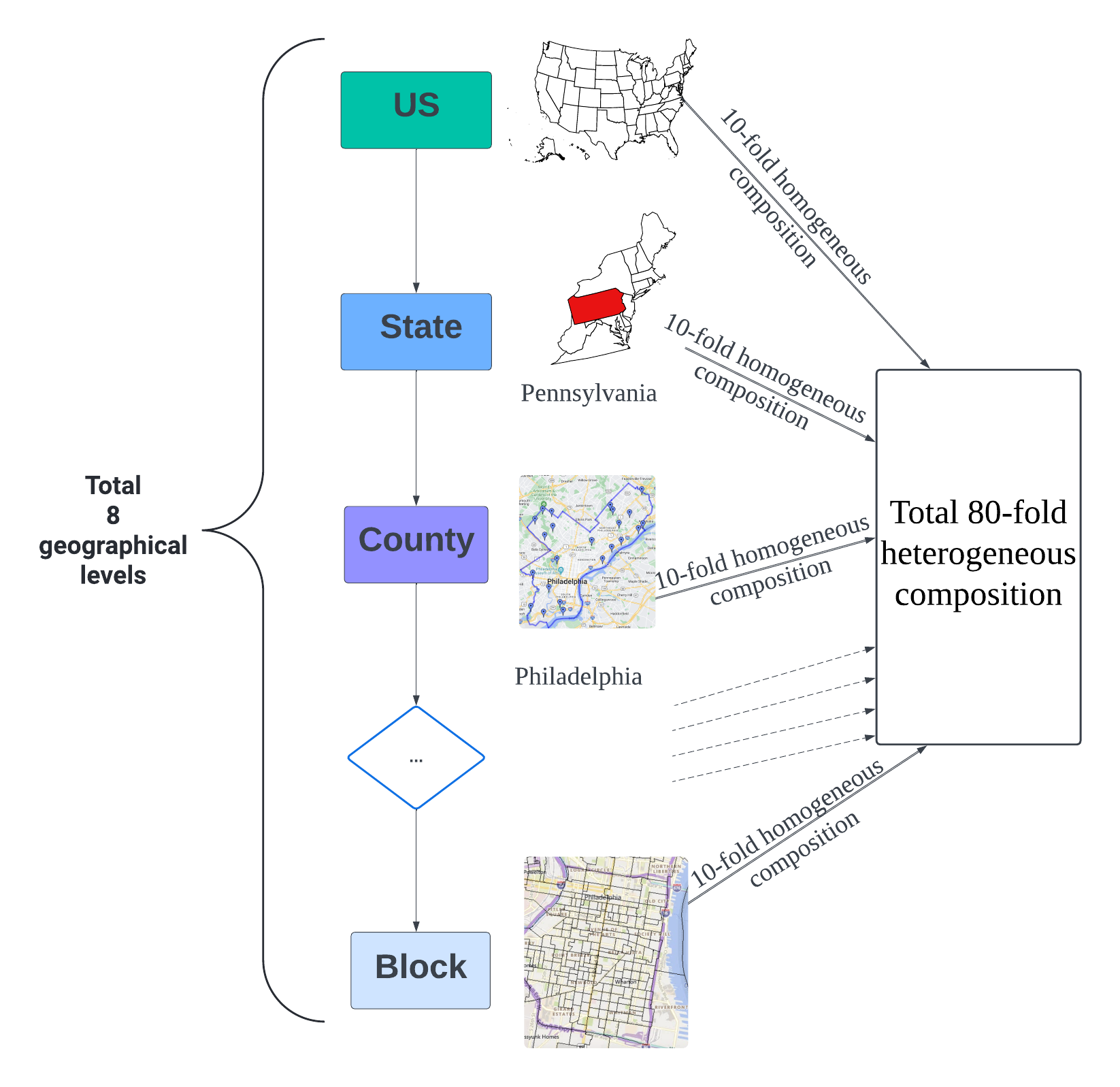}}
\qquad
\subfloat[Privatization and post-processing.\label{fig:DAS-chart}]{\includegraphics[ width=0.36\textwidth]{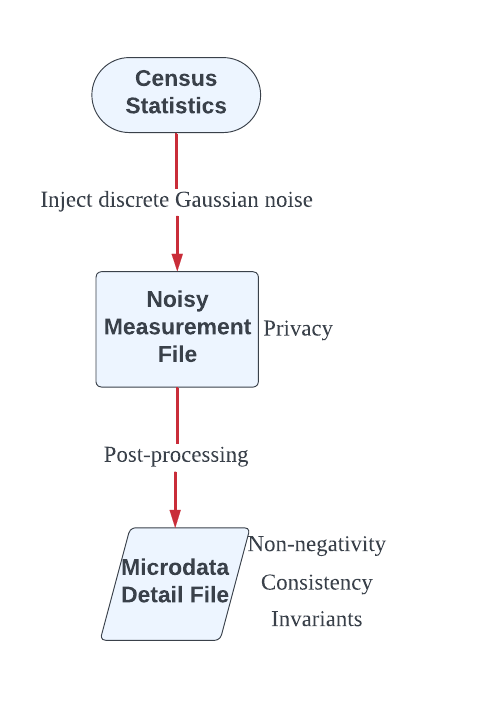}}\qquad
\caption{Overview of the disclosure avoidance system for the {2020 Census Demographic and Housing Characteristics File} \citep{abowd20222020, abowd2022Census}. The omitted geographical levels are tract subset group, tract subset, optimized block group, and population estimates primitive geography (PEPG).}
\end{figure*}

Census data inherently contains sensitive information such as income, race, and age. The direct release of census data, even at the aggregate or summary statistics level, is vulnerable to re-identification and reconstruction attacks, potentially leading to privacy breaches \citep{duncan1989risk,dick2023confidence,Hawes2022census}. A notable example is the reconstruction of 46.5\% of the population from the 2010 U.S.\ Census data \citep{Abowd_2019, abowd2021declaration}. To address the need for privacy and confidentiality in census responses, the U.S.\ Census Bureau adopted differential privacy (DP)---a privacy-preserving technique with a rigorous mathematical foundation \citep{dwork2006calibrating,dwork2006our}---for the 2020 Census. The implementation was carried out through a disclosure avoidance system (DAS) in the form of a TopDown algorithm \citep{abowd20222020, abowd2022Census}. This algorithm processes raw census data and injects noise into key tabulations of confidential information. It generates noisy measurement files (NMFs) as intermediate outputs before post-processing, which ensures non-negativity as well as internal and hierarchical consistency. An illustration of the DAS process is provided in Figure \ref{fig:DAS-chart}.

The injection of noise, whether before or after post-processing, inevitably reduces the accuracy of census data and can lead to undesirable biases against certain subpopulations \citep{Phil2023bureau, Kenny2021use, Kenny2024census}. Generally, privacy protection affects the reliability of policy-making processes based on census data, such as redistricting, and reduces the accuracy of downstream research on census data \citep{anderson2015american, boyd2022Differential, Kenny2023Comment}. Consequently, it is crucial to find out the precise privacy guarantees offered by the DAS for the 2020 Census. If a tighter privacy guarantee than that provided by the Census Bureau can be established, it would allow less noise to be injected while maintaining the same level of privacy, thereby enhancing census accuracy by fully utilizing the nominal privacy budget. This challenge has been posed as an open problem by the Census Bureau \citep{kifer2022bayesian} (see its Section 5.2.1). {It is important to note, however, that the tightness of a privacy guarantee depends significantly on the semantic interpretation of the privacy definition employed \citep{kifer2022bayesian}.}

In this paper, we address this open problem by demonstrating that the 2020 Census offers stronger privacy guarantees than reported by the Census Bureau. {Our analysis recognizes that a mechanism's privacy guarantee can be fully characterized by $f$-DP \citep{Dong2022Gaussian}, which is equivalent to the $(\epsilon,\delta)$-curve, also known as the privacy profile \citep{balle2020privacy}.\footnote{Due to the equivalence between $f$-DP and the privacy profile, we present our privacy analysis in both forms. For numerical illustration, we often compare values of $\epsilon$ for a fixed $\delta$, derived from different approaches.} Our focus is on the privacy guarantees for the 2020 Census Demographic and Housing Characteristics File (DHC), a key data product from the 2020 Census \citep{abowd20222020, abowd2022Census}.} We show that the actual privacy parameter $\epsilon$ is 8.49\% to 13.21\% lower than its nominal value across all eight geographical levels when $\delta$ is not exceedingly small.\footnote{We consider $\delta > 10^{-11}$. For $\delta \leq 10^{-11}$, our obtained $\epsilon$ values remain strictly smaller than the corresponding nominal values.} For example, at the state level, the Bureau's published $\epsilon$ values are 11.07 for $\delta = 10^{-11}$ and 7.79 for $\delta = 10^{-5}$, whereas our analysis indicates reductions to 10.13 and 6.57, respectively.\footnote{While $\delta$ for a single $(\epsilon, \delta)$ pair typically must be smaller than the reciprocal of the data size (approximately $3 \times 10^{-9}$ for the U.S.\ population), this study examines the dependence of $\epsilon$ on $\delta$ as a continuous curve, allowing for cases such as $\delta = 10^{-5}$.} Notably, this improvement requires no modifications to the existing privatization process and incurs no additional cost for the published census data.

Recognizing these underutilized privacy guarantees, one could inject less noise into census tabulations, thereby obtaining more accurate NMFs. Our analysis demonstrates that, at nearly the same privacy level, the noise level employed by the Census Bureau is unnecessarily high. We have developed a hybrid method combining analytic and computational approaches to efficiently determine the optimal level of injected noise that fully leverage the published privacy guarantees. For example, our method results in a 20.88\% reduction in the variance of injected noise for the national level of the 2020 Census. The implementation of our methodology is publicly available on \href{https://github.com/BuxinSu/Census_Privacy/tree/main}{GitHub}.

The benefit of injecting less noise extends to improved estimation properties after post-processing is applied. Using data from the {IPUMS NHGIS Privacy-Protected Demonstration Data \citep{privacyprotected2010}},\footnote{This dataset \citep{privacyprotected2010} encompasses both the non-privatized 2010 Census Summary Files and a privacy-protected version of the 2020 Census DHC File (derived from the 2022-08-25 Demonstration Data release). Notably, the non-privatized version of the 2020 DHC is not publicly available.} across the geographical levels of state, county, tract, and block in Pennsylvania \citep{privacyprotected2010}, our simulation results show that the noise reduction enabled by our analysis decreases the mean squared error (MSE) by approximately 15\% when non-negative post-processing is used. The enhanced accuracy of census would naturally improve the reliability of census-based applications. To illustrate this, we conduct an empirical study using data from the ACS 5-year Census \citep{Census2020ACSST5Y2020.S1501}. Our analysis shows that it can significantly mitigate the distortion in estimates caused by the privacy constraints on the data.


The enhanced privacy analysis of the 2020 U.S.\ Census is made possible by tackling the complex composition structure of the census using the $f$-DP framework \citep{Dong2022Gaussian}. Specifically, the 2020 Census comprises eight geographical levels, with each level containing ten queries (see an illustration in Figure \ref{fig:composition-chart}). For each geographical level, the DAS privatizes queries by injecting integer-valued noise \citep{Canonne2020discrete}. In this paper, we consider the scenario where the noise within each geographical level is independently and identically distributed (i.i.d.). The $f$-DP framework is particularly well-suited for precisely accounting for overall privacy loss when composing many steps, each contributing to the privacy loss. 


A major challenge in applying $f$-DP to the 2020 U.S.\ Census arises from the discreteness of the integer-valued noise used in the DAS, which underlies the technical difficulty of the open problem posed by the Census Bureau \citep{kifer2022bayesian}. The Census Bureau circumvented this challenge by approximating the discrete distribution with its continuous counterpart and using concentrated DP to account for privacy losses \citep{dwork2016concentrated,bun2016concentrated,bun2018composable}. While this approximation is a natural choice for concentrated DP, it introduces looseness in the privacy bounds their method can offer. In contrast, our approach directly addresses the discreteness challenge within the $f$-DP framework by analytically evaluating the main part of the privacy bound while numerically bounding the remainder. Our method, which handles these components separately, presents several technical innovations that might be valuable in other privacy accounting problems where high accuracy is required.

\section{Preliminaries}

To present our main results, we first introduce basic concepts of DP \citep{dwork2006calibrating,dwork2006our}. A randomized mechanism $M$ satisfies $(\epsilon,\delta)$-DP for $\epsilon\geq 0$ and $0\leq \delta \leq 1$ if, for any pair of neighboring datasets $D$ and $D'$---where one can be obtained from the other by adding or removing a single individual record---and any event $S$, we have
\begin{equation}\label{eq:dp_def}
\P(M({D})\in S)\leq \ex^{\epsilon}\cdot\P(M({D}')\in S)  + \delta.
\end{equation}
A smaller value of $\epsilon$ indicates a stronger privacy guarantee. A mechanism's privacy guarantee generally cannot be fully delineated by a single pair of $\epsilon$ and $\delta$ and is instead given by its privacy profile, represented by the $(\epsilon, \delta)$-curve obtained by varying $\epsilon$ or $\delta$ \citep{balle2020privacy}. We refer readers to \cite{kifer2022bayesian} for a semantic interpretation of the parameters $\epsilon$ and $\delta$ in the context of census data.

The Census Bureau injected integer-valued noise following the discrete Gaussian distribution into the tabulations of confidential census data. The discrete Gaussian distribution, denoted by $\mathcal{N}_{\mathbb{Z}}(0,\sigma^2)$, has a probability mass function given by 
\[
p_{\sigma}(x) = \frac{\ex^{-x^2/2\sigma^2}}{{\sum_{i\in\mathbb{Z}}\ex^{-i^2/2\sigma^2}}}
\]
for any $x$ in the set of integers $\mathbb{Z}$ \citep{micciancio2007worst, Canonne2020discrete}, where $\sigma > 0$ is the standard deviation.\footnote{The variance of $\mathcal{N}_{\mathbb{Z}}(0,\sigma^2)$ is very close to $\sigma^2$. See Appendix \ref{sec:variance}.}

An important feature of the DAS used in the 2020 Census lies in its composition structure. As illustrated in Figure~\ref{fig:composition-chart}, the DAS involves eight geographical levels \citep{privacyallocation2022}, for each level releasing ten private queries injected with discrete Gaussian noise, which is assumed to be i.i.d.\ in this paper for simplicity. The challenge lies in quantifying the overall privacy loss---in particular, determining the value of $\epsilon$ in \eqref{eq:dp_def} for a given $\delta$---accumulated across these ten queries.\footnote{While there are 50 states, this compositional structure allows us to consider only one state at a time. This is because an individual record would impact at most one state, as different states represent disjoint subsets of the total U.S.\ population (see more elaboration in \citep{Mcsherry2010privacy,Smith2022making}).} From a technical standpoint, accurately accounting for privacy loss under composition using $(\epsilon, \delta)$-DP alone is difficult \citep{kairouz2017composition}. The Census Bureau addressed this challenge by using divergence-based relaxations of DP \citep{dwork2016concentrated,bun2016concentrated,bun2018composable,mironov2017renyi} in their privacy accounting method. Their privacy guarantees can be converted into $(\epsilon, \delta)$-curves in $(\epsilon, \delta)$-DP.

In contrast, in this paper we employ the more recent $f$-DP framework \citep{Dong2022Gaussian}, which has been shown to be well-suited for privacy analysis with composition structures \citep{bu2020deep,wang2024unified,su2024statistical}. To tackle the discreteness of the integer-valued noise, we have developed several novel techniques to handle distributions supported on lattices. 

To define $f$-DP, consider formulating the problem of distinguishing between $D$ and $D'$ as hypothesis testing: 
\[
H_0: \text{the true dataset is } D \text{ versus } H_1: \text{the true dataset is } D'.
\]
Let $0 \le \phi \le 1$ be any rejection rule and denote by $\alpha_{\phi} = \mathbb{E}_{H_0}[\phi]$ and $\beta_{\phi}= 1 - \mathbb{E}_{H_1}[\phi]$ the type I and type II errors, respectively. The trade-off function $T(M(D), M(D')): [0,1] \rightarrow [0,1]$ between $D$ and $D'$ is defined as
\[
T(M(D), M(D'))(\alpha) = \inf_{\phi}\{\beta_{\phi}: \alpha_{\phi} \le \alpha\}
\]
for any $0 \le \alpha \le 1$ \citep{Dong2022Gaussian}.\footnote{Let $P$ and $Q$ denote the probability distributions of $M(D)$ and $M(D')$, respectively. Formally, the trade-off function $T(M(D), M(D'))$ should be defined through $P$ and $Q$, thereby being expressed as $T(P, Q)$.} We say that a mechanism $M$ satisfies $f$-DP\footnote{We typically require $f:[0,1] \rightarrow [0,1]$ to be a trade-off function for some pair of distributions. It is a trade-off function if and only if $f$ is continuous, convex, non-increasing, and $f(\alpha) \le 1 - \alpha$.} if 
\[
T(M(D), M(D'))(\alpha) \geq f(\alpha)
\]
for any neighboring $D$ and $D'$ and any $\alpha$. 

A larger trade-off function indicates that it is more difficult to distinguish between $H_0$ and $H_1$, meaning the mechanism provides stronger privacy. Mathematically, the $f$-DP guarantee is equivalent to an infinite collection of guarantees offered by an $(\epsilon, \delta)$-curve \citep{Dong2022Gaussian}. However, the former is easier for analytical analysis in several privacy operations such as composition and subsampling \citep{Dong2022Gaussian,wang2024unified}.

{ For a comprehensive comparison between our $f$-DP based approach and the divergence-based DP method used by the U.S.\ Census Bureau, we convert the privacy guarantees provided by both methods to $(\epsilon,\delta)$-curves (see Section~\ref{sec:results} for details).}

\section{Results}
\label{sec:results}

We present the main results of this paper in this section, while deferring technical proofs to Appendix \ref{sec:method}.

\subsection{Improved privacy guarantees at geographical levels}\label{sec:2020census}
\begin{figure}[!t]
    \centering
    \begin{subfigure}[b]{0.24\textwidth}
        \includegraphics[width=\textwidth]{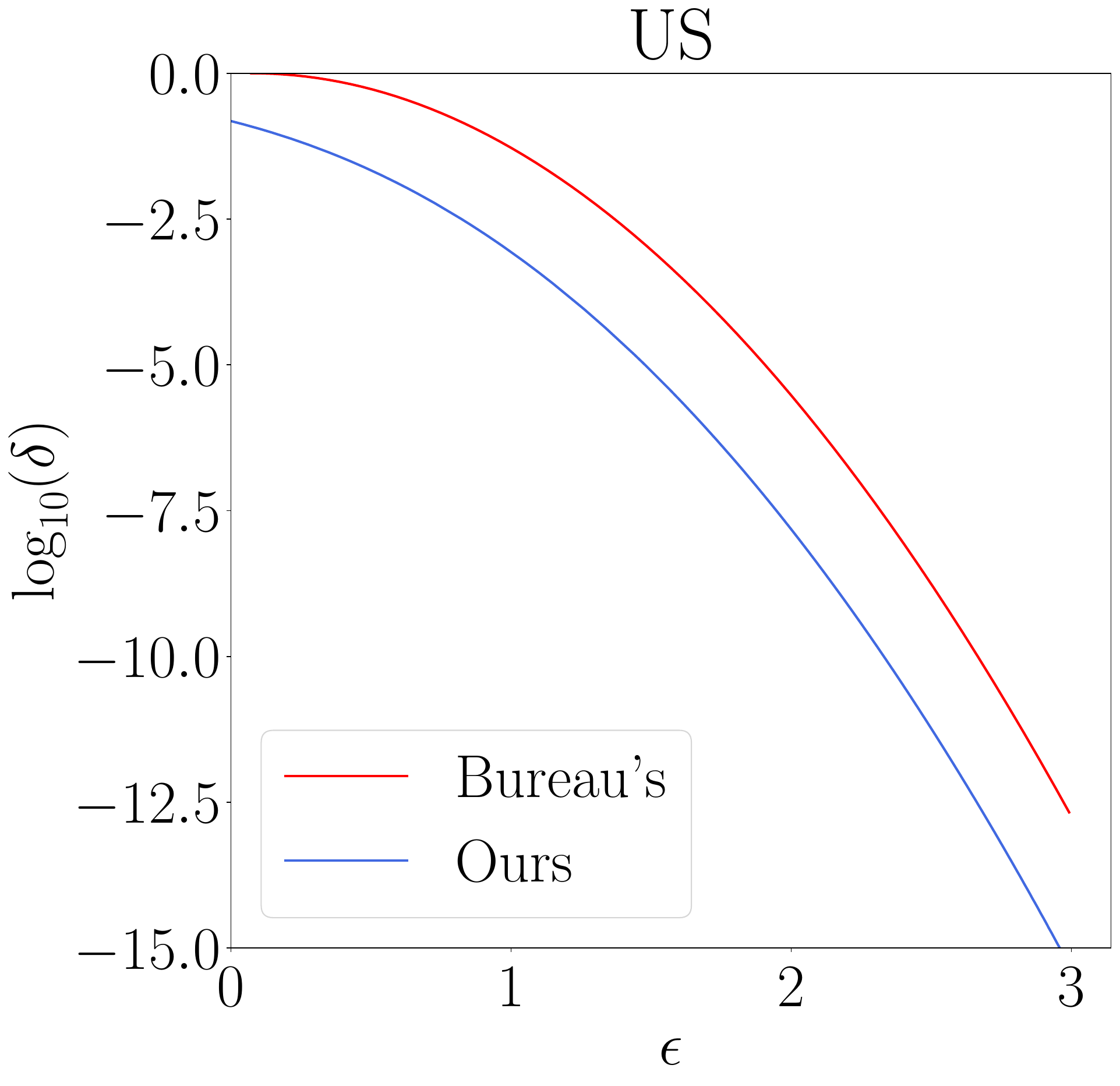}
    \end{subfigure}
    \begin{subfigure}[b]{0.24\textwidth}
        \includegraphics[width=\textwidth]{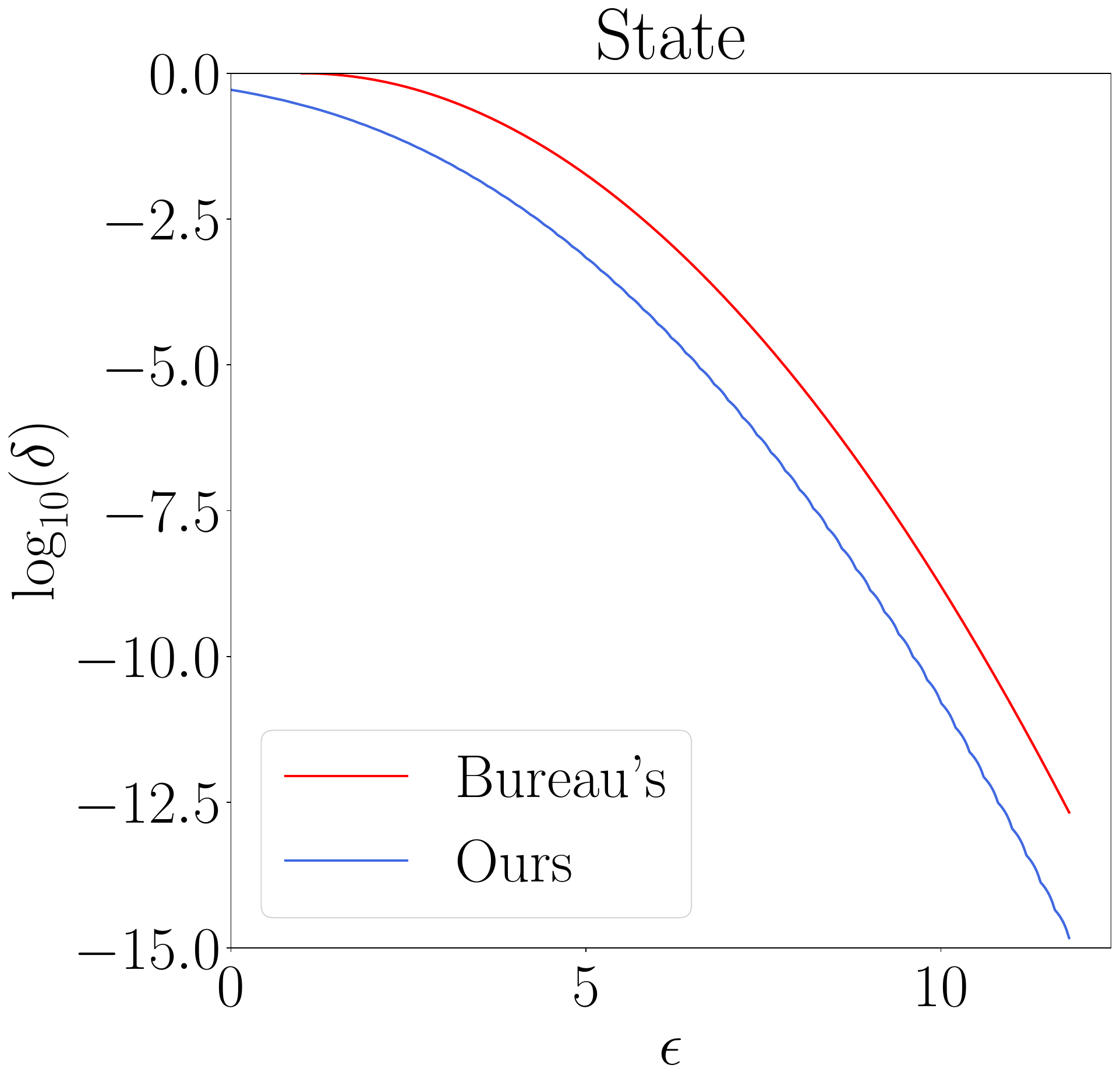}
    \end{subfigure}   
        
    \hfill  

    \begin{subfigure}[b]{0.24\textwidth}
        \includegraphics[width=\textwidth]{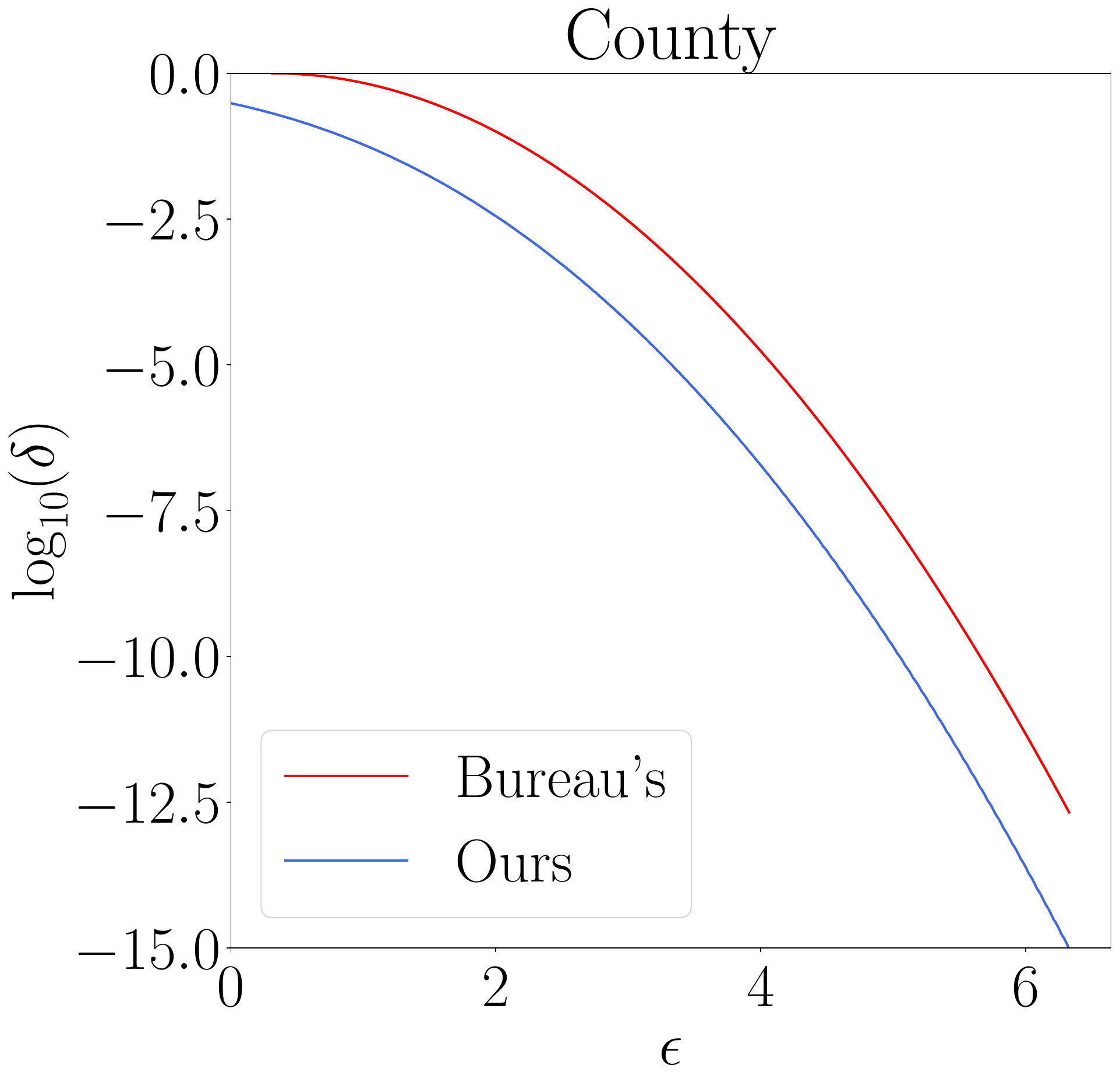}
    \end{subfigure}
    \begin{subfigure}[b]{0.24\textwidth}
        \includegraphics[width=\textwidth]{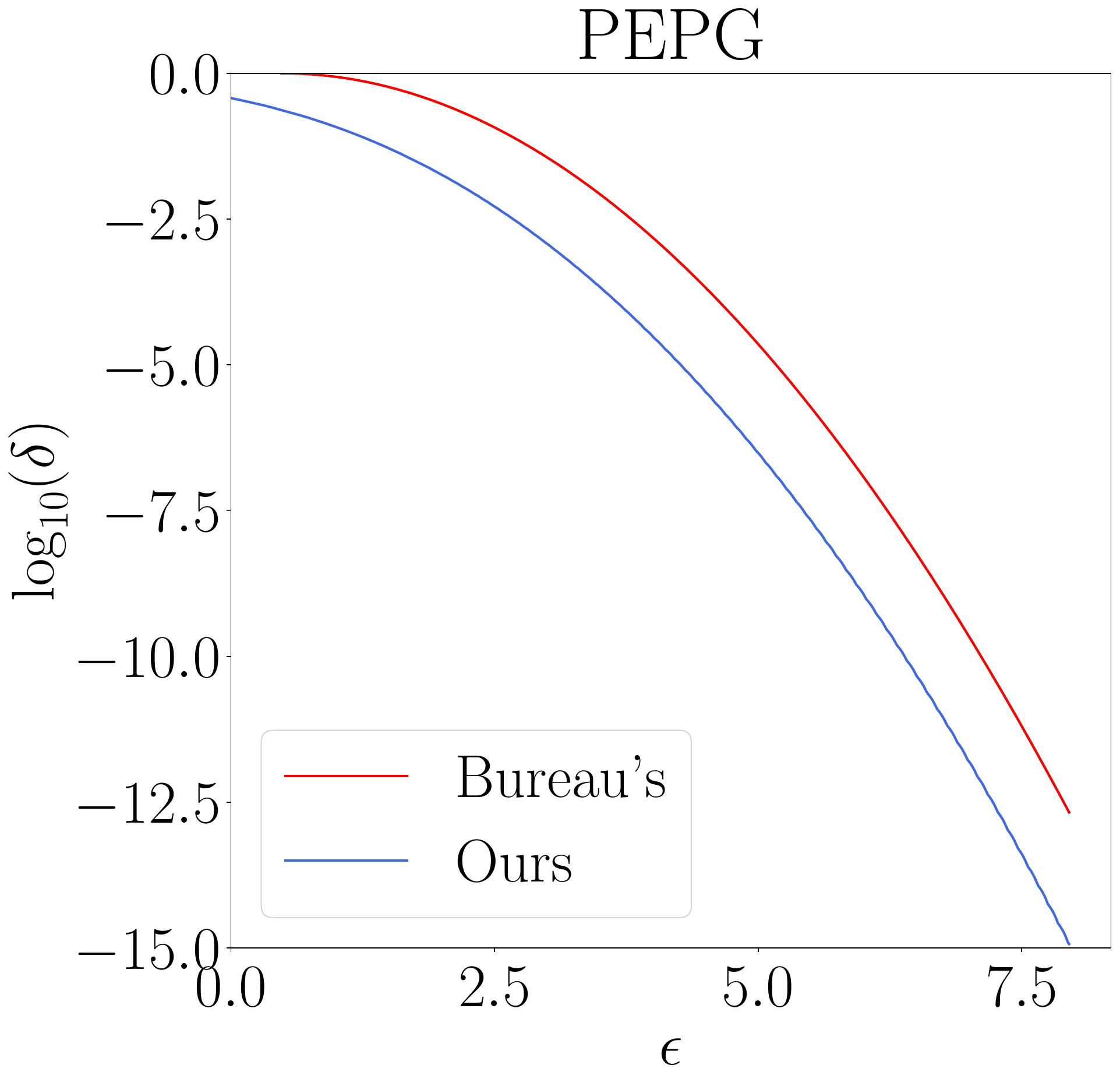}
    \end{subfigure}
    
    \hfill  

    \begin{subfigure}[b]{0.24\textwidth}
        \includegraphics[width=\textwidth]{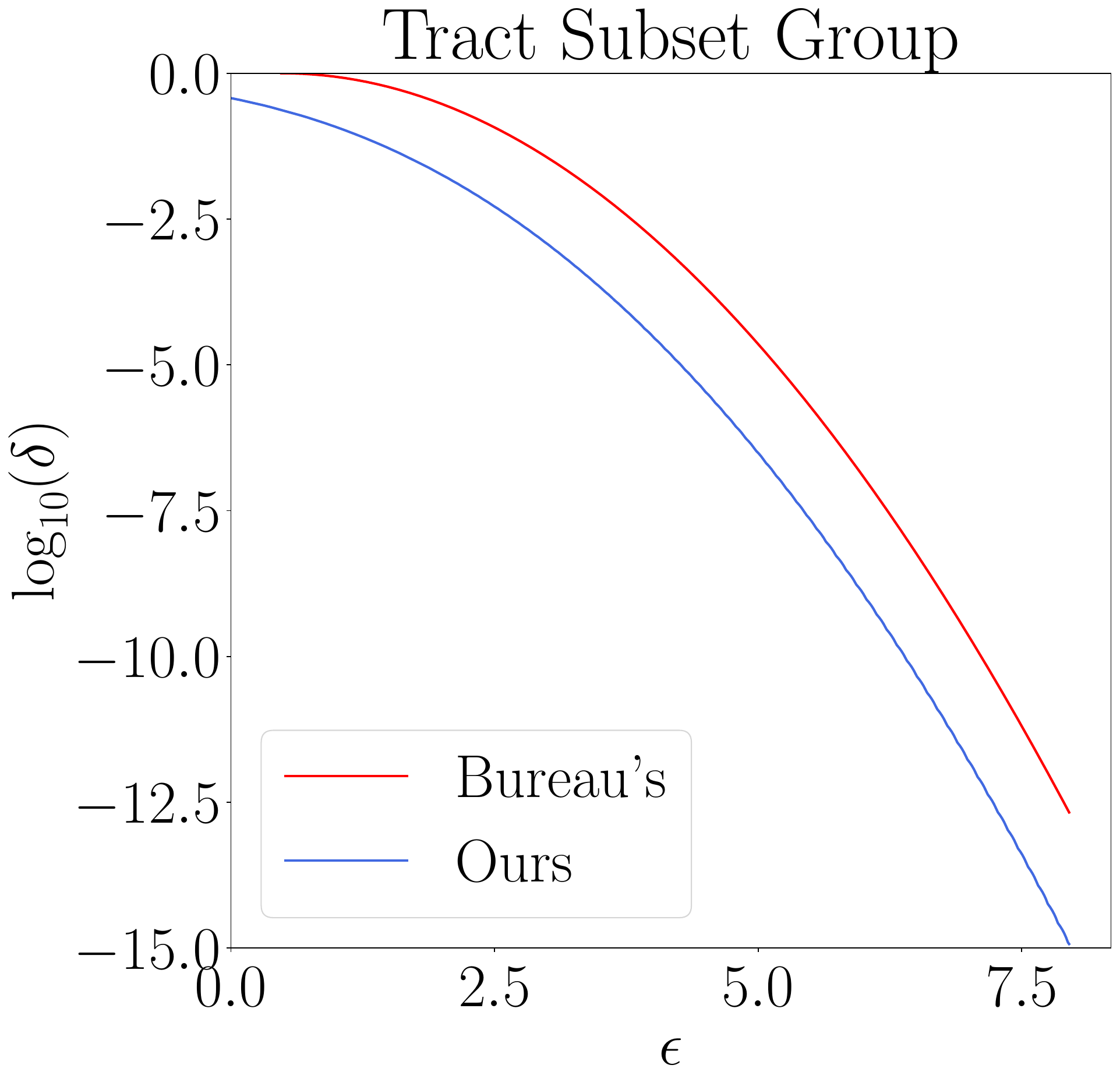}
    \end{subfigure}
    \begin{subfigure}[b]{0.24\textwidth}
        \includegraphics[width=\textwidth]{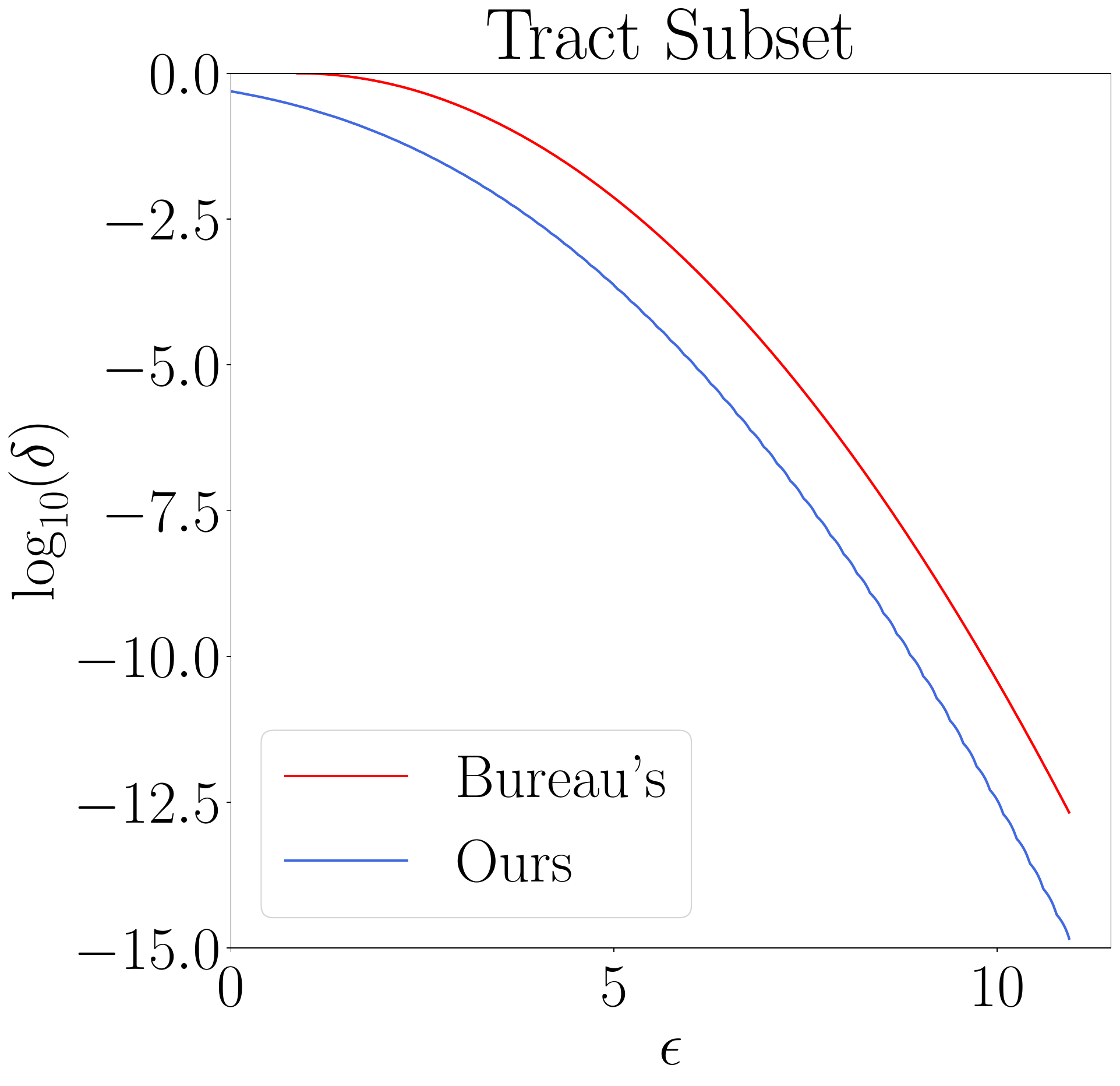}
    \end{subfigure}
        
    \hfill  

    \begin{subfigure}[b]{0.24\textwidth}
        \includegraphics[width=\textwidth]{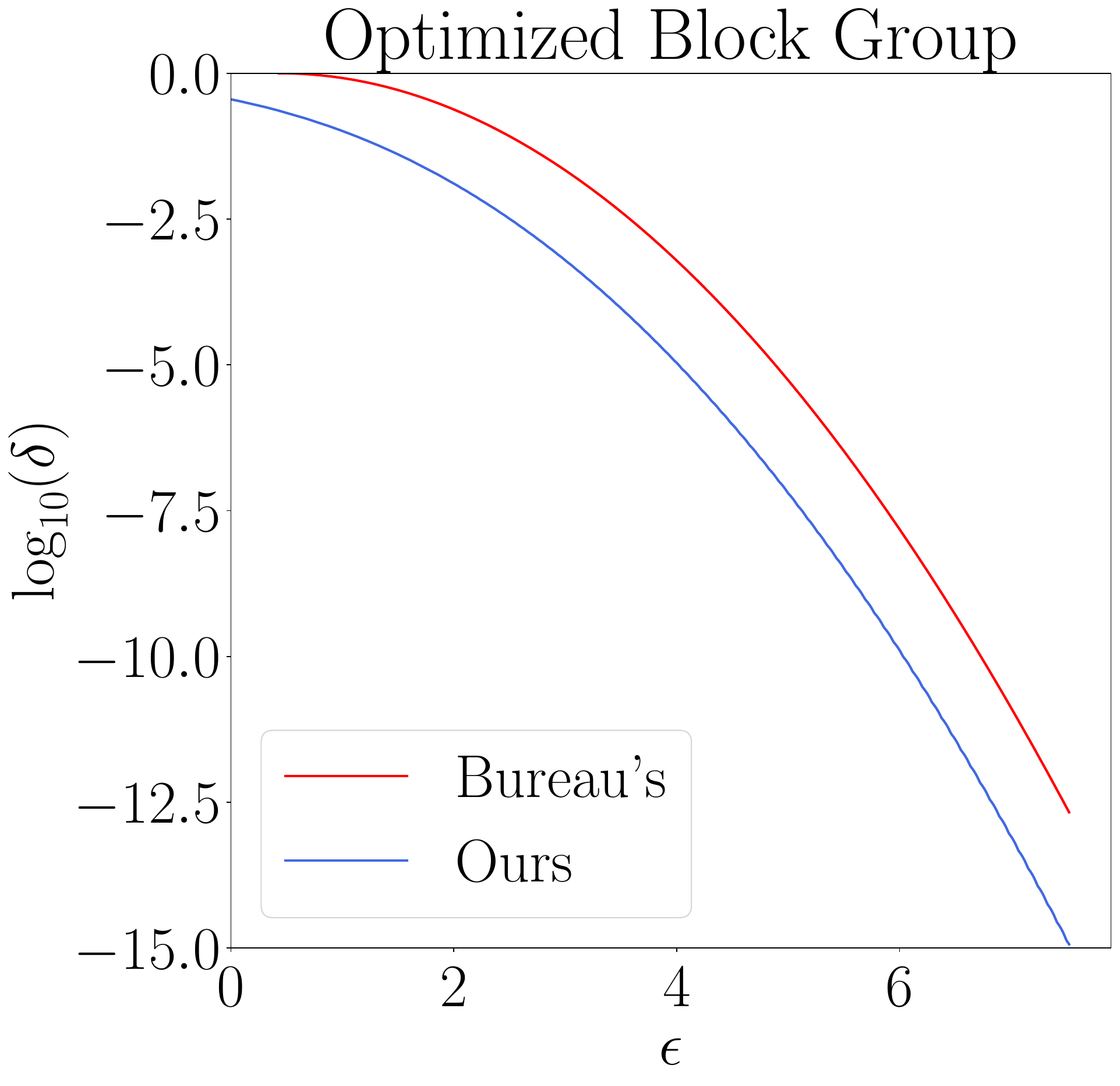}
    \end{subfigure}
    \begin{subfigure}[b]{0.24\textwidth}
        \includegraphics[width=\textwidth]{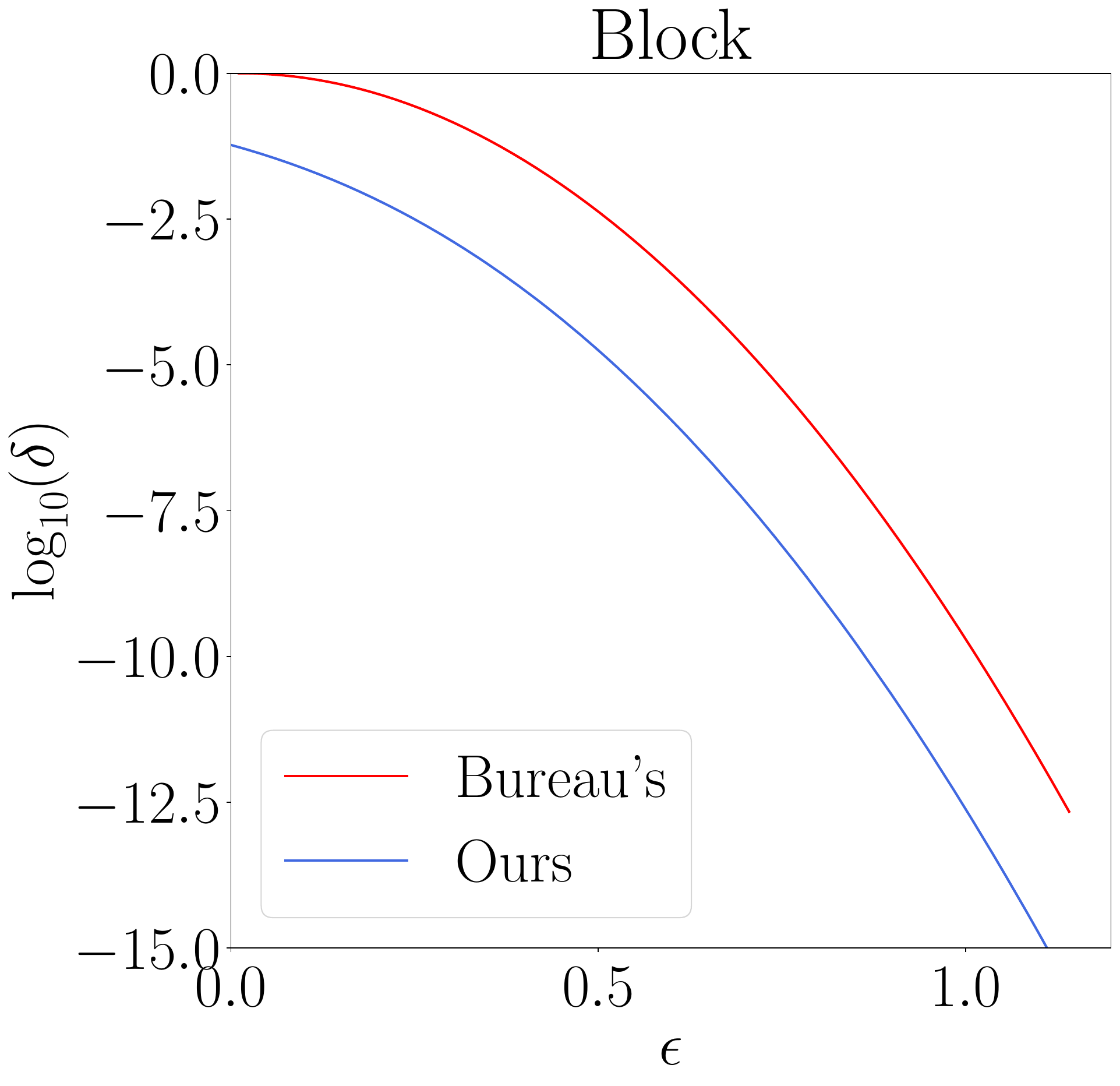}
    \end{subfigure}
 \caption{Comparison of $(\epsilon, \delta)$-curves between our method (blue) and the Census Bureau's accounting method (red) across eight geographical levels of the 2020 {DHC}. The noise configuration follows the privacy-loss budget allocation released by the Bureau on August 25, 2022 \citep{privacyallocation2022}, as detailed in Table~\ref{table:2020 allocation}. The red curves are derived from concentrated DP \citep{bun2016concentrated}, which the Census Bureau used to measure the privacy budget (see Appendix~\ref{sec:method} for details). Our method (blue) achieves a uniformly better trade-off between $\epsilon$ and $\delta$ compared to the Bureau's method. Notably, our method ensures $\delta < 1$ when $\epsilon = 0$, as shown in (\ref{eqn:delta_formula}) in Appendix~\ref{sec:approx}. }
    \label{fig:eps_delta_geo}
\end{figure}

{Figure~\ref{fig:eps_delta_geo} presents the privacy guarantee computed using our method for each geographical level in the form of $(\epsilon, \delta)$-curve, and for comparison, we also present the $(\epsilon, \delta)$-curve derived using the Census Bureau's approach.} From the comparison, under the same noise level published on August 25, 2022 by the Bureau {for the 2020 DHC} \citep{privacyallocation2022}, our $\epsilon$ value is uniformly smaller than the Bureau's value for any $\delta$ in the range between 0 and 1, in particular including $\delta \rightarrow 0$. As this comparison is over the entire $(\epsilon, \delta)$-curves, it demonstrates in a mathematically rigorous sense that the privacy guarantee of the 2020 Census for each geographical level was underestimated using the Bureau's approach.

\begin{figure*}[!htp]
  \centering
  \includegraphics[width=0.7\textwidth]{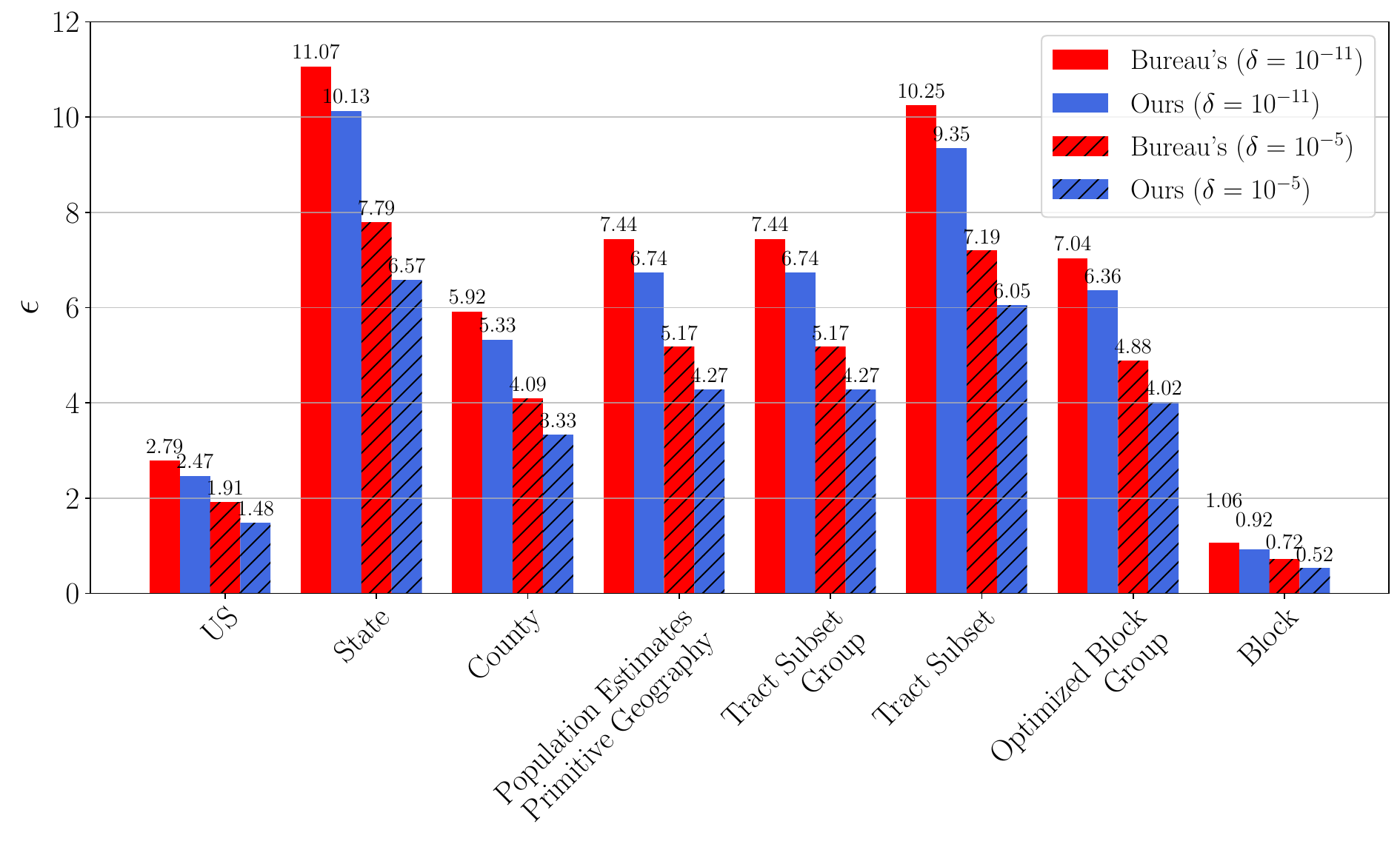}
  \caption{Comparison of $(\epsilon, \delta)$-curves from Figure~\ref{fig:eps_delta_geo} at specific values of $\delta$ for each geographical level of the 2020 {DHC}. The values considered are $\delta = 10^{-11}$ and $\delta = 10^{-5}$. For comparisons at other values of $\delta$, see Figures~\ref{fig:eps_delta_geo_gap} and \ref{fig:eps_delta_geo_percentage} in Appendix~\ref{sec:supp_fig}.}
  \label{fig:improve_epsilon}
\end{figure*}

To obtain a more quantitative understanding of this improvement, we refer to Figure~\ref{fig:improve_epsilon}, which displays the values of $\epsilon$ for $\delta=10^{-11}$ and $\delta=10^{-5}$ for the eight geographical levels. For example, when $\delta=10^{-11}$, the $\epsilon$ parameter is reduced by a range of 8.50\% (at the state level) to 13.76\% (at the block level). For comparisons at other values of $\delta$, see Figures~\ref{fig:eps_delta_geo_gap} and \ref{fig:eps_delta_geo_percentage} in Appendix~\ref{sec:supp_fig}.

\begin{figure}[!htp]
    \centering
    \begin{subfigure}[b]{0.232\textwidth}
        \includegraphics[width=\textwidth]{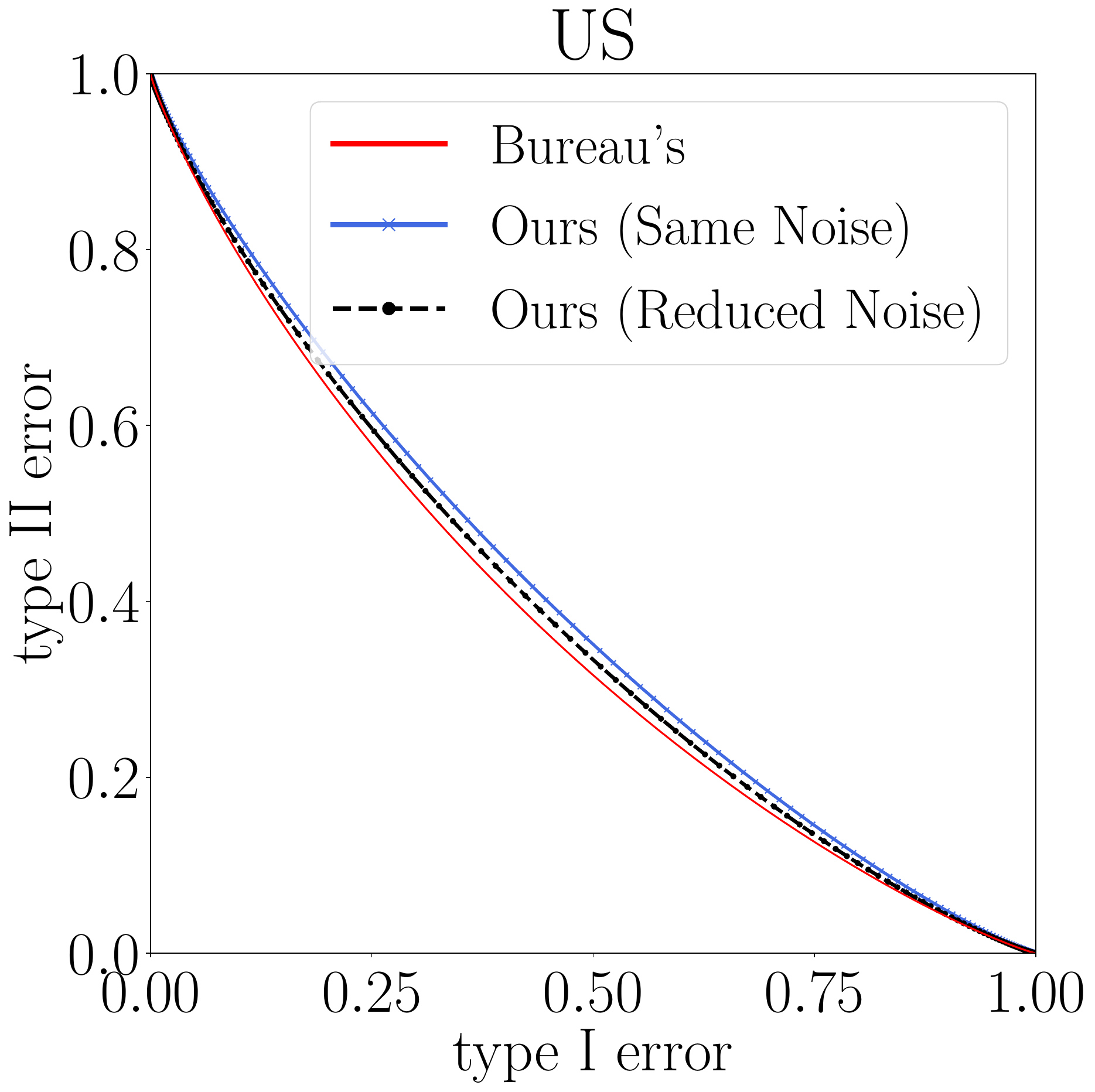}
    \end{subfigure}
    \begin{subfigure}[b]{0.232\textwidth}
        \includegraphics[width=\textwidth]{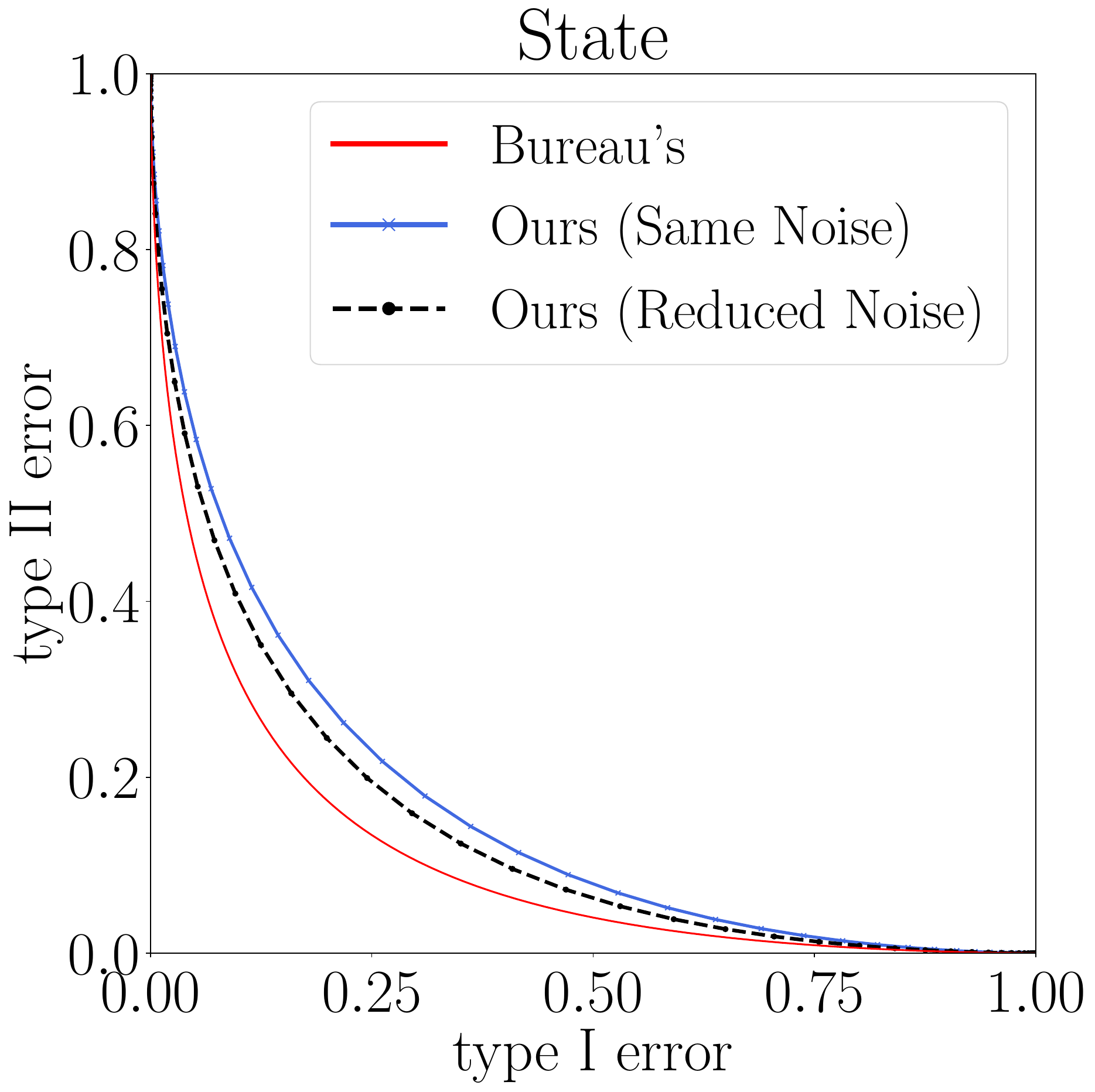}
    \end{subfigure}
    
    \hfill
    
    \begin{subfigure}[b]{0.232\textwidth}
        \includegraphics[width=\textwidth]{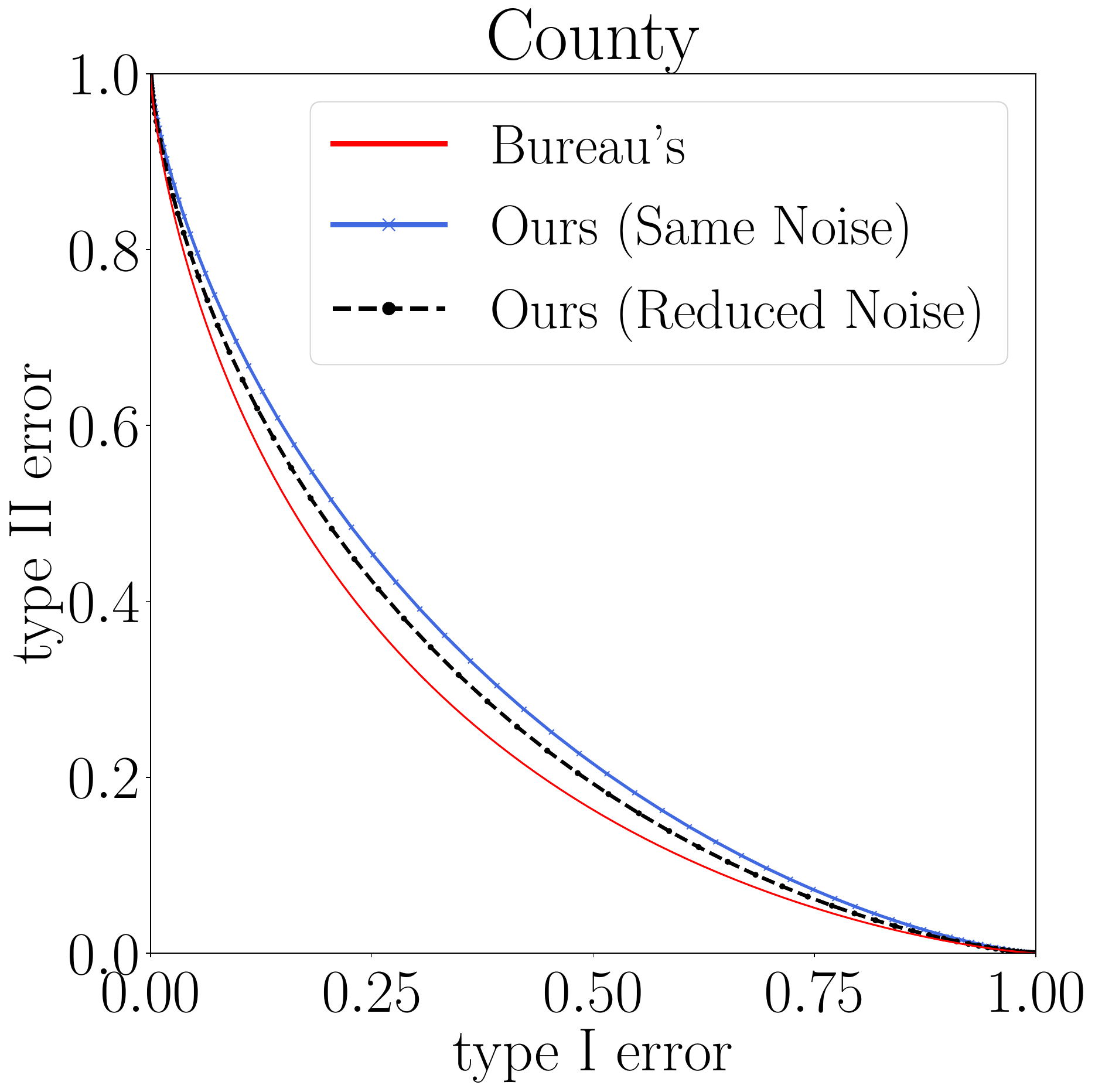}
    \end{subfigure}
    \begin{subfigure}[b]{0.232\textwidth}
        \includegraphics[width=\textwidth]{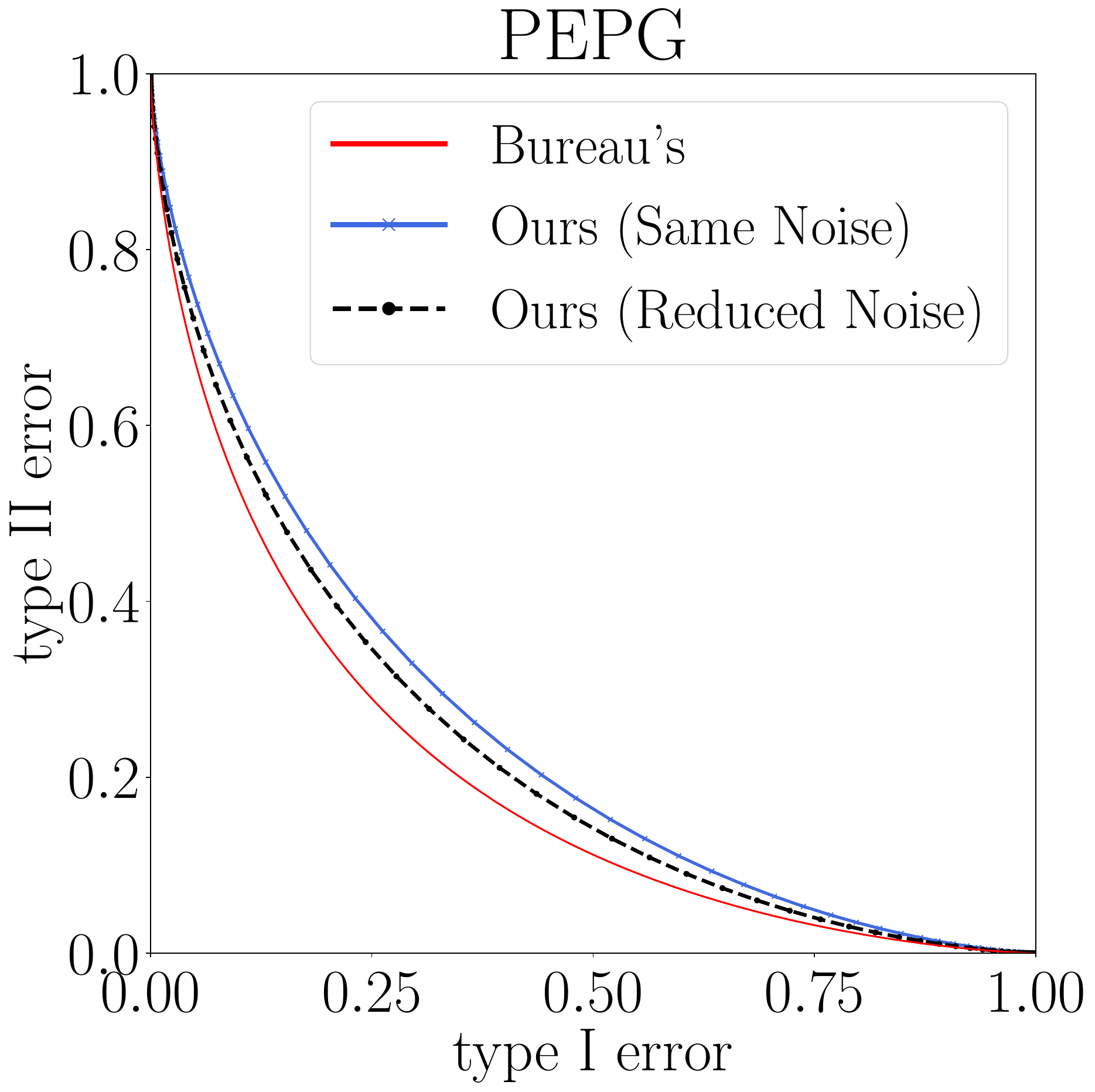}
    \end{subfigure}
    
    \hfill
    
    \begin{subfigure}[b]{0.232\textwidth}
        \includegraphics[width=\textwidth]{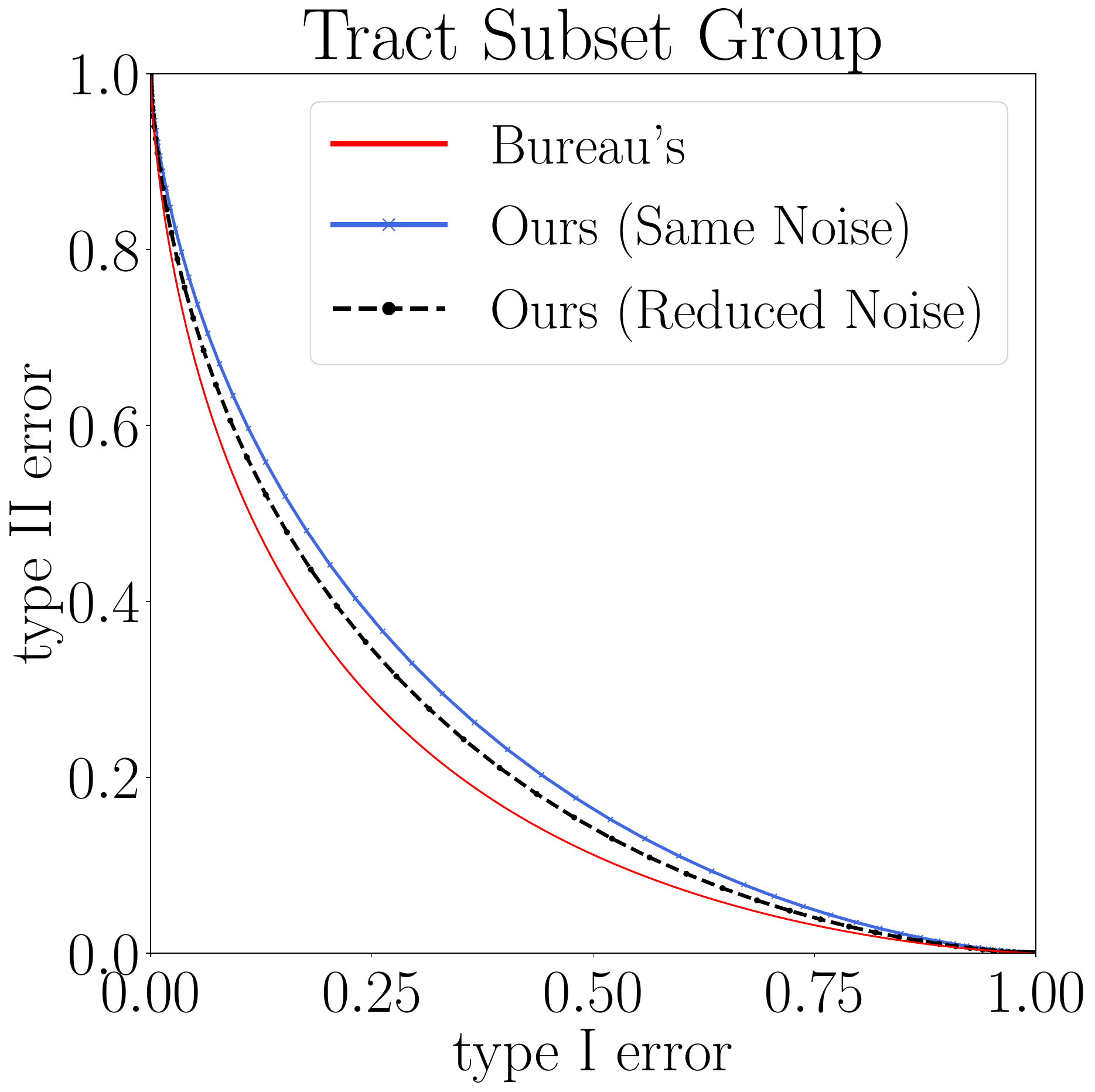}
    \end{subfigure}
    \begin{subfigure}[b]{0.232\textwidth}
        \includegraphics[width=\textwidth]{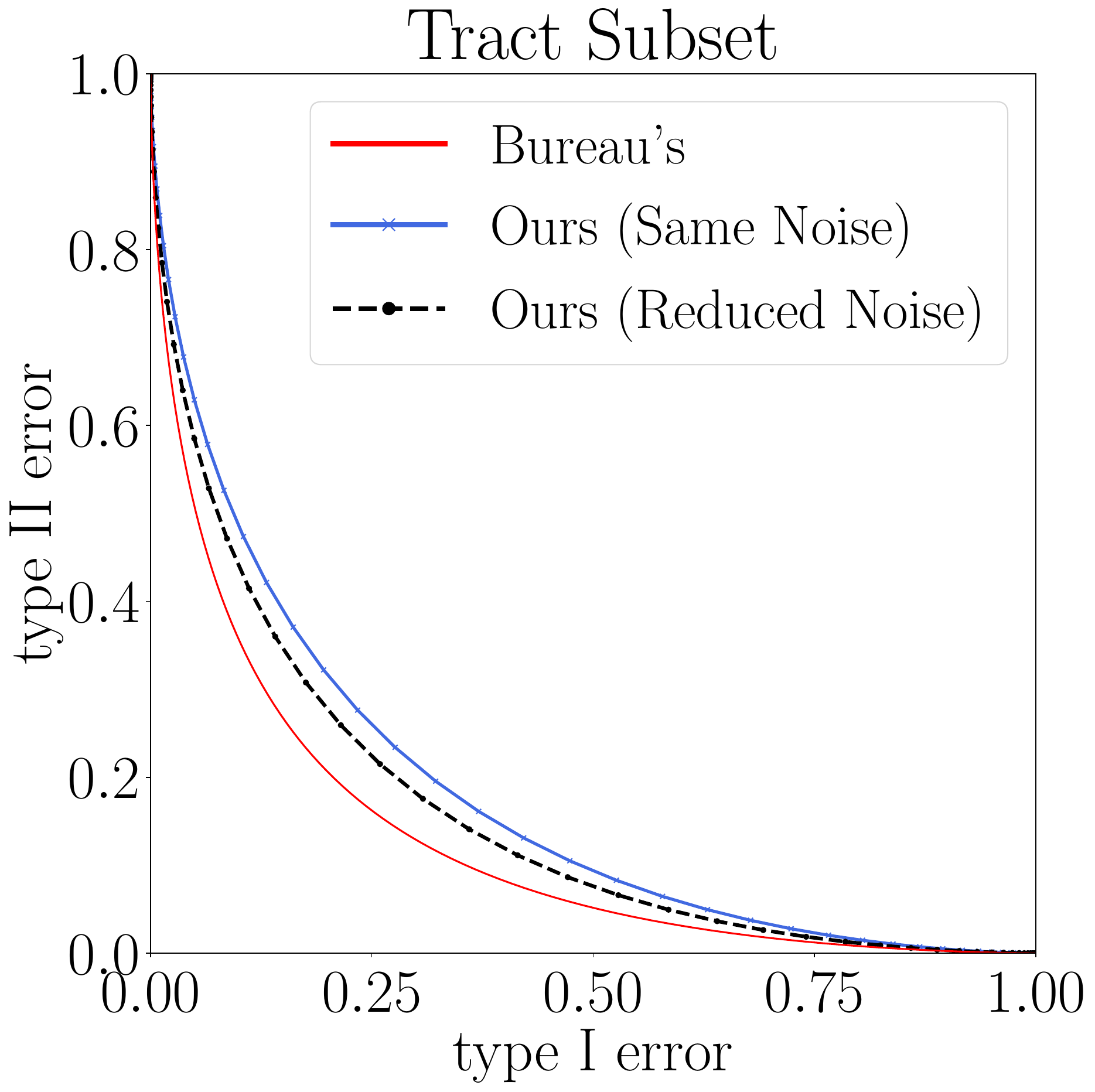}
    \end{subfigure}
        
    \hfill
    
    \begin{subfigure}[b]{0.232\textwidth}
        \includegraphics[width=\textwidth]{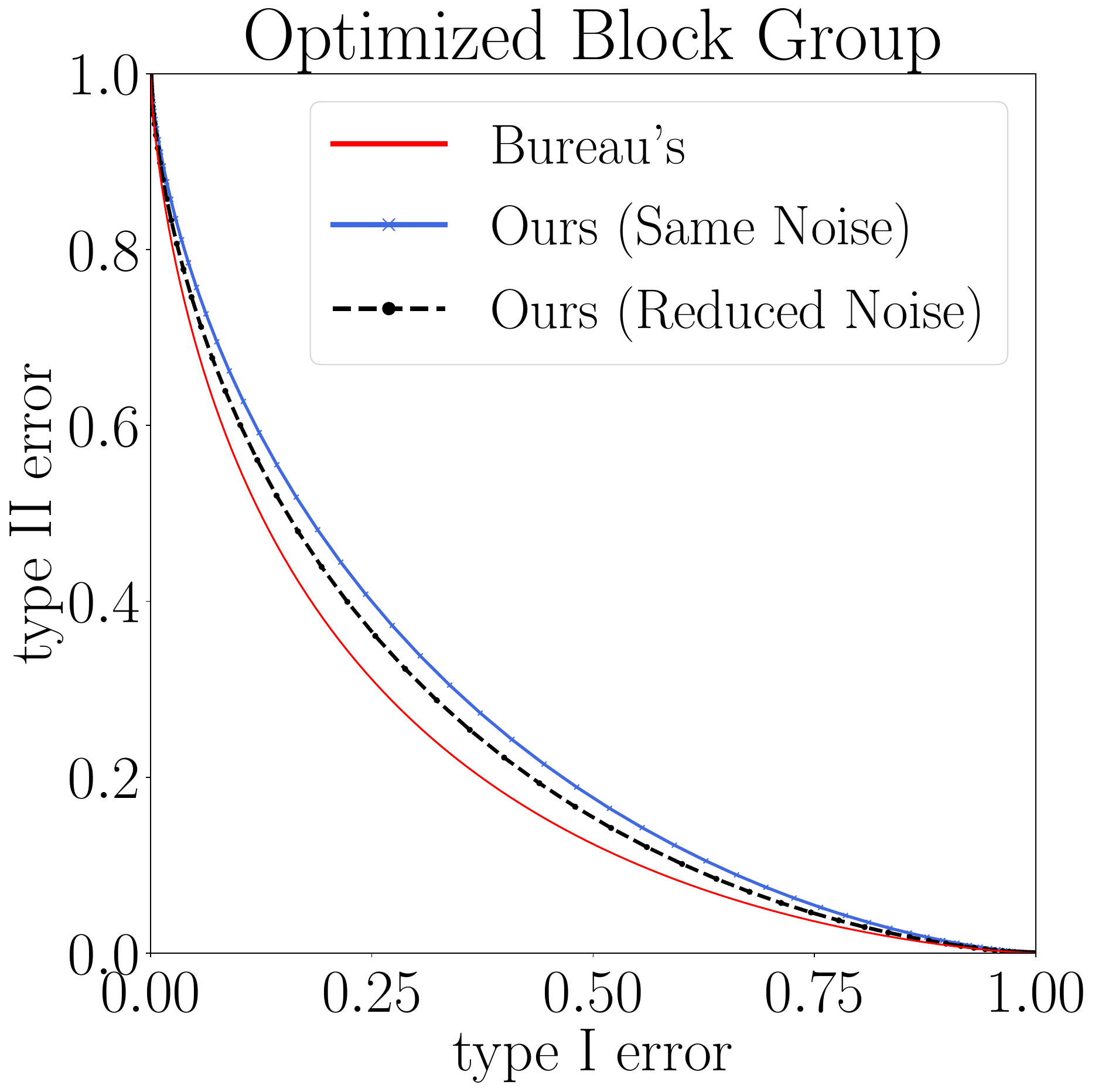}
    \end{subfigure}
    \begin{subfigure}[b]{0.232\textwidth}
        \includegraphics[width=\textwidth]{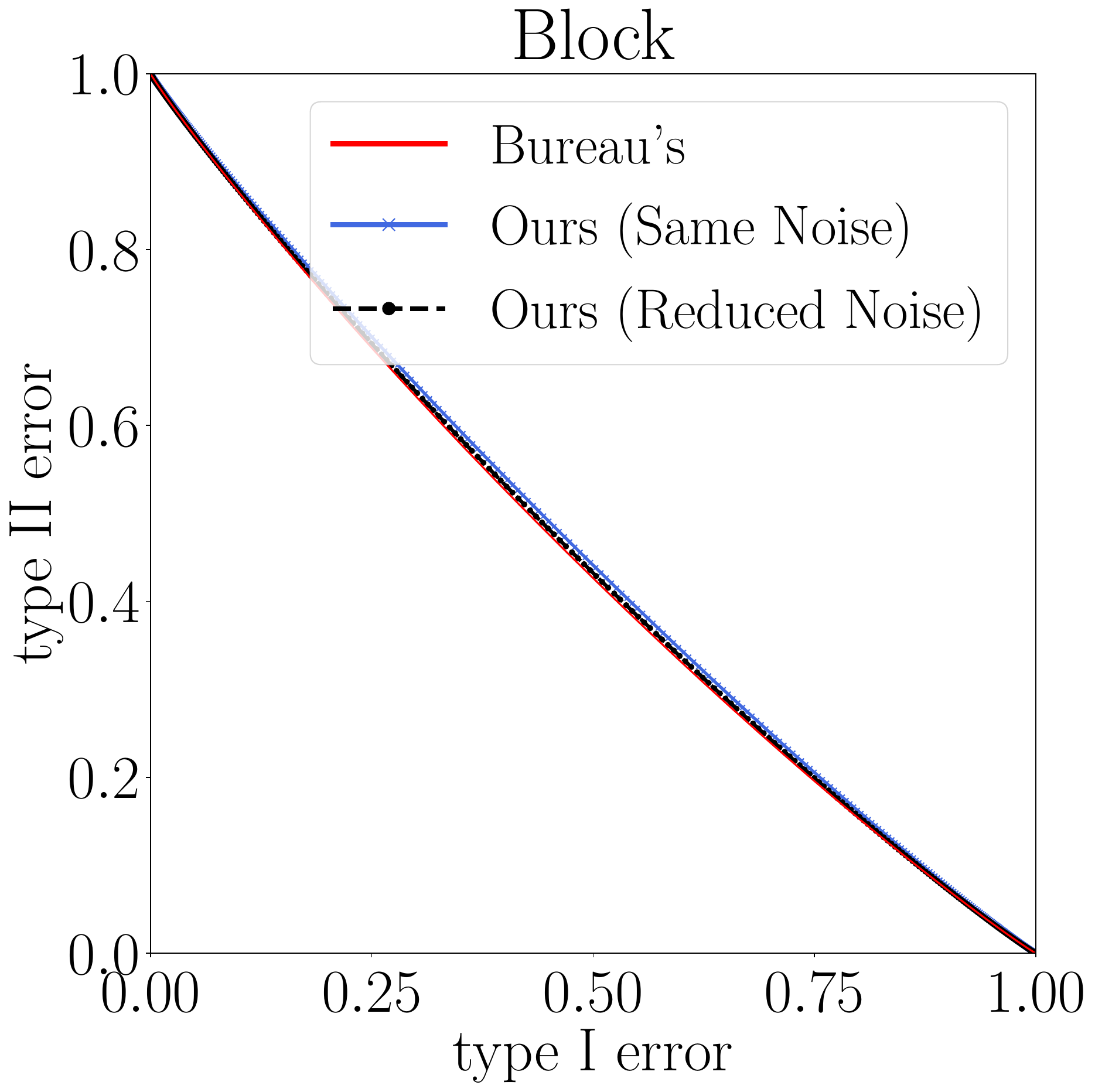}
    \end{subfigure}
    \caption{Comparison of trade-off functions \citep{Dong2022Gaussian} between our method (blue and black) and the Census Bureau's accounting method (red) across eight geographical levels of the 2020 {DHC}. The blue (Ours with Same Noise) and red (Bureau's) curves correspond to the same noise levels as in Figure~\ref{fig:eps_delta_geo} \citep{privacyallocation2022}, while the black (Ours with Reduced Noise) curves reflect noise levels reduced by 15.08\% to 24.82\% (details provided in Table~\ref{table:2020 allocation} in Section~\ref{sec:utility_high}). Zoomed-in views of the trade-off functions are available for the county level in Figure~\ref{fig:tradeoff_alphabeta_zoom_in} in Appendix~\ref{sec:supp_fig}. }
 \label{fig:alpha_beta_geo_more}
\end{figure}

The privacy guarantees of $(\epsilon, \delta)$-curves are equivalent to those of trade-off functions in $f$-DP. Comparisons between our method and the Bureau's in terms of trade-off functions are given in Figure~\ref{fig:alpha_beta_geo_more}. The trade-off function derived using our method lies uniformly above that derived from the Bureau's method, providing stronger privacy guarantees for each geographical level, consistent with the viewpoint of $(\epsilon, \delta)$-curve.

Furthermore, Figure \ref{fig:compare-ACS-5year} in Appendix \ref{sec:residual} evaluates our method's performance on the ACS 5-year estimates data \citep{Census2020ACSDP5Y2020.DP02, Census2020ACSDP5Y2020.DP03, Census2020ACSDP5Y2020.DP04, Census2020ACSDP5Y2020.DP05}. It demonstrates that our method offers even greater advantages compared to the Bureau's approach as the number of folds under composition increases.


\subsection{Enhanced accuracy while maintaining nearly the same privacy guarantee}
\label{sec:utility_high}

Using our new method of privacy accounting, the analysis in Section~\ref{sec:2020census} implies that noise levels can be reduced while still maintaining the original privacy guarantee for each geographical level. The reduced noise level should produce an $(\epsilon, \delta)$-curve that is at least as private as the Bureau's published guarantee over as large a range of $\delta$ as possible. Following this requirement, we determine the reduced noise level such that its $\epsilon$ value matches that of the Bureau's at $\delta = 10^{-11}$ for each geographical level (for example, $\epsilon = 2.79$ at the national level), which is displayed in Table \ref{table:2020 allocation}. The reduction in noise variances is substantial, ranging from 15.08\% (national level) to 24.82\% (block level).

\begin{table}[!htp]
\centering
\renewcommand{\arraystretch}{1.05}
\resizebox{0.46\textwidth}{!}{
\begin{tabular}{ccccc} 
 \hline
 \begin{tabular}{@{}c@{}}Geographical \\ levels \end{tabular} & US & State & County & PEPG  \\
 
 \hline
 \begin{tabular}{@{}c@{}} Bureau's \end{tabular} & $68.49$ & $ 5.00$& $16.12$ & $10.46$ \\

 \hline
 \begin{tabular}{@{}c@{}} Ours \end{tabular} & $54.19$ & $4.25$& $13.28$ & $8.72$ \\ 

  \hline
 \begin{tabular}{@{}c@{}} Reduction \end{tabular} & $20.88 \%$ & $15.08\%$& $17.58\%$ & $16.62\%$ \\ 

 \hline\hline
 \begin{tabular}{@{}c@{}}Geographical \\ levels \end{tabular} & \begin{tabular}{@{}c@{}}Tract Subset \\ Group \end{tabular} & Tract Subset & \begin{tabular}{@{}c@{}}Optimized Block \\ Group \end{tabular} & Block\\
 \hline
 \begin{tabular}{@{}c@{}} Bureau's \end{tabular} & $10.46$ & $ 5.76$& $11.61$ & $456.62$\\
 
 \hline
 \begin{tabular}{@{}c@{}} Ours \end{tabular} & $8.72$ & $4.87$& $9.65$ & $343.27$\\
  \hline
 \begin{tabular}{@{}c@{}} Reduction \end{tabular} & $16.62\%$ & $15.33\%$& $16.89\%$ & $24.82\%$\\
 \hline
\end{tabular}}
\caption{Reduced injected noise for {the 2020 DHC} tabulations while maintaining the same privacy guarantee using our method. The comparison is based on the variance proxy ($\sigma^2$) of the discrete Gaussian noise. The rows corresponding to ``Bureau's'' represent the version of the privacy-loss budget allocation released by the Bureau on August 25, 2022 \citep{privacyallocation2022}.}
\label{table:2020 allocation}
\end{table}

\begin{figure}[!htp]
    \centering
    \begin{subfigure}[b]{0.239\textwidth}
        \includegraphics[width=\textwidth]{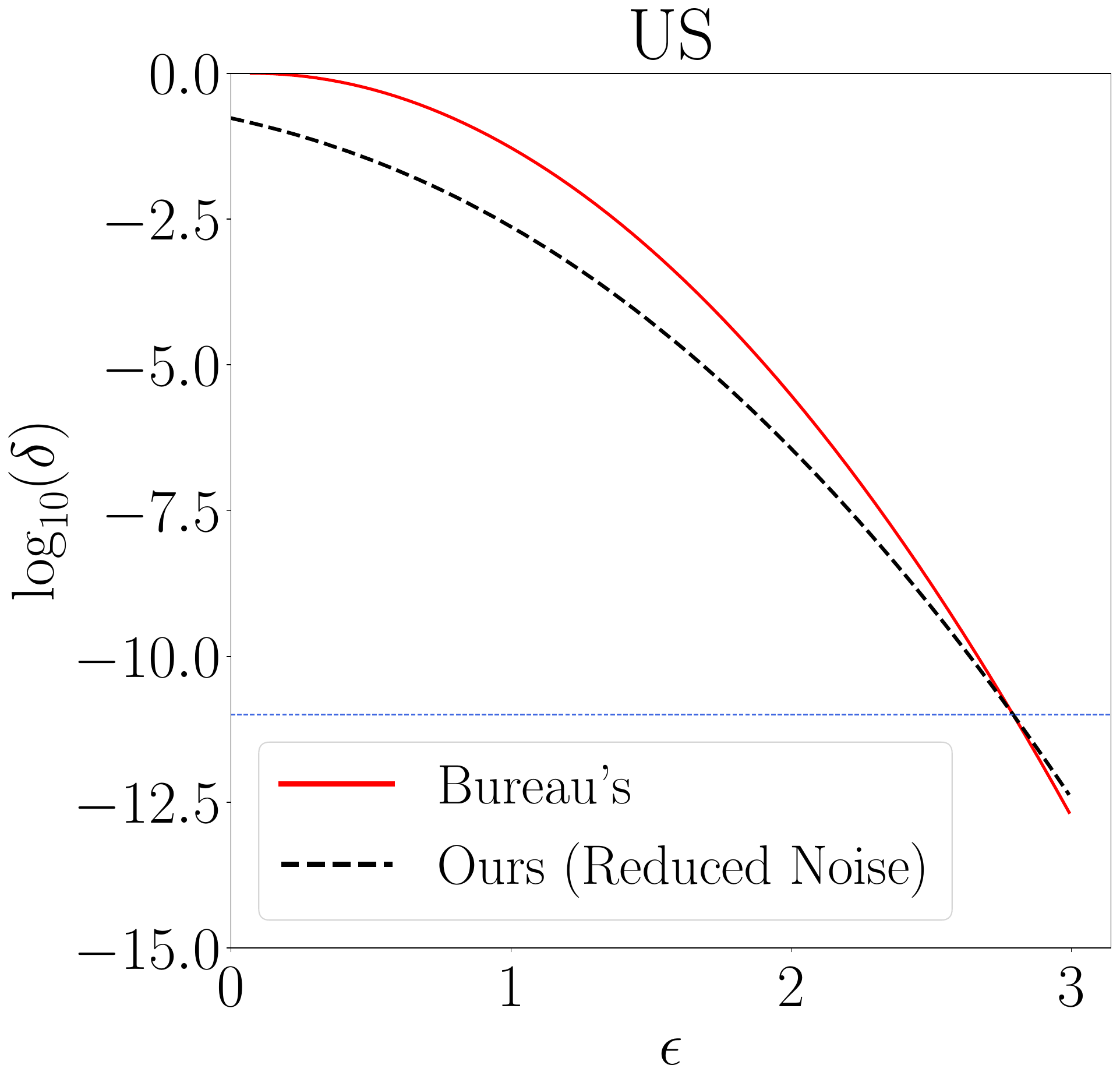}
    \end{subfigure}
    \begin{subfigure}[b]{0.239\textwidth}
        \includegraphics[width=\textwidth]{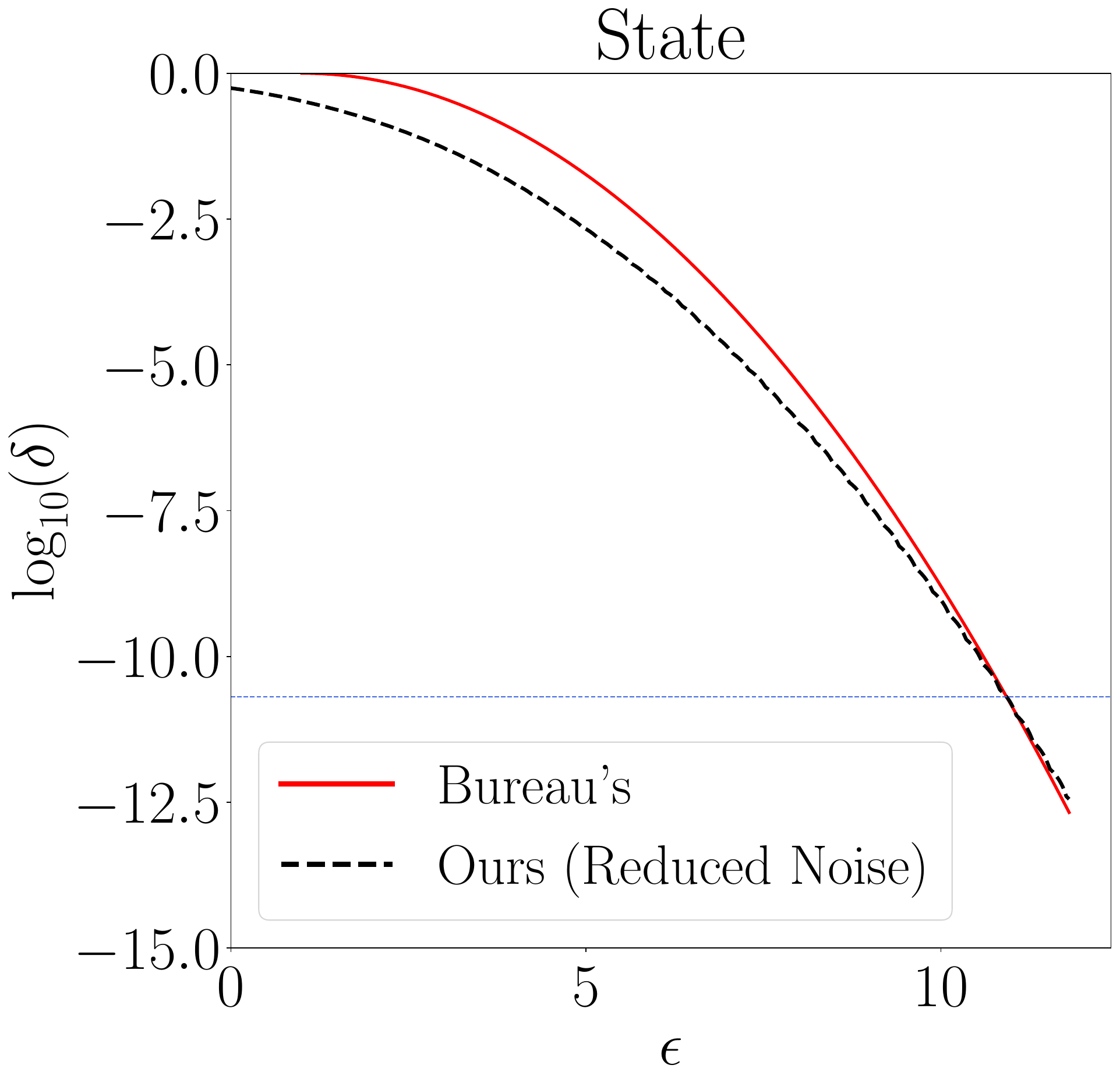}
    \end{subfigure}
        
    \hfill
    
    \begin{subfigure}[b]{0.239\textwidth}
        \includegraphics[width=\textwidth]{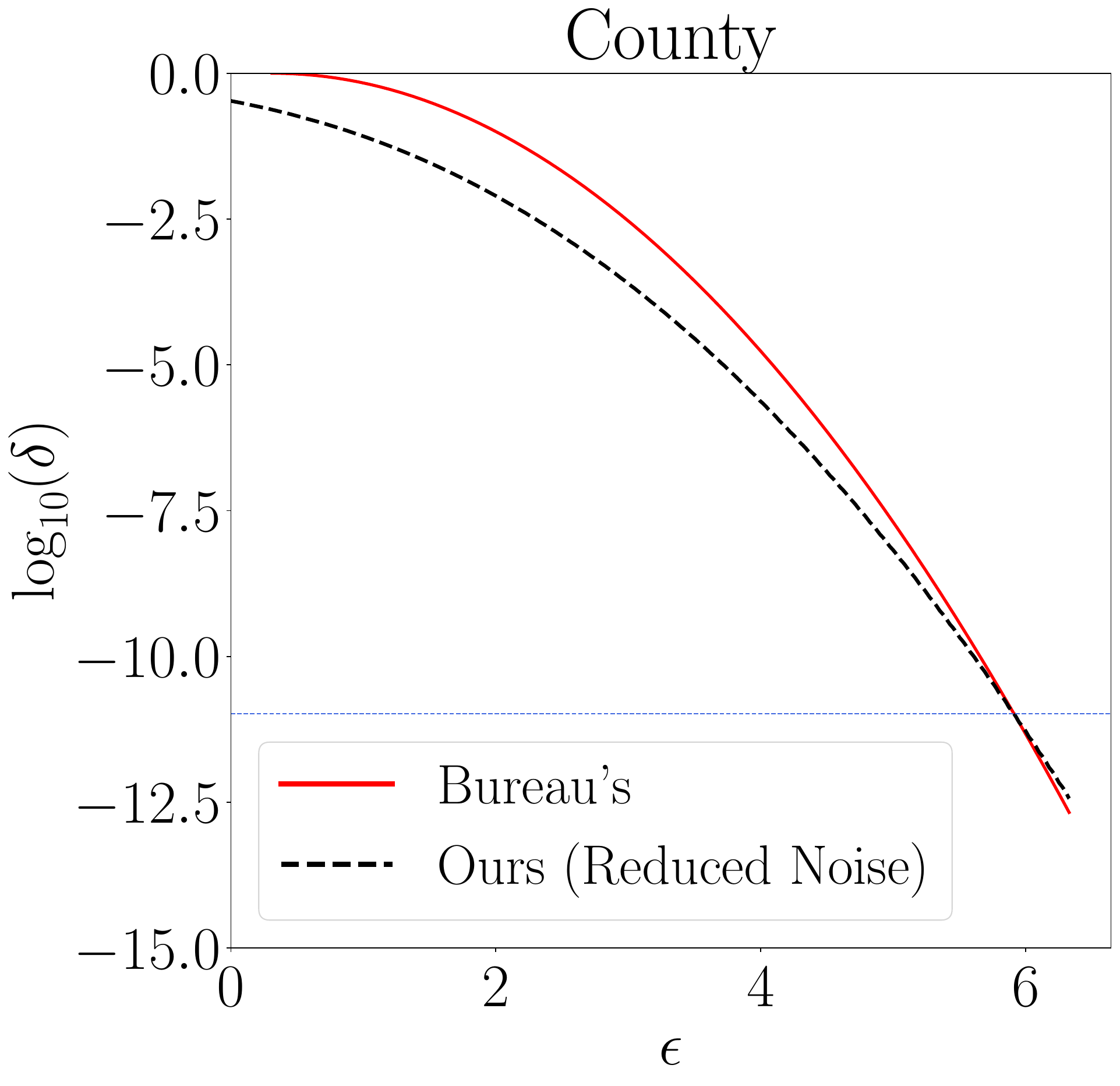}
    \end{subfigure}
    \begin{subfigure}[b]{0.239\textwidth}
        \includegraphics[width=\textwidth]{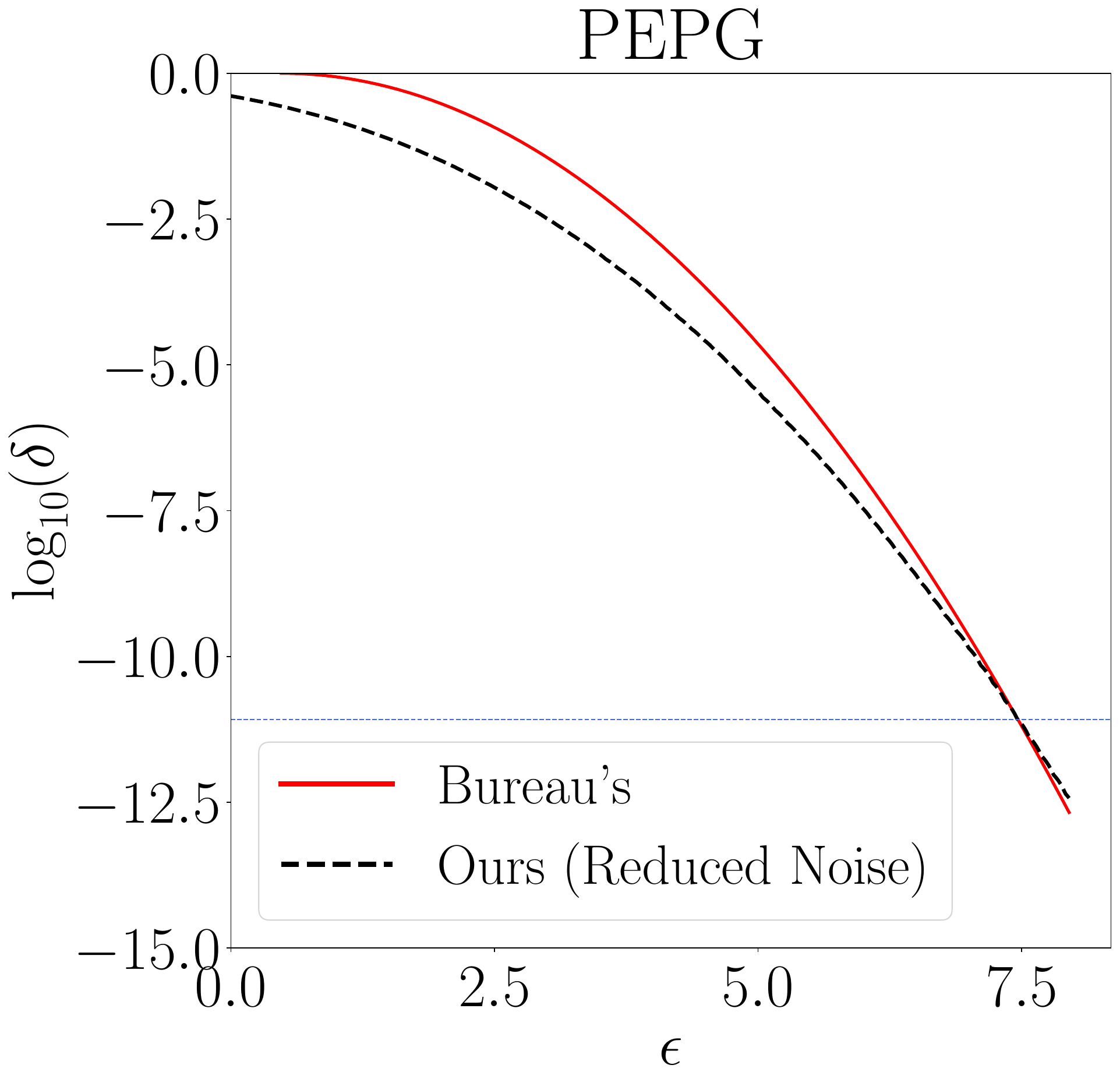}
    \end{subfigure}
    
    \hfill
    
    \begin{subfigure}[b]{0.239\textwidth}
        \includegraphics[width=\textwidth]{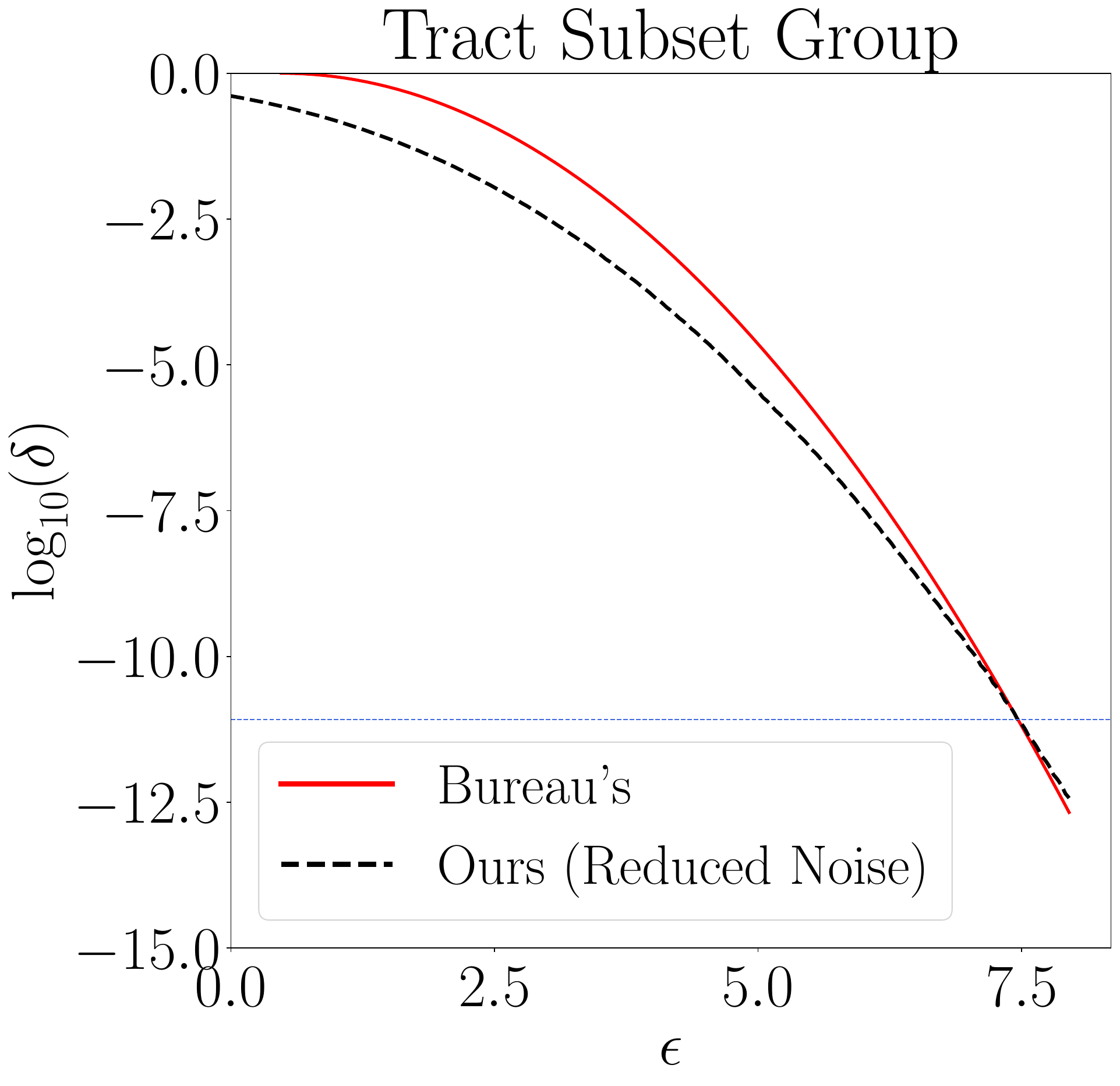}
    \end{subfigure}
    \begin{subfigure}[b]{0.239\textwidth}
        \includegraphics[width=\textwidth]{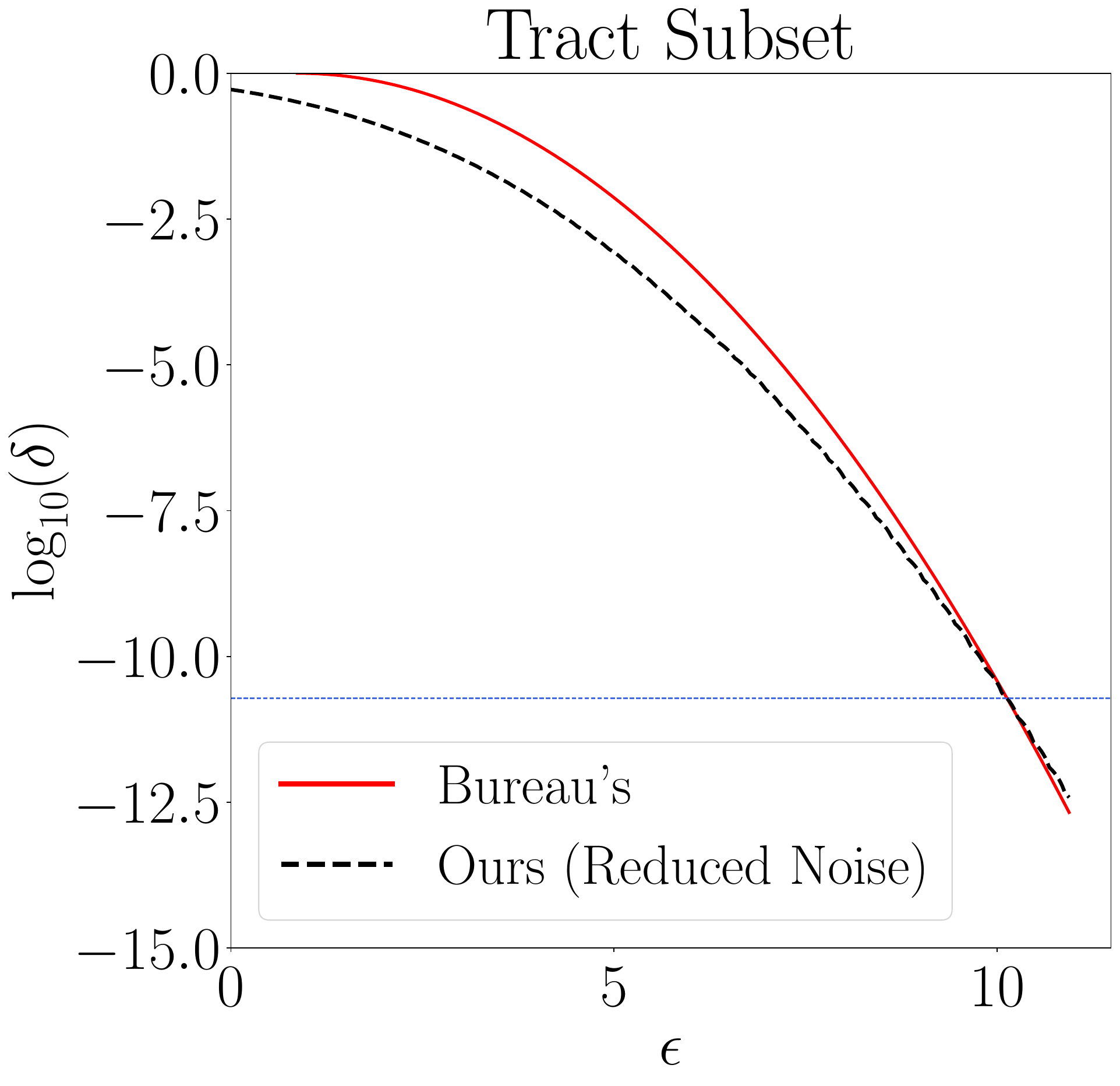}
    \end{subfigure}
        
    \hfill
    
    \begin{subfigure}[b]{0.239\textwidth}
        \includegraphics[width=\textwidth]{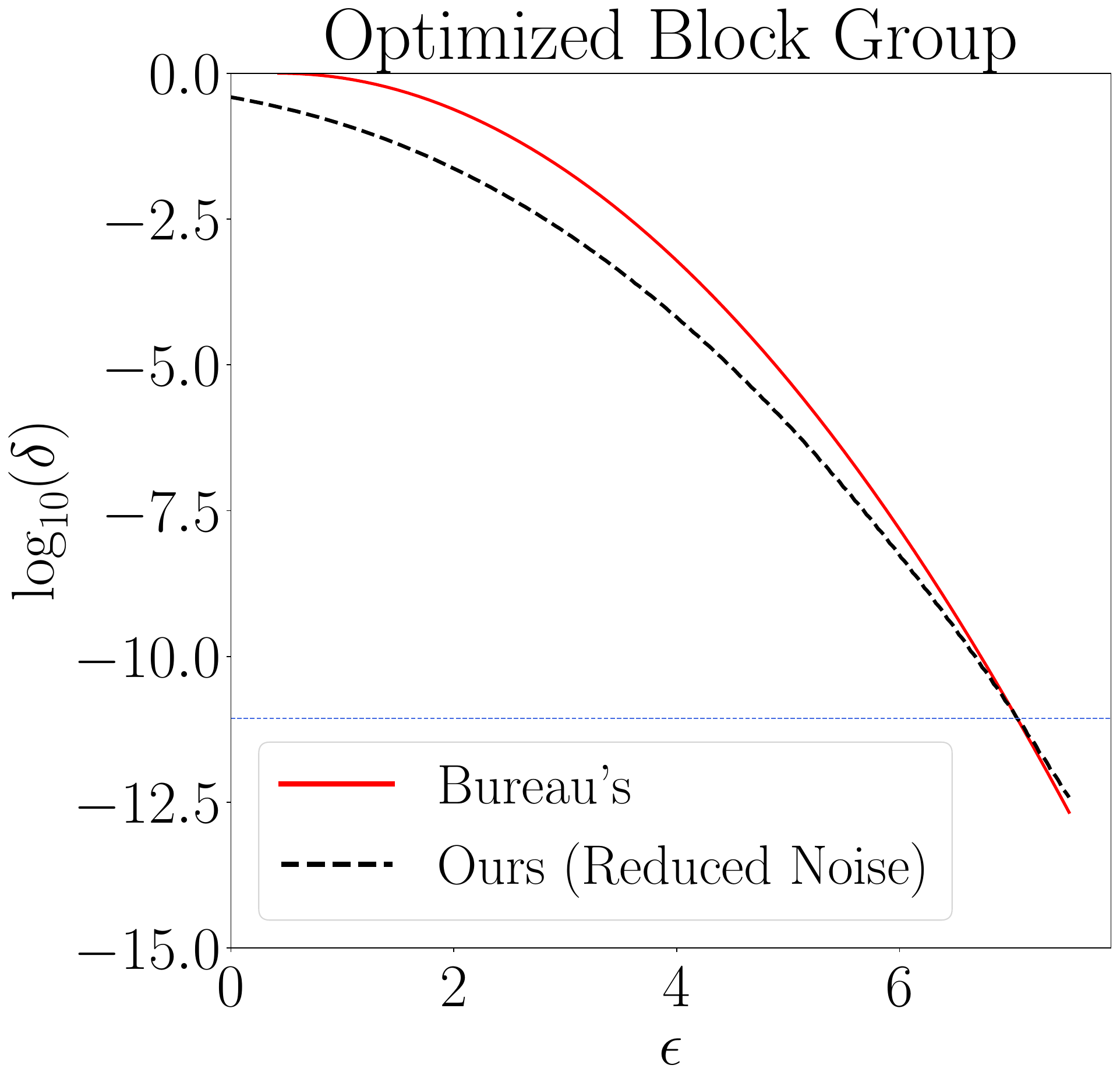}
    \end{subfigure}
    \begin{subfigure}[b]{0.239\textwidth}
        \includegraphics[width=\textwidth]{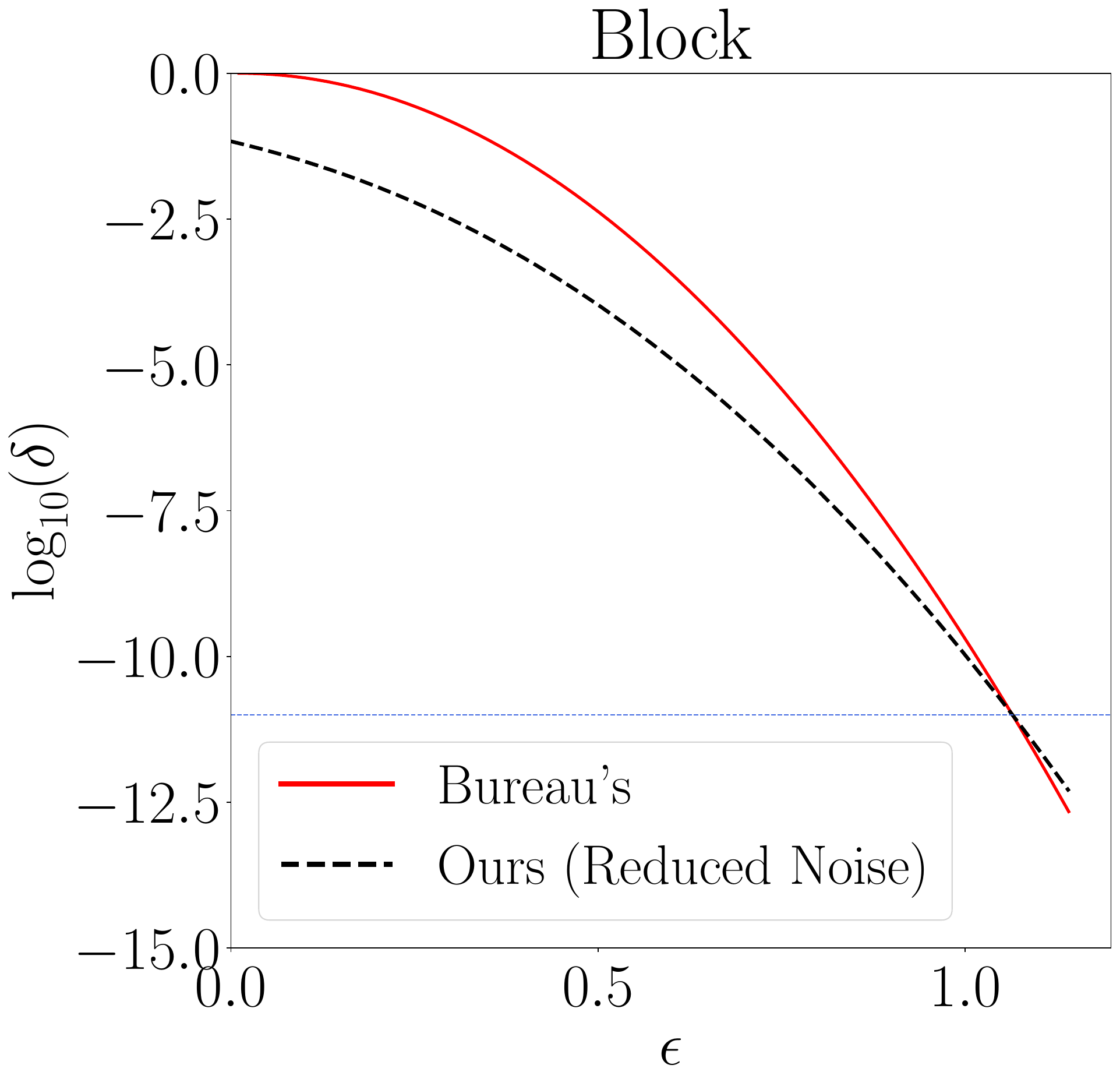}
    \end{subfigure}
 \caption{Trade-off functions with reduced noise levels (our method) and those without noise reduction (Bureau's). The noise levels used by both are shown in Table~\ref{table:2020 allocation}. For each geographical level, the two trade-off functions intersect at $\delta = 10^{-11}$.\vspace{12.5mm}}
    \label{fig:eps_delta_geo_reduce_noise_more}
\end{figure}

Figure~\ref{fig:eps_delta_geo_reduce_noise_more} shows the new $(\epsilon, \delta(\epsilon))$-curves using our method with reduced noise levels, in addition to the Bureau's $(\epsilon, \delta(\epsilon))$-curves using the original (larger) noise levels. For any geographical level, our $\epsilon$ value is smaller than that of the Bureau as long as $\delta > 10^{-11}$, and the gap is significant when $\delta$ is not too small. When $\delta < 10^{-11}$, our $\epsilon$ becomes larger than that of the Bureau's. However, this reversal at such small values of $\delta$ arguably does not affect the interpretation of privacy guarantees. Thus, our privacy guarantee with a smaller noise level is practically at least as strong as the Bureau's with a significantly larger noise level for each geographical level. In fact, one can match the value of $\epsilon$ at any value of $\delta$---for example, taking a value even smaller than $10^{-11}$---and the comparison remains the same. That being said, an interesting research direction is to understand in a principled manner how this reversal at a small value of $\delta$ affects privacy interpretation.

We also present the trade-off functions with noise reduction in Figure~\ref{fig:alpha_beta_geo_more}. The new trade-off functions (black) lie above those derived using the Bureau's method without noise reduction, thereby offering stronger privacy guarantees, except for very small regions near the endpoints that are difficult to discern visually. Interested readers can refer to Figure~\ref{fig:tradeoff_alphabeta_zoom_in} in Appendix~\ref{sec:supp_fig}, which illustrates how these two trade-off functions intersect in the case of County.






\begin{figure}[t]
    \centering
    \includegraphics[width=0.5\textwidth]{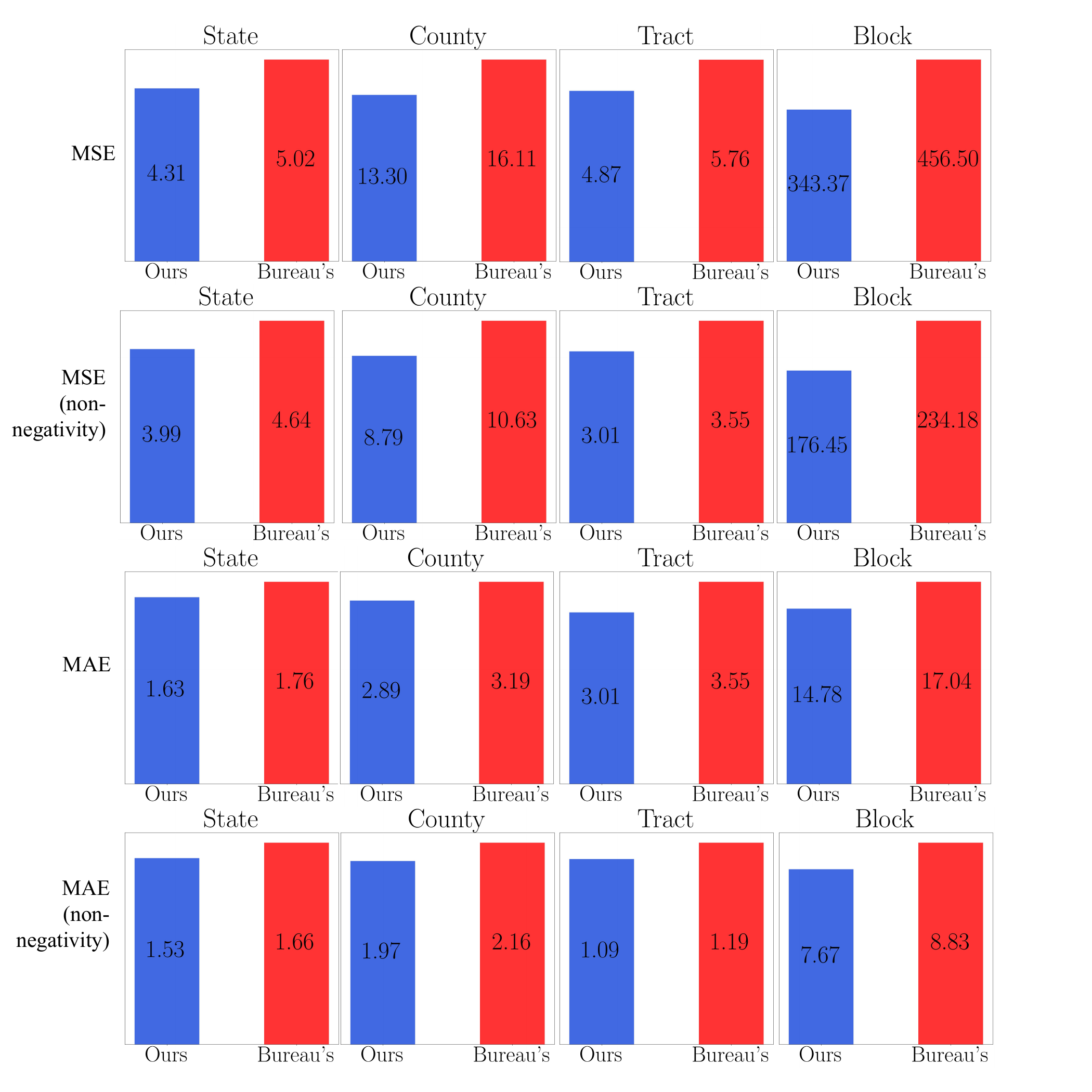}
    \caption{Enhanced accuracy of the 2010 U.S.\ Census in Pennsylvania, measured by MSE and MAE, with and without non-negativity post-processing. Results are grouped by geographical level. For illustration, only the simplest non-negative post-processing is applied. Noise variances for the discrete Gaussian distribution are as specified in Table~\ref{table:2020 allocation}. Our method (blue) consistently shows lower errors compared to the Bureau's approach (red) for all geographical levels.}
    \label{fig:bias_geo_level}
\end{figure}

Formally, the Bureau releases census data by post-processing the NMF through the DAS. To examine how these reduced noise levels translate into improved estimation performance through post-processing, we conduct an analysis using the demographic and housing characteristics file from the 2010 Decennial Census \citep{privacyprotected2010}. Figure~\ref{fig:bias_geo_level} shows our results for the geographical levels of state, county, tract, and block in Pennsylvania, comparing the MSE\footnote{Let $y_{2010}$ and $y_{2020}$ denote the non-privatized 2010 and 2020 Census Summary Files, respectively. Similarly, let $y'_{2010}$ and $y'_{2020}$ denote the simulated privacy-protected Summary Files generated from $y_{2010}$ and $y_{2020}$. To evaluate the performance of a privacy protection mechanism, we compute the MSE and MAE using the difference $(y'_{2010} - y_{2010})^2$ and $|y'_{2010} - y_{2010}|$, respectively. These metrics can be directly and precisely calculated using the publicly available 2010 Census Summary Files. In contrast, the conventional definitions of MSE and MAE, given by $\mathbb{E}(y'_{2020} - y_{2020})^2$ and $\mathbb{E}|y'_{2020} - y_{2020}|$, are not directly observable in practice.} and the mean absolute error (MAE) with the simplest possible post-processing of preserving non-negativity. Without post-processing, our method reduces the MSE by 14.14\%, 17.44\%, 15.45\%, and 24.78\%, and the MAE by 7.39\%, 9.40\%, 8.47\%, and 13.26\% for the state, county, tract, and block levels, respectively. With non-negative post-processing, the MSE is reduced by 14.01\%, 17.31\%, 15.21\%, and 24.65\%, and the MAE by 7.83\%, 8.80\%, 8.40\%, and 13.14\%, correspondingly. These results consistently demonstrate that our method can reduce the error introduced by DP constraints, thereby enhancing utility across all geographical levels. Notably, the most significant improvement occurs at the block level, where the noise---and thus the privacy protection---is the greatest.


\subsection{Improved overall privacy guarantee}
\label{sec:enhanced_overall}

\begin{figure}[!htp]
    \centering
    \includegraphics[height=0.25\textwidth,width=0.3\textwidth]{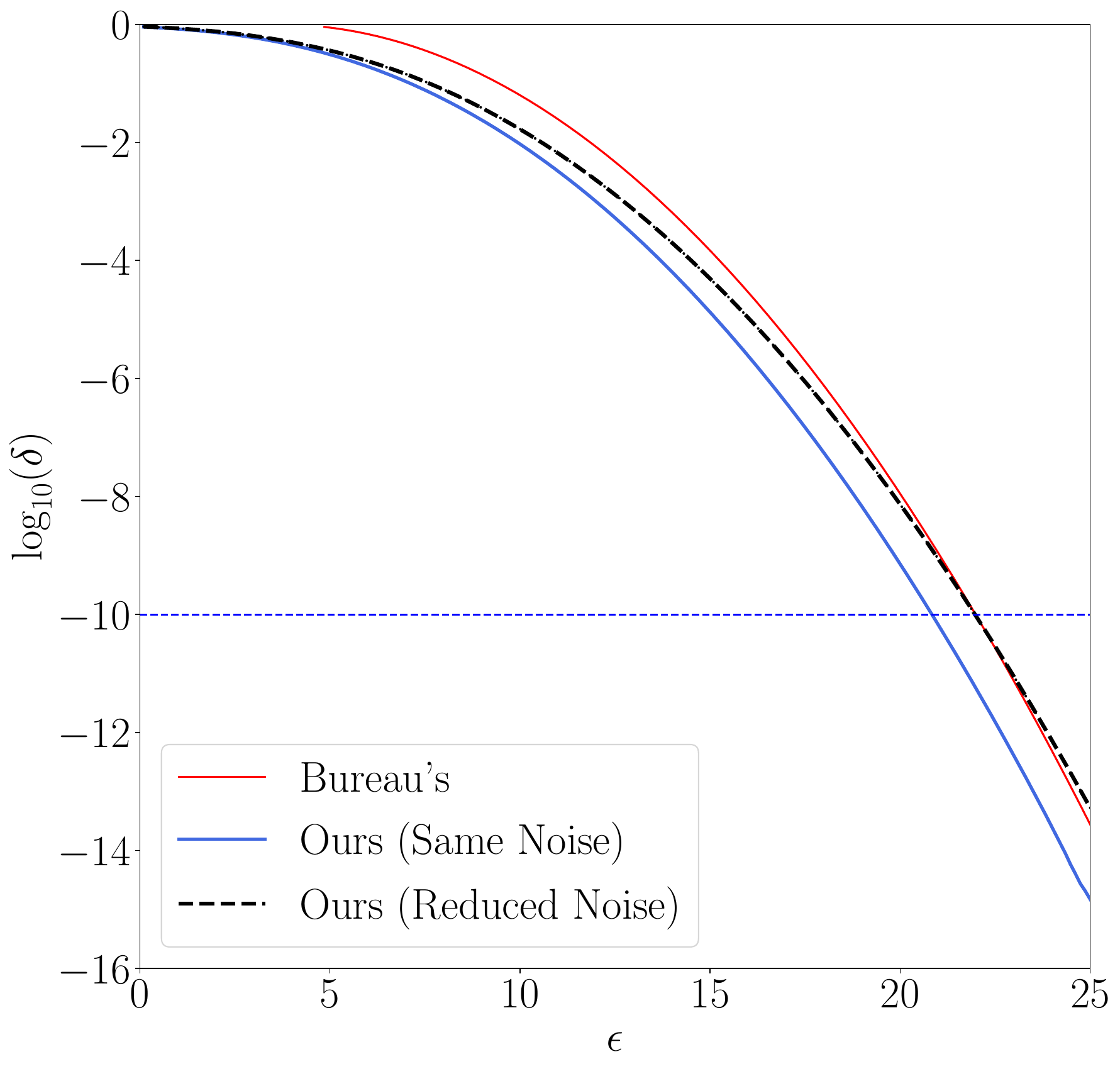}
    \caption{$(\epsilon, \delta(\epsilon))$-curves under composition of all eight geographical levels of the 2020 U.S.\ Census. The black curve uses variance proxy that is reduced by 8.59\%. The comparison in terms of trade-off function is shown in Figure \ref{fig:tradeoff_alphabeta} in Appendix~\ref{sec:supp_fig}.}
    \label{fig:tradeoff_epsdelta}
\end{figure}

We now consider the overall privacy guarantee across all eight geographical levels  {for the 2020 DHC}. Using the noise levels in Table \ref{table:2020 allocation} (the row corresponding to the Bureau's approach), we present in Figure~\ref{fig:tradeoff_epsdelta} the $(\epsilon, \delta)$-curves for both our accounting method and the Bureau's method under the composition of the eight geographical levels. Our method gives a smaller value of $\epsilon$ at any value of $\delta$ than the Bureau's method, thereby providing a stronger overall privacy guarantee under composition.




As with Section~\ref{sec:utility_high}, we can reduce the noise level such that the overall privacy parameter $\epsilon$ matches that of the Bureau's method at a certain value of $\delta$, say, $10^{-10}$. Our analysis demonstrates that this allows for reducing the variance proxy parameter $\sigma^2$ across all geographical levels by 8.59\%. As shown in Figure~\ref{fig:tradeoff_epsdelta}, the new $(\epsilon, \delta)$-curve with noise reduction has a smaller $\epsilon$ than the curve computed using the Bureau's method with the original noise level for $\delta > 10^{-10}$. Since $10^{-10}$ is smaller than the reciprocal of the U.S.\ population, our method arguably provides at least the same level of privacy guarantee while injecting less noise into the census data.




We employ a two-step process to compose the total privacy cost over $8 \times 10$ folds under composition. The first step, as outlined in Sections \ref{sec:2020census} and \ref{sec:utility_high}, accounts for the privacy guarantee at each geographical level. For the second step, which aggregates across different levels, a technical challenge arises due to the heterogeneity of injected noise across these levels. We overcome this challenge by leveraging a probabilistic characterization of discrete Gaussian distributions, enabling us to maintain high precision when aggregating across geographical levels. Full technical details are provided in Appendix \ref{sec:inid-overall}.

\subsection{Mitigating distortion in downstream analyses}
We examine how our $f$-DP based accounting method improves the accuracy and reliability of downstream analyses using census data. The underlying intuition is that, while maintaining the same privacy guarantee, the use of $f$-DP accounting allows for reduced noise added to census counts. To illustrate this, we analyze 
{ 1) MSE between the non-privatized 2010 Census Summary Files and the simulated privacy-protected Summary Files after non-negative postprocessing, across nine racial query; 2) the relationship between earnings and education level using the 2020 ACS 5-year estimates \citep{Census2020ACSST5Y2020.S1501,Muller23}.}

\subsubsection{Impact on racial group counts}
\label{sec:2010_downstream}

\begin{figure}[!htp]
    \centering
    \includegraphics[width=0.53\textwidth]{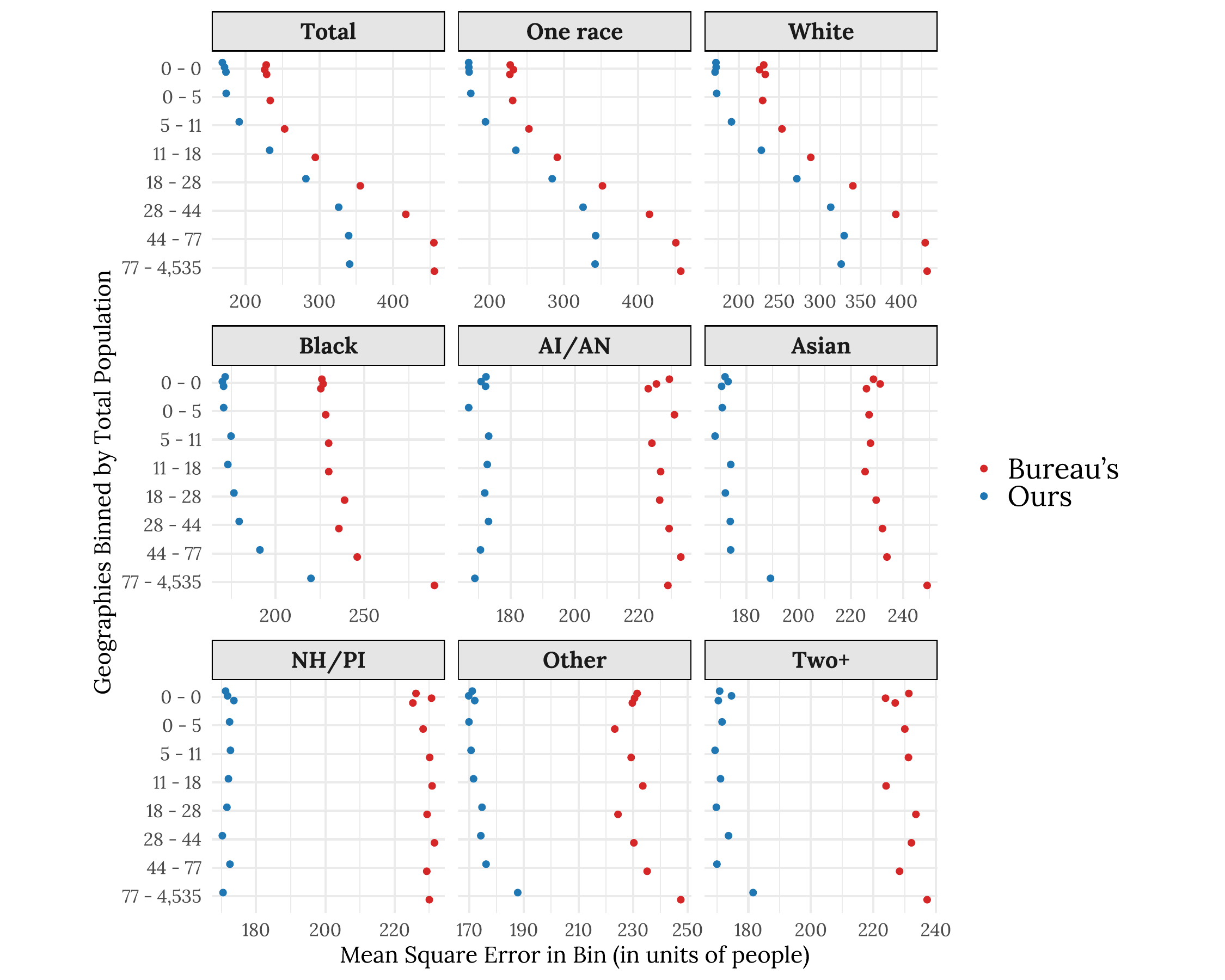}
    \caption{MSE for block-level population counts by racial groups in Pennsylvania. The MSE measures the average magnitude of deviation between the non-privatized 2010 Census Summary Files and the simulated privacy-protected Summary Files after non-negative postprocessing. The abbreviations ``AI/AN'' and ``NH/PI'' denote ``Population of one race: American Indian and Alaska Native alone'' and ``Population of one race: Native Hawaiian and Other Pacific Islander alone'', respectively.}
    \label{fig:DHC_simulation}
\end{figure}

{ We analyze the MSE between the non-privatized 2010 Census Summary Files and our simulated privacy-protected Summary Files \citep{imai2023bias}. For our clean non-privatized data, we use population counts from the 2010 Census across 421,545 blocks in Pennsylvania, with nine racial queries. To simulate the privacy protection mechanism, we add discrete Gaussian noise $\mathcal{N}_{\ZZ}(0, \sigma^2)$, where the proxy variance $\sigma^2$ is derived from the block-level allocations detailed in Table~\ref{table:2020 allocation}, followed by a non-negative postprocessing step. Figure~\ref{fig:DHC_simulation} clearly indicates that our proposed allocation method consistently achieves lower MSE compared to the Bureau’s official implementation. Moreover, the MSE reduction is notably more substantial for blocks with larger populations, as demonstrated in the first three figures shown in Figure~\ref{fig:DHC_simulation}.}

\subsubsection{Impact on education-level and earnings analysis}

\begin{figure*}[!htp]
  \centering
  \includegraphics[width=0.70\textwidth]{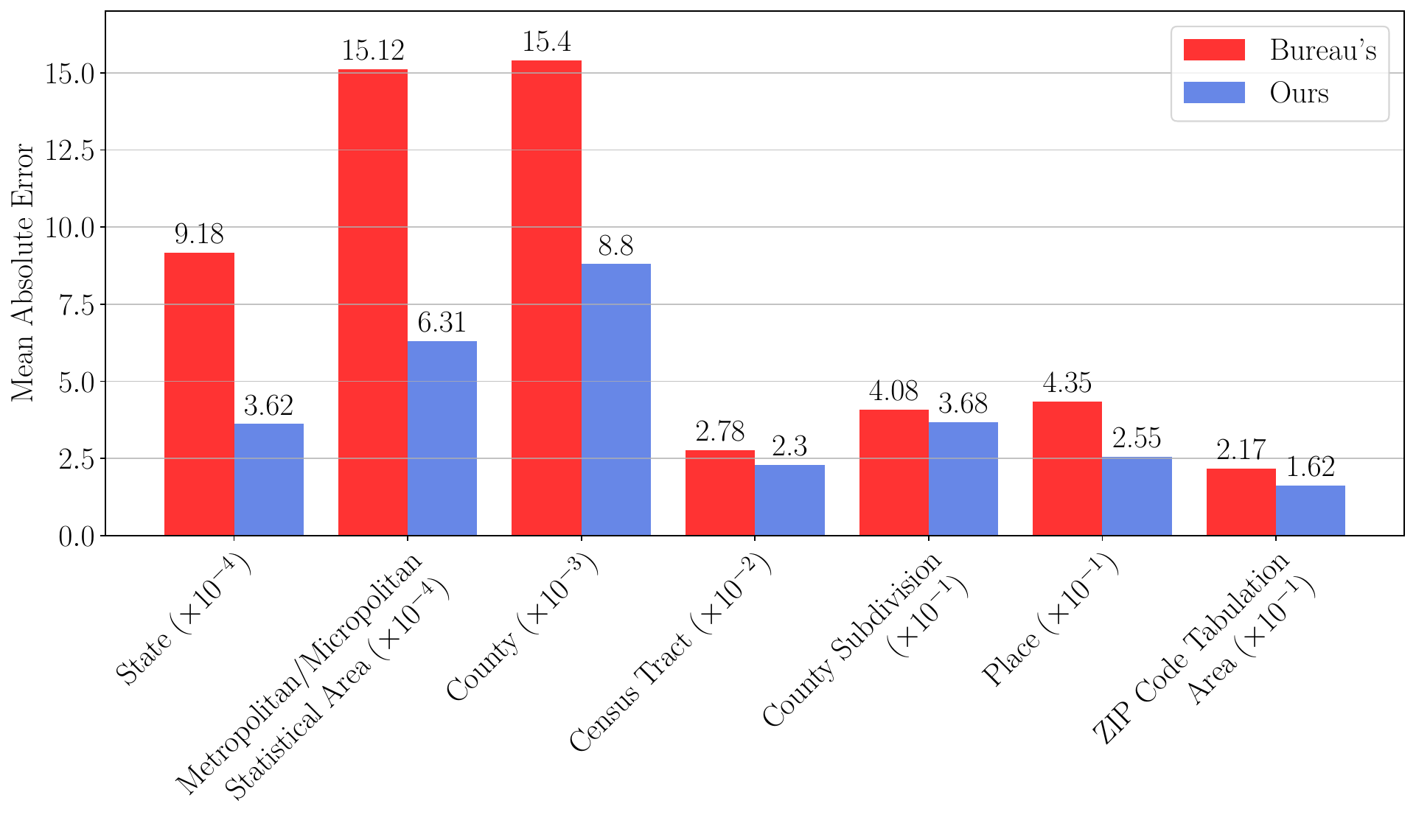}
  \caption{Reduced distortion in estimating the slope coefficient due to privacy constraints for downstream analysis of private census data, measured in terms of MAE \eqref{eq:kbeta}. Noise variances follow the configuration specified in Table \ref{table:2020 allocation}.}
  \label{fig:down_task}
\end{figure*}

We fit a simple linear regression model of the form $y = \beta x + \alpha$, where $y$ represents the median earnings in a geographical area (state, metropolitan/micropolitan statistical area, county, census tract, county subdivision, place, or ZIP code tabulation area), and $x$ denotes the proportion of individuals with a bachelor's degree or higher in the same area. Let $\hat\beta$ be the estimate of the slope coefficient. We then add discrete Gaussian noise with variance proxy parameter $\sigma^2$ to each of the six education-level categories in the ACS 5-year estimates: less than 9th grade, 9th to 12th grade (no diploma), high school graduate (including equivalency), some college (no degree), associate's degree, and bachelor's degree or higher, followed by non-negativity post-processing. Let $\hat{\beta}_{\sigma}$ be the slope coefficient obtained by regressing the median earnings $y$ on the proportion $x$ computed from the noise-added data.

For a given privacy parameter, we determine the proxy standard deviation $\sigma$ using either our $f$-DP accounting method or the Census Bureau's method. Our approach consistently results in a smaller $\sigma$ across all possible scenarios. To assess the accuracy of the slope coefficient derived from privatized data, we calculate the MAE between the original slope estimate $\widehat{\beta}$ and its privatized counterpart $\widehat{\beta}_\sigma$:\footnote{It is worth noting that the analyst does not account for the noise distribution in the privatization process. However, the effectiveness of analysis can often be enhanced by incorporating the distribution (see examples in \cite{cumings2024full, drechsler2022nonparametric, awan2024best}).} 
\begin{equation}\label{eq:kbeta}
\frac{1}{K}\sum_{i=1}^K\left|\widehat{\beta}_\sigma^{(i)} - \widehat{\beta} \right|,
\end{equation} 
where $K$ is the number of independent trials, and $\widehat{\beta}_\sigma^{(i)}$ is the estimate from the $i$-th run. Setting $K = 3$, Figure~\ref{fig:down_task} demonstrates that our method significantly reduces the distortion of the privatized estimates relative to the original estimate at every geographical level, from the state level down to the ZIP code tabulation area. At the state level, for instance, our method reduces the MAE by 60.57\% compared to the Bureau's method.

\section{Discussion}
\label{sec:discussion}

In this paper, we have analyzed the privacy guarantees of the 2020 U.S.\ Census in comparison with the privacy levels published by the Census Bureau. Our analysis demonstrates that the actual privacy guarantee is significantly stronger than that provided by the Bureau's existing approach, as evidenced by our uniformly smaller $\epsilon$ value for any $\delta$. This discovery of underestimated privacy by the Census Bureau was made possible through a novel application of the $f$-DP framework to the compositional structure of the U.S.\ Census, addressing an open problem posed by the Census Bureau \citep{kifer2022bayesian}.

Our analysis indicates that less noise can be injected into census data while maintaining nearly the same privacy guarantee. We have empirically demonstrated that our method would enhance the accuracy of census data and substantially reduce distortion due to privacy constraints in downstream analyses. Given the widespread use of census data across social science \citep{Sullivan2020Coming}, political science \citep{ansolabehere2008end,cohen2021census}, and economics \citep{autor2003rise,bureau2023guidance}, we anticipate numerous opportunities to leverage this improved privacy-utility trade-off established in our work.



An important future direction is the development of numerically accurate and computationally efficient accounting methods for heterogeneous privacy budget allocation, where the injected discrete Gaussian noise has varying variance under composition. This research direction is motivated by the observation that released NMFs in products such as the demographic and housing characteristics (DHC) file often exhibit heterogeneous composition structures, even within the same geographical level \citep{Cumings2024geographic}. A significant challenge in this direction is that composition structures in these applications are often complex and not fully detailed in their reports \citep{Census2023implementation}. Furthermore, even when the composition structure is known, both the accuracy and computational efficiency of our $f$-DP based accounting method deteriorate for heterogeneous allocations, as evidenced by the comparison between results in Sections~\ref{sec:2020census} and \ref{sec:enhanced_overall}. This degradation arises from the need to address precision issues in floating-point arithmetic (see further discussion in Appendix \ref{sec:numerical_difficulty}). Neither purely analytical accounting methods \citep{kairouz2021distributed,zhu2022optimal} nor purely numerical methods \citep{Koskela2020computing,gopi2021numerical} alone resolve these issues. For instance, numerical methods are computationally infeasible for achieving the same level of accuracy as our method for the 2020 U.S.\ Census, regardless of whether the allocation is homogeneous or heterogeneous (see elaboration in Appendix \ref{rmk:computation_time}). A potential research avenue is to integrate these two types of accounting methods, which we leave for future work.

{\small
\subsection*{Acknowledgments}

We thank Simson Garfinkel for valuable information on the implementation of the DAS and Thomas Steinke for insightful comments on an early version of the manuscript. We are also grateful to Jeremy Hsu for reporting our findings in \textit{New Scientist} and to Daniel Kifer and Philip Leclerc for beneficial discussions on the composition structure of the U.S.\ Census and its privacy interpretation. Their feedback allowed us to improve the presentation of this paper on top of the version posted on arXiv in October 2024. This work was supported in part by NSF DMS-2310679, a Meta Faculty Research Award, and Wharton AI for Business.




\bibliography{References}
}

\newpage
\appendix
\onecolumn


\section{Technical proofs and details}
\label{sec:method}

This section presents our main methodology and key tools for deriving the privacy profile of the U.S. Census. Section \ref{sec:background} provides an overview of differential privacy. Section \ref{sec:useful_facts} summarizes technical facts about discrete Gaussian distributions used to bound approximation errors. In Section \ref{sec:fDP-DGM}, we present $f$-DP guarantees for the discrete Gaussian mechanisms and discuss the challenge of deriving the exact privacy profile under composition. Section \ref{sec:approx} describes our approach to approximating the privacy profile for homogeneous DGMs, with the approximation error analyzed in Section  \ref{sec:residual}. 
Section \ref{sec:inid-overall} and \ref{sec:overall_trade_off} examine the composition of heterogeneous DGMs used to allocate the privacy budget across eight geographical levels.

\subsection{Preliminaries on differential privacy}
\label{sec:background}

In this section, we discuss the basics of differential privacy \citep{dwork2006calibrating,dwork2006our} and its application in protecting the U.S. Census using the discrete Gaussian mechanism \citep{Canonne2020discrete}.


Let $\mathcal{X}$ represent the sample space, and let $D \subset \mathcal{X}^m$ be a dataset containing $m$ data records.
Consider a deterministic query $M:\mathcal{X}^m\rightarrow \mathbb{Z}^d$ that takes only integer values.
To ensure privacy, the discrete Gaussian mechanism (DGM), which adds discrete Gaussian noise to $M$, is employed within the DAS.
Recall the discrete Gaussian distribution given in Section \ref{sec:results}.
The DGM takes each query $M(D)$ as an input and outputs the privatized query 
\begin{align}
\label{eq:DG-mechanisms}
  \widetilde{M}(D) = M(D) + \mathcal{N}_{\mathbb{Z}}(0,\sigma^2).
\end{align}



The privacy budget for the 2020 U.S.\ Census is measured using zero-Concentrated Differential Privacy (zCDP) \citep{bun2016concentrated}, which is based on R\'enyi divergence. For two distributions, $P$ and $Q$, with probability density functions $p$ and $q$, respectively, the R\'enyi divergence of order $\alpha > 1$ is defined as
$
    R_{\alpha}(P\|Q) = \frac{1}{\alpha - 1} \log \int p(x)^{\alpha} q(x)^{1 - \alpha}dx.
$
$R_{1}(P\|Q)$ or $R_{\infty}(P\|Q)$ is the limit of $R_{\alpha}(P\|Q)$ as $\alpha$ tends to $1$ or infinity, respectively.
Based on the R\'enyi divergence, one has the following definition of zCDP. Here, the R\'enyi divergence between two random variables is understood as the divergence between their respective distributions.

\begin{definition}[zCDP, \citep{bun2016concentrated}]
    A randomized mechanism $\widetilde{M}$ is said to satisfy $\rho$-zCDP if 
    \begin{align*}
        R_{\alpha}(\widetilde{M}(D)\|\widetilde{M}(D')) \leq \rho \alpha, \qquad \text{for all } \alpha > 1,
    \end{align*}
and for any neighboring datasets $D$ and $D'$.
\end{definition}

The privacy-loss budget allocation released on August 25, 2022 by the Bureau \citep{privacyallocation2022} has $\rho = 3.65$.
Another key aspect related to the privacy budget is the sensitivity of each query.
For a query $M$ taking values in $\mathbb{R}^{d}$, the $l_2$-sensitivity of $M$ is defined as
\begin{align*}
    \Delta_M = \sup_{D,D'} \left\{\left\| M(D) - M(D')\right\|_{\ell_2} \right\},
\end{align*}
where $\|\cdot\|_{\ell_2}$ is the $\ell_2$-norm of a vector and the supremum is taken over all datasets $D$ and $D'$ that differ in at most one data record.

In the implementation of the TopDown algorithm \citep{Census2023implementation}, an add/delete sensitivity of 1 is used. This reflects the difference between two neighboring datasets when a single data record is added or removed. The underlying theory of the DAS \citep{abowd20222020} considers a broader sensitivity model that also accounts for replacing one record with another, resulting in a sensitivity of up to $\sqrt{2}$ for coarsened counting queries with binary categories (e.g., “18 and older” vs. “17 and younger”). We adopt the add/delete sensitivity of 1. Our method can be naturally extended to the replacement case (yielding a sensitivity of $\sqrt{2}$) by treating the binary categories as the 2-fold composition of two counting queries.

According to \cite{Canonne2020discrete}, the discrete Gaussian mechanism is $\rho$-zCDP if we take $\sigma^2 = \Delta_M^2 / 2\rho.$
zCDP is currently adopted by the Bureau to count the privacy budget of the 2020 Census.
A better zCDP guarantee for the discrete Gaussian is also investigated by \cite{kairouz2021distributed}.
The Bureau obtained the privacy budget $(\epsilon,\delta)$ for the Census by converting $\rho$-zCDP to $(\epsilon,\delta)$-DP using the following equation from \cite{bun2018composable}:
\begin{align*}
    \epsilon = \rho + 2\sqrt{-\rho\log\delta}.
\end{align*}

However, zCDP may provide a loose privacy profile due to the inherent looseness of the R\'enyi divergence \citep{balle2020hypothesis}. 
In this paper, we aim to provide an $f$-DP guarantee for the DGM, which is known to be tight \citep{Dong2022Gaussian,wang2024unified}.
Under the setting of $f$-DP, the distinguishability between $\widetilde{M}(D)$ and $\widetilde{M}(D')$ can be quantified using hypothesis testing \citep{kairouz2017composition,Dong2022Gaussian}.
Consider a hypothesis testing problem $H_0: P \hbox{ v.s. } H_1:Q$ and a rejection rule $\phi\in[0,1].$ We define the type I error as $\alpha_{\phi} = \mathbb{E}_{P}[\phi]$, which is the probability of incorrectly rejecting the null hypothesis $H_0$. 
The type II error $\beta_{\phi}= 1 - \mathbb{E}_{Q}[\phi]$ is the probability that we accept the alternative $H_1$ wrongly.

The trade-off function $T(P,Q)$ is the minimal type II error at a given level $\alpha$ of the type I error, that is,
\begin{align*}
    T(P,Q)(\alpha) = \inf_{\phi}\{\beta_{\phi}: \alpha_{\phi}\leq \alpha\},
\end{align*}
where the infimum is taken over all rejection rule $\phi.$
According to the Neyman–Pearson lemma \citep[cf.,][]{Lehmann2005testing}, the infimum is achieved by the likelihood ratio test.
For any two random variables $\xi$ and $\zeta$, we define $T(\xi,\zeta)$ as the trade-off function between the respective distributions.

\begin{definition}[$f$-DP]
    We say a mechanism $\widetilde{M}$ satisfies $f$-DP if $
    T(\widetilde{M}(D), \widetilde{M}(D'))(\alpha) \geq f(\alpha)
$
for any $\alpha\in [0,1]$ and any neighboring datasets $D$ and $D'$.
\end{definition}
$f$-DP is equivalent to $(\epsilon,\delta(\epsilon))$-DP for all $\epsilon>0$ according to Proposition 2.12 in \cite{Dong2022Gaussian}.

\subsection{Useful facts for discrete Gaussian distributions}
\label{sec:useful_facts}
In this section, we introduce several useful facts about discrete Gaussian distributions, including the sub-Gaussian tail bound and properties of the characteristic functions. {These properties will be used to derive the privacy profile for the composition of discrete Gaussian mechanisms in Sections \ref{sec:fDP-DGM}, \ref{sec:approx}, \ref{sec:residual}, and \ref{sec:inid-overall}.} Proofs for these results are deferred to Section \ref{proof:useful_facts}.

\paragraph{Sub-Gaussian properties of discrete Gaussian distributions.} 

Let $X_i \sim\overset{\text{i.i.d.}}{} \mathcal{N}_{\ZZ}(0,\sigma^2)$ and define $S_n = \frac{1}{B_n}\sum_{i=1}^n X_i,$ where $B_n = \sqrt{n}\sigma$.
According to Corollary 17 in \cite{Canonne2020discrete}, $X_i$ is sub-Gaussian with variance proxy $\sigma^2.$
Therefore, $S_n$ is the sum of $n$ i.i.d.\ sub-Gaussian random variables, each with a variance proxy of $\sigma^2$, and is thus sub-Gaussian with variance proxy $n\sigma^2$. Specifically, it holds
\begin{align}
\label{eq:sub-Gaussian-sum}
    \PP\left(S_n > m_1 \right) = \PP\left(\sum_{i=1}^{n} X_i > m_1 B_n \right)  \leq \ex^{- \frac{(m_1 B_n)^2}{2 n \sigma^2}} = \ex^{- \frac{m_1^2}{2}},
\end{align}
for any $m_1>0.$

\paragraph{Variance of discrete Gaussian distributions.}
\label{sec:variance}
For any variance proxy $\sigma^2$ used by the Bureau in Table \ref{table:2020 allocation} (the row corresponding to Bureau's), the variance of $\mathcal{N}_{\ZZ}(0, \sigma^2)$ is close to $\sigma^2$. 
\begin{fact}
    The variance of $\mathcal{N}_{\ZZ}(0, \sigma^2)$ is bounded as follows: 
    for any $4.25 \leq \sigma^2 \leq 5.00$, 
    \begin{align*}
        \sigma^2 - 5.8 \times 10^{-34} < \Var(\mathcal{N}_{\ZZ}(0, \sigma^2)) < \sigma^2;
    \end{align*}
    for any $5.00 < \sigma^2 \leq 10.00$,
    \begin{align*}
        \sigma^2 - 2.8 \times 10^{-40} < \Var(\mathcal{N}_{\ZZ}(0, \sigma^2)) < \sigma^2;
    \end{align*}
    for any $\sigma^2>10.00 $, 
    \begin{align*}
        \sigma^2 - 1.5 \times 10^{-82} < \Var(\mathcal{N}_{\ZZ}(0, \sigma^2)) < \sigma^2.
    \end{align*}
\end{fact}

{The following characteristic function of $S_n$ will be used to investigated the privacy profile of the DGM.} 
\paragraph{Characteristic functions of \texorpdfstring{$S_n$}{}.}
The characteristic function of $S_n$ can be represented as follows:
\begin{align} \label{eqn:char_func}
\begin{split}
    f_{S_n}(t) = \EE \ex^{it S_n} =\ & \left( \frac{\sum_{u=- \infty}^{\infty} \ex^{- u^2/2 \sigma^2} \ex^{i \cdot t/B_n \cdot u}}{\sum_{u=- \infty}^{\infty} \ex^{- u^2/2 \sigma^2}} \right)^n\\
    \overset{(a)}{=}\ & \left( \frac{\sum_{u=- \infty}^{\infty} \ex^{- \sigma^2 (t/B_n - 2 \pi u)^2/2}}{\sum_{u=- \infty}^{\infty} \ex^{- 2 \sigma^2 \pi^2 u^2}} \right)^n\\
    =\ & \left( \ex^{-\frac{t^2}{2 n}} \cdot \frac{\theta_3\left( -i \sigma \pi t / \sqrt{n}, \ex^{-2 \sigma^2 \pi^2} \right)}{\theta_3\left(0,\ex^{-2\sigma^2\pi^2}\right)} \right)^n\\
    =\ & \ex^{-t^2/2} \left( \frac{\theta_3\left( -i \sigma \pi t / \sqrt{n}, \ex^{-2 \sigma^2 \pi^2} \right)}{\theta_3\left(0,\ex^{-2\sigma^2\pi^2}\right)}\right)^{n},
\end{split}
\end{align}
where $\theta_{3}(u,q) = 1 + 2 \sum_{k=1}^\infty q^{k^2}\cos(2 k u)$ is a theta function and (a) holds due to Poisson summation formula.

{To characterize the maximum value and monotonicity of $f_{S_n}(t)$, {which will be used to bound the characteristic function}, we use the following lemma.} 
\begin{lemma} \label{lemma:comparison}
For any $0 \leq \mu < \nu \leq \frac{1}{2}$, we have
$
        \sum_{x \in \ZZ} \ex^{-\frac{(x - \mu)^2}{2 \sigma^2}} > \sum_{x \in \ZZ} \ex^{-\frac{(x - \nu)^2}{2 \sigma^2}}.
$
\end{lemma}

As a direct consequence of Lemma \ref{lemma:comparison}, 
the derivative of $\sum_{x \in \ZZ} \ex^{-\frac{(x - \mu)^2}{2 \sigma^2}}$ with respect to $\mu \in (0, 1/2)$ is negative. That is, it holds
\begin{equation} \label{eqn:dec_char_fn}
    \frac{d}{d \mu} \sum_{x \in \ZZ} \ex^{-\frac{(x - \mu)^2}{2 \sigma^2}} < 0, \qquad \hbox{ for } \mu\in(0,1/2).
\end{equation}


{ The following proposition concerns the expectation of discrete Gaussian random variables. The expectation is well-defined only when $\mu$ is a half-integer.}
\begin{proposition}[Correction to Fact 18 in \cite{Canonne2020discrete}]
\label{prop:correction}
    The DGM is unbiased in the sense that
$
        \EE\ \mathcal{N}_{\ZZ} (\mu, \sigma^2) = \mu
$
    if and only if $\mu \in \frac{1}{2} \ZZ$.
\end{proposition}
\begin{proof}
By Lemma \ref{lemma:comparison}, we have
    \begin{align*}
        \EE \mathcal{N}_{\ZZ}(\mu, \sigma^2) - \mu = \sum_{x \in \ZZ} (x - \mu) \ex^{- \frac{(x - \mu)^2}{2 \sigma^2}} = \sigma^2 \cdot \frac{d}{d \mu} \sum_{x \in \ZZ} \ex^{-\frac{(x - \mu)^2}{2 \sigma^2}} < 0,
    \end{align*}
for any $0 < \mu < 1/2$. 
This completes the proof.
\end{proof}
{Using Lemma \ref{lemma:comparison}, we derive the following properties of $f_{S_n}$ which will be essential in proving Fact \ref{fact:iid_approx}.}
\begin{proposition} \label{prop:char_monotone}
    The characteristic function $f_{S_n}(t)$ is periodic with period $2 \pi B_n$. Moreover, $f_{S_n}(t)$ is strictly increasing on $(- \pi B_n, 0)$ and is strictly decreasing on $(0, \pi B_n)$.
    Consequently, $f_{S_n}(t)$ achieves its maximum at $t=0$ with a maximum value of $f_{S_n}(0) = 1$. 
\end{proposition}

\subsection{\texorpdfstring{$f$-DP}{} guarantees for discrete Gaussian mechanisms}
\label{sec:fDP-DGM}

{ In this subsection, we consider the $f$-DP guarantee under a small number of compositions. Specifically, we study $n$-fold composition with $n \leq 2$. In this case, the trade-off function can be computed efficiently due to the small value of $n$. However, for large $n$, computing a closed-form expression for the trade-off function becomes computationally expensive. Moreover, as we will show, specifying the trade-off function for heterogeneous compositions is much more involved than for the homogeneous case.}

From an $f$-DP perspective, for two neighboring datasets $D$ and $D'$, the privacy of the discrete Gaussian mechanisms is to test
\begin{align*}
    H_0: \widetilde{M}(D)=M(D) + \mathcal{N}_{\ZZ}(0,\sigma^2) \qquad \hbox{v.s.} \qquad H_1: \widetilde{M}(D')=M(D') + \mathcal{N}_{\ZZ}(0,\sigma^2).
\end{align*}
For integer-valued queries $M(D)$ and $M(D')$, $f$-DP provides a tight privacy profile of the DGM (without composition).
First, we have 
\begin{align*}
    T(\mathcal{N}_{\ZZ}(M(D),\sigma^2),\mathcal{N}_{\ZZ}(M(D'),\sigma^2)) &= T(\mathcal{N}_{\ZZ}(0,\sigma^2),\mathcal{N}_{\ZZ}(M(D')-M(D),\sigma^2)) 
    \\
    &\geq T(\mathcal{N}_{\ZZ}(0,\sigma^2),\mathcal{N}_{\ZZ}(\mu,\sigma^2)),
\end{align*}
where $\mu\in\mathbb{Z}$ is the sensitivity of $M$.
Therefore, it is sufficient to evaluate $T(X,X+\mu)$ for $X\sim \mathcal{N}_{\ZZ}(0,\sigma^2).$
Let $\Phi_{\ZZ,\sigma}$ and $\phi_{\ZZ,\sigma}$ be the cumulative distribution function (cdf) and probability mass function (pmf) of $\mathcal{N}_\ZZ(0,\sigma^2),$ respectively. 
We then have the following $f$-DP guarantee for the discrete Gaussian mechanisms.

\begin{theorem}
\label{thm:fDP-GDM-no-composition}
For $\mu\in\mathbb{Z}$ and $X\sim\mathcal{N}_{\mathbb{Z}}(0,\sigma^2)$, we have
\begin{align*}
    T(X,X+\mu)(\alpha) = \Phi_{\mathbb{Z},\sigma}(\Phi_{\mathbb{Z},\sigma}^{-1}(1-\alpha) - \mu) - \frac{\varphi_{\mathbb{Z},\sigma}(t_{\alpha}-\mu)}{\varphi_{\mathbb{Z},\sigma}(t_{\alpha})}\left(\alpha + \Phi_{\mathbb{Z},\sigma}(t_{\alpha}) -1 \right),
\end{align*}
where $t_\alpha = \Phi_{\mathbb{Z},\sigma}^{-1}(1-\alpha)\in\mathbb{Z}.$
In particular, for each knot $\alpha$ such that $1-\alpha = \Phi_{\mathbb{Z},\sigma}(\Phi_{\mathbb{Z},\sigma}^{-1}(1-\alpha))$ (i.e., $1-\alpha\in\Phi_{\mathbb{Z},\sigma}(\mathbb{Z})$), it holds
\begin{align*}
        T(X,X+\mu)(\alpha) = \Phi_{\mathbb{Z},\sigma}\left(\Phi_{\mathbb{Z},\sigma}^{-1}(1-\alpha) - \mu\right).
\end{align*}
\end{theorem}

\begin{proof}
For simplicity, we prove the case $\sigma=1$ and the general case can be derived similarly.
We have
\begin{align*}
    \alpha(t) = \mathbb{P}[X>t] + c\varphi_{\mathbb{Z}}(t)
    = 1 - \Phi_{\mathbb{Z}}(t_{\alpha}) + c_{\alpha}\varphi_{\mathbb{Z}}(t_\alpha),
\end{align*}
where $t_{\alpha} = \Phi_{\mathbb{Z}}^{-1}(1-\alpha)$ and $c_{\alpha} = \frac{\alpha + \Phi_{\mathbb{Z}}(t_\alpha) - 1}{\varphi(t)}.$
Then, it holds
\begin{align*}
    \beta(\alpha) &= \mathbb{P}[X\leq t_\alpha - \mu] - c_\alpha\varphi_{\mathbb{Z}}(t_\alpha-\mu)
    \\
    &=\Phi_{\mathbb{Z}}(\Phi_{\mathbb{Z}}^{-1}(1-\alpha) - \mu) - \frac{\varphi_{\mathbb{Z}}(t_{\alpha}-\mu)}{\varphi_{\mathbb{Z}}(t_{\alpha})}\left(\alpha + \Phi_{\ZZ}(t_{\alpha}) -1 \right).
\end{align*}
\end{proof}

By converting $f$-DP to $(\epsilon,\delta)$-DP using Proposition 2.12 in \cite{Dong2022Gaussian} and Theorem \ref{thm:fDP-GDM-no-composition}, we obtain that $\widetilde{M}(D)$ is $(\epsilon,\delta)$-DP with 
\begin{align*}
    \delta(\epsilon) = \mathbb{P}\left[X >  \frac{\epsilon\sigma^2}{\mu} - \frac{\mu}{2}\right] - \ex^{\epsilon}\mathbb{P}\left[X >  \frac{\epsilon\sigma^2}{\mu} + \frac{\mu}{2}\right],
\end{align*}
where the probability is taken with respect to $X\sim\mathcal{N}_\ZZ(0,\sigma^2).$
A similar result appears in Theorem 7 of \cite{Canonne2020discrete}. In practical applications, estimating the privacy profile for census data remains challenging due to the effects of composition, as illustrated in Figure \ref{fig:composition-chart}.

Mathematically, an $n$-fold composition of the DGM is represented by the sequence $(\widetilde{M}_i(D))_{i=1}^n$, where the associated hypothesis testing problem is given by: \begin{align*} H_0: \left(\widetilde{M}_i(D) \right)_{i=1}^n \qquad \text{vs.} \qquad H_1: \left(\widetilde{M}_i(D')\right)_{i=1}^n. \end{align*} 
Here each $\widetilde{M}_i(D) = M_i(D) + \mathcal{N}_{\ZZ}(0,\sigma_i^2)$, with $\sigma_i^2$ being a variance proxy of $\mathcal{N}_{\ZZ}(0,\sigma_i^2)$ and $M_i$ a given query. The $f$-DP guarantee corresponds to the hypothesis test: 
\begin{align*} H_0: \prod_{i=1}^n P_i \qquad \text{vs.} \qquad H_1: \prod_{i=1}^n Q_i, 
\end{align*} 
where $P_i$ is the distribution of $\mathcal{N}_{\ZZ}(0,\sigma_i^2)$ and $Q_i$ is that of $\mathcal{N}_{\ZZ}(\mu_i,\sigma_i^2)$, with $\mu_i$ being the sensitivity of $M_i$. {For 2-fold homogeneous composition with $\mu_i\equiv \mu$ and $\sigma_i \equiv \sigma$, the $f$-DP guarantee is given in the following theorem.}

\begin{theorem}
\label{thm:closed-trade-off}
    For the hypothesis testing problem
    \begin{align*}
        H_0: \mathcal{N}_{\ZZ}(0,\sigma^2) \times \mathcal{N}_{\ZZ}(0,\sigma^2) \text{ v.s. }  H_1: \mathcal{N}_{\ZZ}(\mu,\sigma^2) \times \mathcal{N}_{\ZZ}(\mu,\sigma^2)
    \end{align*}
    with $\mu\in\mathbb{Z}$, we have the following closed-form representation of the type I and type II errors of the likelihood ratio test.
    The trade-off function is piecewise linear, where each knot has the form
    \begin{align*}
        \alpha(t) = \frac{c_{0,\sqrt{2}\sigma}}{c_{0,\sigma}^2}\left(\sum_{i>t/2,i\in\ZZ}  \ex^{-i^2/\sigma^2}\right) + \frac{c_{-1/2,\sqrt{2}\sigma}}{c_{0,\sigma}^2}\left(\sum_{i> \frac{t-1}{2}, i\in\ZZ} \ex^{-(i+1/2)^2/\sigma^2} \right),
\end{align*}
and the corresponding type II error is given by
\begin{align*}
    \beta(t) = \frac{c_{0,\sqrt{2}\sigma}}{c_{0,\sigma}^2}\left(\sum_{i\leq t/2 - \mu}  \ex^{-i^2/\sigma^2}\right) + \frac{c_{-1/2,\sqrt{2}\sigma}}{c_{0,\sigma}^2}\left(\sum_{i\leq \frac{t-1}{2} - \mu} \ex^{-(i+1/2)^2/\sigma^2} \right),
\end{align*}
where $c_{\mu,\sigma'} = \sum_{i=-\infty}^{\infty} \ex^{-(i-\mu)^2/\sigma'^2}$ for any $\mu\in\mathbb{Z}/2$ and $\sigma'>0.$
As a result, the 2-fold homogeneous composition of the DGM is $(\epsilon,\delta)$-DP with
\begin{align*}
    \delta(\epsilon) = 1 + \max_{t}\left\{-\ex^{\epsilon}\alpha(t) - \beta(t) \right\}.
\end{align*}
\end{theorem}

\noindent\textbf{Remark.}
The obtained type I and type II errors, as well as the privacy profile, are tight. According to Theorem \ref{thm:closed-trade-off}, to precisely specify the trade-off function, we need to partition the support of $X_1 + X_2$ into two segments— (2$\mathbb{Z}$ and $2\mathbb{Z} + 1$). Each segment is associated with a coefficient, $c_{0,\sqrt{2}\sigma}$ or $c_{-1/2,\sqrt{2}\sigma}$, and corresponds to a Gaussian distribution over the lattices 2$\mathbb{Z}$ or $2\mathbb{Z} + 1$.
However, the resulting tight type I and type II errors are complicated, involving the constants $c_{0,\sqrt{2}\sigma}$ and $c_{-1/2,\sqrt{2}\sigma}.$ 
It is noteworthy that $c_{0,\sqrt{2}\sigma} \neq c_{-\frac{1}{2},\sqrt{2}\sigma} $,  differing from what is stated in Fact 18 of \cite{Canonne2020discrete}. For a correction to Fact 18 in \cite{Canonne2020discrete}, please refer to Proposition \ref{prop:correction}.
When $\sigma$ is large, the difference between  $c_{0,\sqrt{2}\sigma}$ and $c_{-1/2,\sqrt{2}\sigma}$ is negligible, which motivates our approximation of the $n$-fold composition of the DGM using a distribution supported on $\mathbb{Z}/(\sqrt{n}\sigma)$ in Section \ref{sec:approx}.
In fact, if we replace $c_{-1/2,\sqrt{2}\sigma}$ by $c_{0,\sqrt{2}\sigma}$, then the type I error can be approximated by
\begin{align*}
    \frac{c_{0,\sqrt{2}\sigma}}{c_{0,\sigma}^2} \sum_{y > t, y\in\mathbb{Z}} \phi\left( \frac{y}{\sqrt{2}\sigma}\right),
\end{align*}
where $\phi$ is the pdf of the standard Gaussian distribution. Thus, we approximate the distribution of $(X_1 + X_2)/(\sqrt{2}\sigma)$ by a measure $\nu$ supported on $\mathbb{Z}/\sqrt{2}\sigma$ (not a probability measure) with $\nu[ Y = i/\sqrt{2}\sigma] =  \frac{1}{\sqrt{2}\sigma}\phi(i/\sqrt{2}\sigma)$ for some measurable function $Y\sim \nu.$ 


We can extend Theorem \ref{thm:closed-trade-off} to i.i.d.\ $n$-fold composition.
In fact, for $n\geq 2$, we have each knot of the type I error is 
\begin{align}
\label{eq:n-fold-typeI}
    \alpha(t) = \mathbb{P}\left[ \sum_i {X_i} > t\right] = \sum_{k=0}^{n-1}c_{n,k} \sum_{y\in n\ZZ + k,
    y>t}  \ex^{-\frac{y^2}{2\sigma^2}},
\end{align}
and the type II error is 
\begin{align}
\label{eq:n-fold-typeII}
    \beta(t) = \mathbb{P}\left[ \sum_i{X_i} \leq t - n\mu\right] = \sum_{k=0}^{n-1}c_{n,k} \sum_{\substack{y\in n\ZZ + k,\\y \leq t - n\mu}}  \ex^{-\frac{-y^2}{2\sigma^2}},
\end{align}
where $c_{n,k} = \ex^{-\frac{k(n-1)}{2n\sigma^2}} \cdot \sum_{u_i \in \mathbb{Z}}\ex^{-\frac{\sum_{i=1}^n u_i^2 + 2\sum_{i=1}^k u_i + (\sum_{i=1}^{n-1}u_i)^2}{2\sigma^2}}$ is a finite constant.

\eqref{eq:n-fold-typeI} and \eqref{eq:n-fold-typeII} reveal the structure underlying the summation of discrete Gaussian random variables. 
Specifically, the support $\mathbb{Z}$ of $\sum_{i} X_i$ is partitioned into $n$ segments, with each segment corresponding to a Gaussian distribution over lattices. Each segment is associated with a distinct coefficient $c_{n,k}$, reflecting the distribution’s structure across these lattice-based partitions.

Note that $c_{n,k}$ is summing a discrete function across lattices of $(n-1)$ dimensions.
Computing this constant $c_{n,k}$ is complicated, consequently complicating the practical application of the closed-form expressions found in \eqref{eq:n-fold-typeI} and \eqref{eq:n-fold-typeII}. To address this challenge, an efficient approximation method is introduced in Section \ref{sec:approx}.
Similar to the case $n=2$, in Section \ref{sec:approx}, we approximate the distribution of $\sum_{i=1}^nX_i/\sqrt{n}\sigma$ using a univariate random function $Y\sim \nu$ with $\nu[Y = i/\sqrt{n}\sigma] = \frac{1}{\sqrt{n}\sigma}\phi(i/\sqrt{n}\sigma).$

The independent but not identically distributed (i.n.i.d.) case, where $X_1 \sim\mathcal{N}(0,\sigma_1^2)$ and $X_2\sim\mathcal{N}(0,\sigma_2^2)$, is much more complicated.
In fact, for $n=2$, the corresponding type I error becomes 
\begin{align*}
    \alpha(t) = \mathbb{P}\left[\sigma_2^2 X_1 + \sigma_1^2 X_2 > t\right].
\end{align*}
The support of $\sigma_2^2 X_1 + \sigma_1^2 X_2$ can be estimated only when $\sigma_1^2, \sigma_2^2\in a\mathbb{Z}$ for some $a\in\mathbb{R}$ as a result of the Chinese remainder theorem.
For simplicity, we consider $a=1$ and the support of $\sigma_2^2 X_1 + \sigma_1^2 X_2$ is $\mathrm{gcd}(\sigma_1^2,\sigma_2^2) \times\mathbb{Z}$ according to the Chinese remainder theorem.
For this i.n.i.d.\ case, the approximation of the privacy profile is derived in Section \ref{sec:inid-overall}.

\subsection{Approximate the privacy profiles of discrete Gaussian mechanisms}
\label{sec:approx}

{ In this subsection, we examine the privacy profile of homogeneous composition within the same geographical level, as illustrated in Figure \ref{fig:composition-chart}. The results presented here are used to derive the privacy profiles within each geographical level, specifically in Figure \ref{fig:eps_delta_geo}, Figure \ref{fig:improve_epsilon}, and Figure \ref{fig:eps_delta_geo_reduce_noise_more}. Our approach to deriving the privacy profile ( $(\epsilon,\delta)$-curve) is based on $f$-DP. The corresponding trade-off function in the $f$-DP framework is deferred to Section \ref{sec:overall_trade_off}.}

To convert $f$-DP to $(\epsilon,\delta)$-DP, we adopt Proposition 3.2 in \cite{wang2022analytical}.
Let $X_i \sim \mathcal{N}_{\ZZ}(0,\sigma_i^2)$ and $Y_i \sim \mathcal{N}_{\ZZ}(\mu_i,\sigma_i^2).$ 
Define $\xi_i = \log \frac{q(Y_i)}{p(Y_i)}$ and $\zeta_i = \log \frac{q(X_i)}{p(X_i)}$ 
with $p$ and $q$ being the probability density function of $P$ and $Q$, respectively.

\begin{lemma}[Proposition 3.2 in \cite{wang2022analytical}]
\label{lemma:privacy-profile}
    The $n$-fold composition of the DGM $(\widetilde{M}_i(D))_{i=1}^n$ is $(\epsilon,\delta)$-DP with 
    \begin{align*}
        \delta(\epsilon) = 
        \mathbb{P}\left[\sum_{i=1}^n \xi_i > \epsilon\right] - \ex^{\epsilon}\mathbb{P}\left[\sum_{i=1}^n \zeta_i > \epsilon\right]. 
    \end{align*}
\end{lemma}

For the i.i.d.\ case where $\sigma_i \equiv \sigma$, we obtain the following privacy profile for the $n$-fold composition of the DGM.

\begin{proposition}[Privacy profile for the i.i.d.\ composition of the DGM]
\label{prop:IID-profile}
    For the i.i.d.\ case with $\sigma_i \equiv\sigma$, the $n$-fold composition of the DGM $(\widetilde{M}_i(D))_{i=1}^n$ is $(\epsilon,\delta)$-DP with 
    \begin{align} \label{eqn:delta_formula}
    \delta(\epsilon) &= \mathbb{P}\left[S_n > \frac{1}{B_n}\left(\frac{2\epsilon\sigma^2}{2} - \frac{n}{2} \right) \right] - \ex^{\epsilon} \mathbb{P}\left[S_n > \frac{1}{B_n}\left(\frac{2\epsilon\sigma^2}{2} + \frac{n}{2} \right) \right],
    \end{align}
where $S_n = \frac{1}{B_n}\sum_{i=1}^n X_i,$ $B_n = \sqrt{n}\sigma$. 
\end{proposition}

Proposition \ref{prop:IID-profile} is a straightforward conclusion of Lemma \ref{lemma:privacy-profile} and \eqref{eq:n-fold-typeI}.
{ Even though Proposition \ref{prop:IID-profile} provides a closed-form representation of the privacy profile, it requires evaluating a summation over an $n$-dimensional lattice supported on the entire space $\mathbb{Z}^n$, which is computationally inefficient.
As motivated by the case $n = 2$ in Section \ref{sec:fDP-DGM}, one may approximate the distribution of $S_n$ by a $1$-dimensional distribution.}
Note that the support is scaled to $\mathbb{Z}/B_n$ and we approximate the composition by the following $1$-dimensional distribution over lattices $\mathbb{Z}/B_n$.

\begin{theorem}
\label{thm:privacy-profile}
    Consider $M(D) = (M_i(D))_{i=1}^n$ with $M_i(D)\in\mathbb{R}$ being a counting query with sensitivity 1. Let $\widetilde{M}_i(D) = M_i(D) + \mathcal{N}_{\mathbb{Z}}(0,\sigma^2)$ for any $\sigma^2\in\mathbb{R}$. Then, we have $\widetilde{M}(D) = (\widetilde{M}_i(D))_{i=1}^n$ is $(\epsilon,\delta)$-DP with
        \begin{align*}
        \delta(\epsilon) = \frac{1}{B_n}\sum_{\frac{i}{B_n} > \frac{2 \sigma^2 \epsilon - n}{2B_n}}^{U_1} \phi\left(\frac{i}{B_n} \right) - 
        \ex^{\epsilon}\left(\frac{1}{B_n} \sum_{\frac{i}{B_n} > \frac{2 \sigma^2 \epsilon + n}{2 B_n}}^{U_2} \phi \left(\frac{i}{B_n} \right) + R_2(n,\sigma,\epsilon) \right) + R_1(n,\sigma,\epsilon),
    \end{align*}
where $U_1 = \max\{20, \frac{2 \sigma^2 \epsilon - n}{2B_n}\}$ and $U_2 = \max\{20, \frac{2 \sigma^2 \epsilon + n}{2B_n}\}.$ $R_1(n,\sigma,\epsilon)$ and $R_2(n,\sigma,\epsilon)$ are  residual terms computed by the Fourier transform with
\begin{align*}
    &R_{1}(n,\sigma,\epsilon) \leq \sum_{\frac{i}{B_n}>\frac{2\sigma^2\epsilon - n}{2B_n}}^{U_1} r_{n,\sigma}\left(\frac{i}{B_n}\right) + \PP\left[ S_n > U_1\right],\qquad\hbox{and}
    \\
    &R_{2}(n,\sigma,\epsilon) \leq \sum_{\frac{i}{B_n}>\frac{2\sigma^2\epsilon + n}{2B_n}}^{U_2} r_{n,\sigma}\left(\frac{i}{B_n}\right) + \PP\left[ S_n > U_2 \right].
\end{align*}
Here, $r_{n,\sigma}$ has a closed-form representation
\begin{align*}
    r_{n,\sigma}(y) = \frac{1}{2 \pi B_n} \left| \int_{-\pi B_n}^{\pi B_n} \ex^{-t^2/2} \ex^{-i t y} \left(\frac{\theta_3\left( -i \sigma \pi t / \sqrt{n}, \ex^{-2 \sigma^2 \pi^2} \right)}{\theta_3\left(0,\ex^{-2\sigma^2\pi^2}\right)}\right)^n  dt - \int^{\infty}_{\infty} \ex^{-t^2/2} \ex^{-i t y} dt\right|,
\end{align*}
with $\theta_{3}(u,q) = 1 + 2 \sum_{k=1}^\infty q^{k^2}\cos(2 k u)$ being a theta function.
\end{theorem}

\begin{figure*}[!htp]
    \centering
    \begin{subfigure}[b]{0.45\textwidth}
        \includegraphics[width=\textwidth]{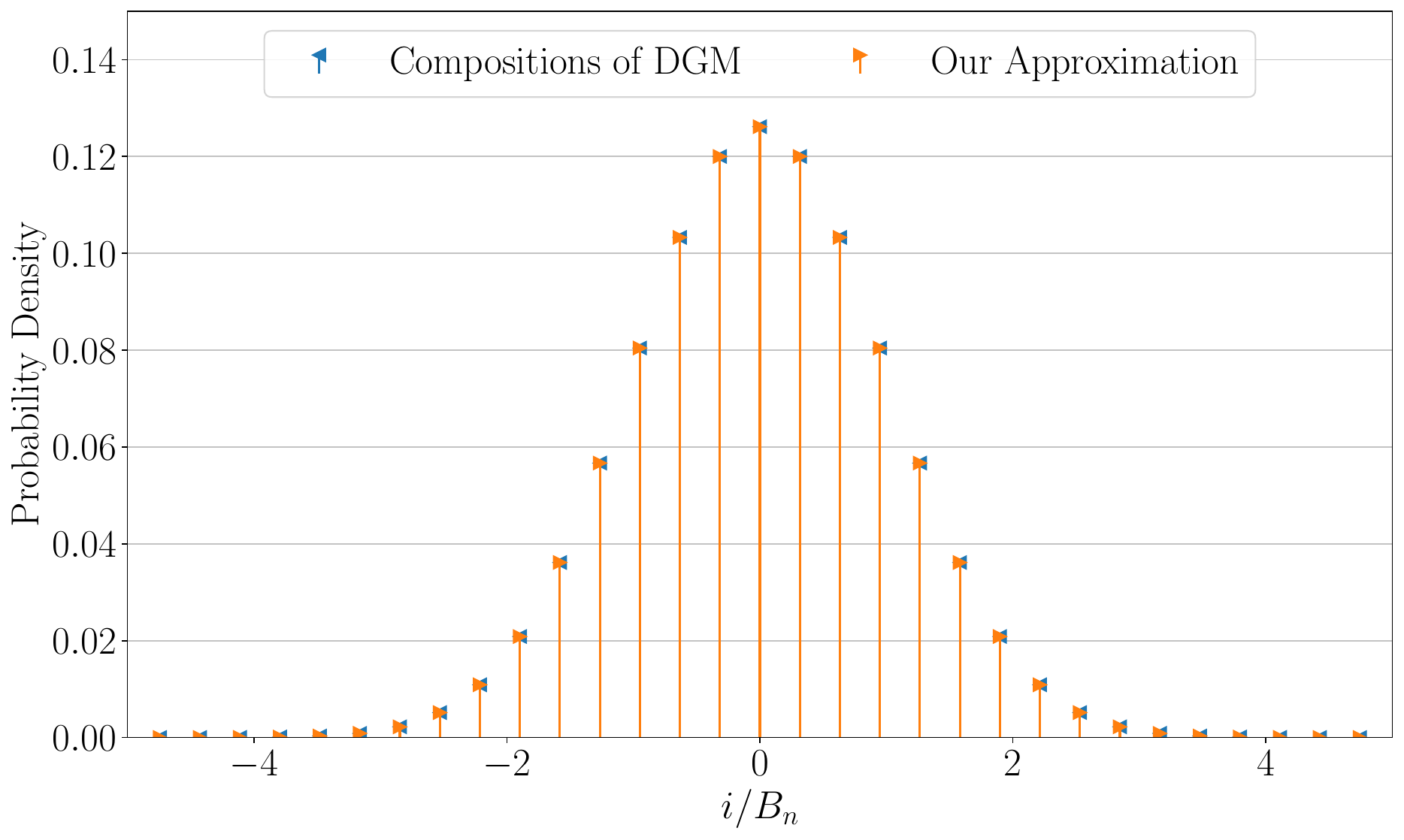}
    \end{subfigure}
    \begin{subfigure}[b]{0.45\textwidth}
        \includegraphics[width=\textwidth]{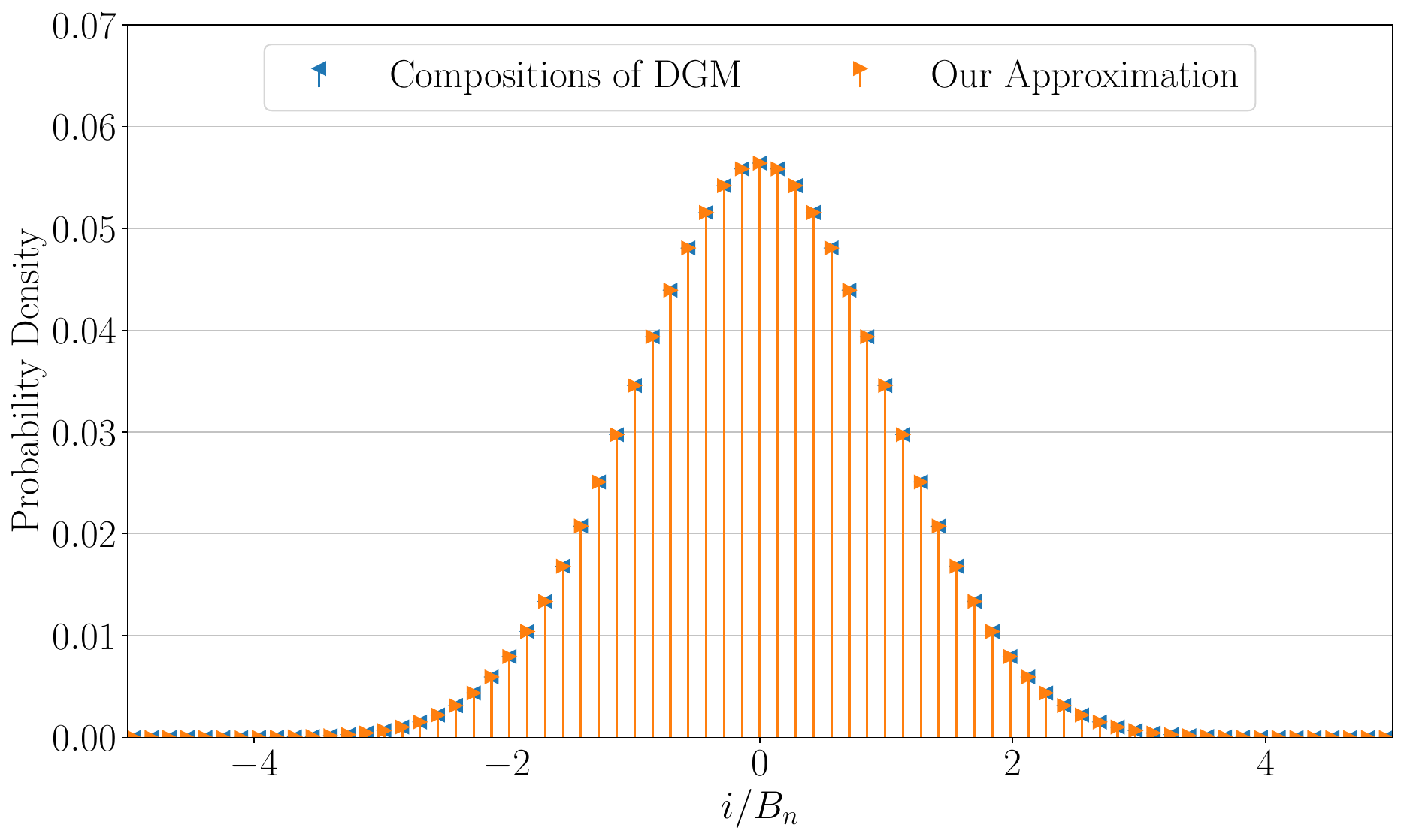}
    \end{subfigure}
    \caption{Comparisons of pmf and approximation with $\sigma^2 = 5$ (left) and $\sigma^2 =25$ (right). 
    }
    \label{fig:approxiamtion_comparison}
\end{figure*}

In Theorem \ref{thm:privacy-profile}, we approximate the probability mass function of $S_n$ using a function (not a probability mass function) $\frac{1}{B_n}\phi\left(\frac{i}{B_n} \right)$. The error associated with this approximation is examined in detail in Section \ref{sec:residual}.
For the case $n=2$, where the pmf of $S_n$ can be computed directly, we compare $\frac{1}{B_n}\phi\left(\frac{i}{B_n} \right)$ with the actual pmf of $S_n$ in Figure \ref{fig:approxiamtion_comparison}.
From Figure \ref{fig:approxiamtion_comparison}, we observe that our approximation should be powerful intuitively.
We provide a numerical estimate of the residuals $R_1(n,\sigma,\epsilon)$ and $R_2(n,\sigma,\epsilon)$ in Section \ref{sec:residual}.
Approximating the $n$-fold composition using a 1-dimension distribution is also investigated by \cite{genise2020improved, kairouz2021distributed}, where they approximate the $n$-dimensional discrete Gaussian distribution using $W_n \sim \mathcal{N}_{\ZZ}(0,n\sigma^2).$
Moreover, an analytical upper bound on the approximation error is given in Corollary 12 of \cite{kairouz2021distributed}.

\begin{proof}[Proof of Theorem \ref{thm:privacy-profile}]

According to Lemma \ref{lemma:privacy-profile} and Proposition \ref{prop:IID-profile}, we have, for any $y\in\mathbb{Z}/B_n,$
\begin{align*}
    r_{n,\sigma}(y) = \mathbb{P}\left[ S_n = y\right] - \frac{1}{B_n}\phi(y).
\end{align*}
For $\phi(y),$ we have the following Fourier inversion formula:
\begin{align*}
    \phi(y) = \frac{1}{2 \pi} \int^{\infty}_{\infty} \ex^{-t^2/2} \ex^{-i t y} dt.
\end{align*}
Using the inversion formula for discrete distribution (cf., Exercise 3.3.2 (iii) in \cite{durrett2019probability}), we have
\begin{align*}
    \mathbb{P}[S_n = y] =\frac{1}{2 \pi B_n} \int_{-\pi B_n}^{\pi B_n} \ex^{-i \zeta y} f_{S_n}(\zeta) d\zeta,
\end{align*}
where $f_{S_n}$ is the characteristic function of $S_n$, i.e., $f_{S_n}(y) = \EE_{S_n} \ex^{iy S_n}.$
\eqref{eqn:char_func} shows that, 
\begin{align*}
\begin{split}
    f_{S_n}(t) = \EE \ex^{it S_n} 
    = \ex^{-t^2/2} \left( \frac{\theta_3\left( -i \sigma \pi t / \sqrt{n}, \ex^{-2 \sigma^2 \pi^2} \right)}{\theta_3\left(0,\ex^{-2\sigma^2\pi^2}\right)}\right)^{n}.
\end{split}
\end{align*}
This completes the proof of Theorem \ref{thm:privacy-profile}. 
\end{proof}

\subsection{Estimate the residual}
\label{sec:residual}
This subsection is to numerically estimate the residual terms $R_1(n,\sigma,\epsilon)$ and $R_2(n,\sigma,\epsilon)$ in Theorem \ref{thm:privacy-profile}. Technical details are deferred to Section \ref{sec:appe-estimate-residual}.

Numerically, we found that the residual term can indeed be extremely small for applications such as the Census. To estimate the residual term $R_1(n, \sigma, \epsilon)$, we decompose it as follows:
\begin{align*}
  R_1(n,\sigma,\epsilon) = \sum_{\frac{i}{B_n}>\frac{2\sigma^2\epsilon - n}{2B_n}}^{U_1} r_{n,\sigma}\left(\frac{i}{B_n}\right) + \PP\left[ S_n > U_1\right]
  =: \mathcal{E}_{11} + \mathcal{E}_{12}.
\end{align*}

For the error term $\mathcal{E}_{11}$, we note that $r_{n,\sigma}$ achieves an extremely small error that makes $\mathcal{E}_{11}$ negligible.
We have listed the estimates of $r_{n,\sigma}\left({i}/{B_n}\right)$ in Table \ref{table: iid error terms} which hold uniformly for all $i \in \mathbb{Z}.$
In the applications of the 2020 Census, where $n$ is at least $10$ and the smallest $\sigma^2$ is $4.99$, an error of $3.0 \times 10^{-37}$ is insignificant compared to the bureau’s choice of $\delta = 10^{-10}$. The details of the numerical bounds can be found in Section \ref{sec:appe-estimate-residual}.

For the second error term $\mathcal{E}_{12}$, we bound it using the sub-Gaussian tail bound \eqref{eq:sub-Gaussian-sum} and obtain 
$$
    \mathcal{E}_{12} = \PP\left[ S_n > U_1\right] \leq \ex^{-20^2/2}.
$$
Note that this term is numerically smaller than $1.4 \times 10^{-87}$.

\begin{table*}[ht]
\centering
\begin{tabular}{ |p{2.5cm}||p{2cm}|p{2cm}|p{2cm}|p{2cm}|  }
 \hline
 \multicolumn{5}{|c|}{Estimate of the residual $r_{n,\sigma}$} \\
 \hline
 $n$-fold Compositions of $\mathcal{N}_{\ZZ}(0, \sigma^2)$ & $\sigma^2 = 1$ &$\sigma^2 = 5$ &$\sigma^2 = 10$ & $\sigma^2 \geq 16$ \\
 \hline
 $n = 5$   & $5 \times 10^{-6}$  & $3 \times 10^{-32}$ & $2 \times 10^{-65}$ & $\ll 10^{-100}$
 \\
 $n = 9$   & $5 \times 10^{-7}$  & $1 \times 10^{-36}$ & $4 \times 10^{-74}$ & $\ll 10^{-100}$
 \\
 $n = 10$  &$4 \times 10^{-7}$ &$3 \times 10^{-37}$ & $3 \times 10^{-75}$ & $\ll 10^{-100}$
 \\
 $n = 18$   & $2 \times 10^{-7}$  & $2 \times 10^{-39}$ & $2 \times 10^{-79}$ & $\ll 10^{-100}$
 \\
 $n = 20$  &$9 \times 10^{-8}$ &$8 \times 10^{-40}$ & $4 \times 10^{-80}$ & $\ll 10^{-100}$
 \\
 $n = 27$  &$7 \times 10^{-8}$ &$2 \times 10^{-40}$ & $2 \times 10^{-81}$ & $\ll 10^{-100}$
 \\
 $n = 50$  &$4 \times 10^{-8}$ &$2 \times 10^{-41}$ & $3 \times 10^{-83}$ & $\ll 10^{-100}$
 \\
 $n = 100$ &$2 \times 10^{-8}$ &$6 \times 10^{-42}$ & $3 \times 10^{-84}$ & $\ll 10^{-100}$
 \\
 \hline
\end{tabular}
\caption{We bound the residual term $\sup_{y\in \mathbb{Z}/B_n}r_{n,\sigma}(y)$ numerically.}
\label{table: iid error terms}
\end{table*}

\begin{figure*}[!htp]
     \centering
    \begin{subfigure}[b]{0.45\textwidth}
        \includegraphics[width=\textwidth]{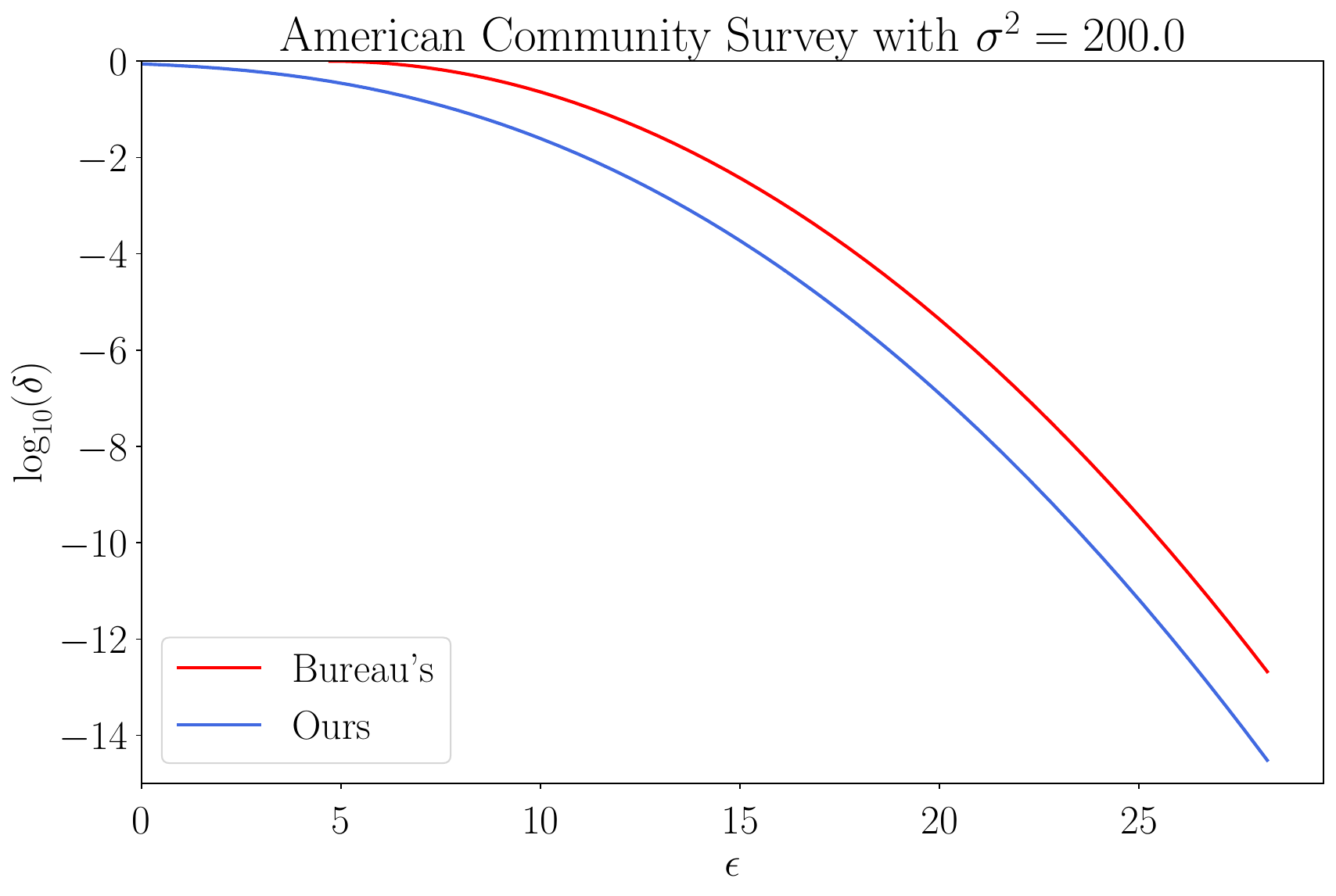}
    \end{subfigure}
    \begin{subfigure}[b]{0.45\textwidth}
        \includegraphics[width=\textwidth]{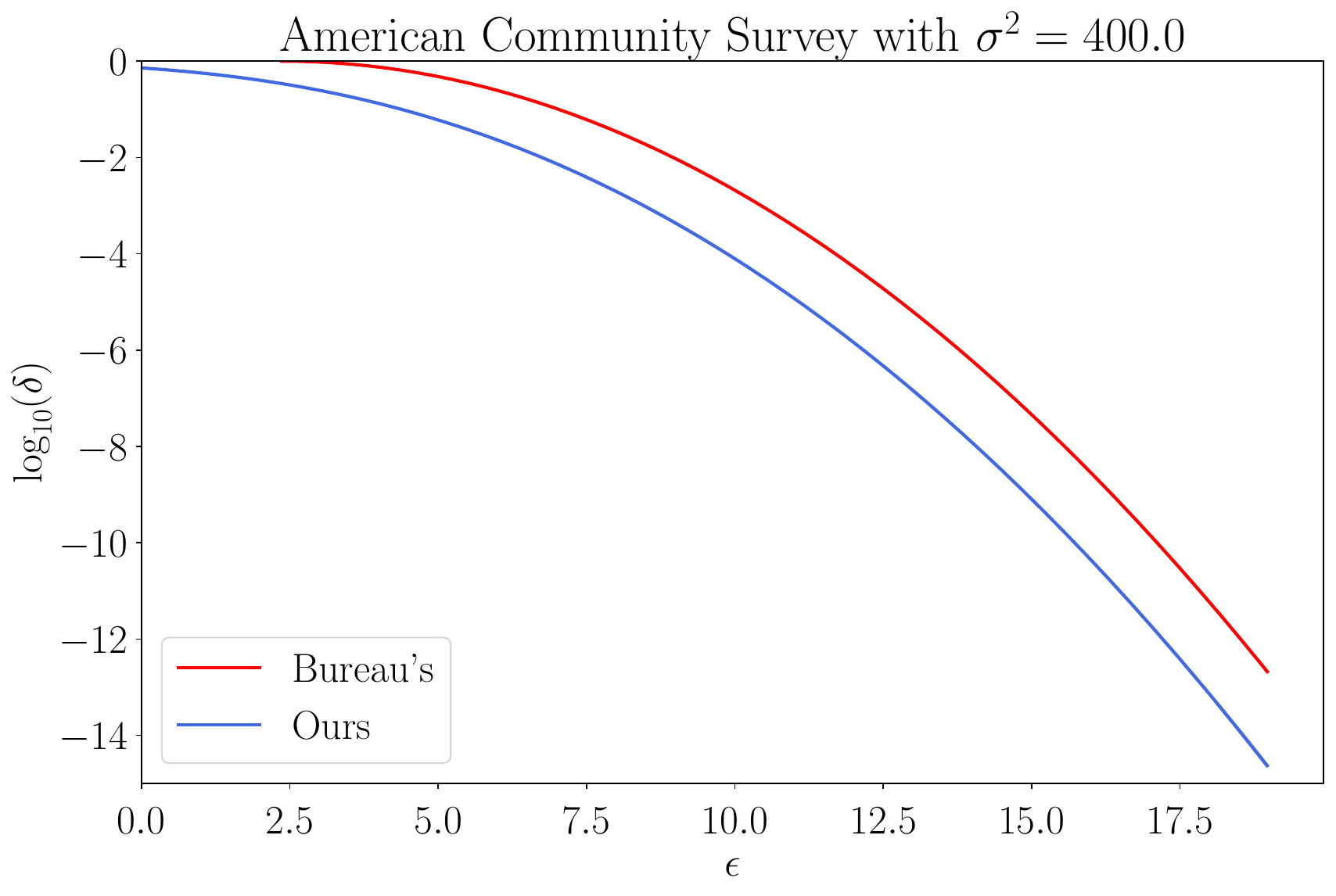}
    \end{subfigure}
    \caption{Comparisons with zCDP using American Community Survey 5-year data, a smaller $(\epsilon,\delta)$-curve means the privatized dataset is more private.}
    \label{fig:compare-ACS-5year} 
\end{figure*}

The other error term $R_2(n,\sigma,\epsilon)$ can be estimated similarly as follows:
\begin{align*}
    R_2(n,\sigma,\epsilon)  = \sum_{\frac{i}{B_n}>\frac{2\sigma^2\epsilon + n}{2B_n}}^{U_2} r_{n,\sigma}\left(\frac{i}{B_n}\right) + \PP\left[ S_n > U_2 \right]
    =: \mathcal{E}_{21} + \mathcal{E}_{22}.
\end{align*}
The two terms $\mathcal{E}_{21}$ and $\mathcal{E}_{22}$ can be estimated similarly.
Precisely, $\mathcal{E}_{22}$ is bounded using the sub-Gaussian tail bound and $\mathcal{E}_{21}$ is bounded using Table \ref{table: iid error terms}.
Then, the overall privacy budget is counted as 
\begin{align}
 \left| \delta(\epsilon) - \left\{ \frac{1}{B_n}\sum_{\frac{i}{B_n} > \frac{2 \sigma^2 \epsilon - n}{2B_n}}^{\frac{2 \sigma^2 \epsilon + n}{2B_n}} \phi\left(\frac{i}{B_n} \right) -  \ex^{\epsilon}\left(\frac{1}{B_n} \sum_{\frac{i}{B_n} > \frac{2 \sigma^2 \epsilon + n}{2 B_n}}^{ \frac{2 \sigma^2 \epsilon + n}{B_n}} \phi \left(\frac{i}{B_n} \right) \right) \right\} \right| 
 \leq \mathcal{E}_{11} + \mathcal{E}_{12} + \ex^{\epsilon}\left(\mathcal{E}_{21} + \mathcal{E}_{22} \right). \label{eq:approxi_profile}
\end{align}
Based on the upper bound in \eqref{eq:approxi_profile} and the error estimate in Table \ref{table: iid error terms}, we obtain the privacy budget $\epsilon$ in Figure \ref{fig:improve_epsilon} by solving $\delta(\epsilon) = 10^{-11}$ using binary search.
In addition to the number of folds of the composition in Table \ref{table: iid error terms}, Figure \ref{fig:compare-ACS-5year} compares our method using the ACS 5-year estimates with $n = 1890$. It shows that the our method enjoys greater advantage when the number of folds under composition is larger.

\subsection{Counting the overall \texorpdfstring{$(\epsilon,\delta(\epsilon))$-curve}{} of Allocation 2022-08-25}
\label{sec:inid-overall}

This section is to count the overall privacy budget among all 8 geographical levels that corresponds to total heterogeneous composition of the DGMs, which contains independent but not identically distributed discrete Gaussian noise for different geographical levels.
{The results presented here are used to derive the privacy profiles among the overall 8 geographical levels, specifically in Figure \ref{fig:tradeoff_epsdelta}.}
The technical details of all this section is postponed to Section \ref{proof:inid-overall}.

\begin{table}[!htp]
\centering
\begin{tabular}{ccccccc} 
 \hline
 $a_1$ & $a_2$ & $a_3$ & $a_4$ & $a_5$ & $a_6$ & $a_7$  \\
 
 \hline
 2.0\% & 27.40\% & 8.50\% & 13.10\% & 23.80\% & 11.80\% & 0.3\% \\
 
 \hline\hline
 \hline
 $n_1$ & $n_2$ & $n_3$ & $n_4$ & $n_5$ & $n_6$ & $n_7$  \\
 
 \hline
 10 & 10 & 10 & 20 & 10 & 10 & 10 \\
 
 \hline
\end{tabular}
\caption{Actual allocation of the $a_i$ and number of folds of composition $n_i$ for each geographical level in Privacy-loss Budget Allocation 2022-08-25 \citep{privacyallocation2022}.}
\label{table:real-allocation}
\end{table}

The allocation adopted by the bureau (the row corresponding to Bureau's) in Table \ref{table:2020 allocation}, is from the file released on 2022-08-25 \citep{privacyallocation2022}.
{ As we can see, the composition structure of the noise allocation in Table \ref{table:2020 allocation} aligns with our depiction in Figure \ref{fig:composition-chart}. Specifically, the noise within the same geographical level is i.i.d., while the noise across different geographical levels is i.n.i.d.\ As discussed in Section \ref{sec:fDP-DGM}, even in the case of 2-fold composition, the privacy profile for the i.n.i.d.\ setting is significantly more complicated than that of the i.i.d.\ case, and the result in Section \ref{sec:approx} for the i.i.d.\ case cannot be directly extended to the i.n.i.d.\ case. For general $n$-fold composition, we clarify the privacy profile as follows.
}
To count the overall privacy budget of the Allocation 2022-08-25, we divide the the $80$-fold i.n.i.d.\ composition into $k$ groups and each group $i$ is $n_i$-fold i.i.d.\ composition with $n_i$ being given in Table \ref{table:real-allocation}.
Let $\rho = 3.65$ be the total $\rho$-zCDP budget in Privacy-loss Budget Allocation 2022-08-25 and let $a_i$ be the allocation of the budget $\rho$ in the $i$-th geographical level. 
Moreover, denote $n = 10$, for each geographical level, each query is added an i.i.d.\ $\mathcal{N}_{\ZZ}(0, \sigma_{i}^2)$ with $\sigma^2_{i} = \frac{n}{2 a_i \rho}$. 
Then, the privacy profile is 
\begin{align} \label{eqn:inid_profile}
\begin{split}
    \delta(\epsilon) &= \mathbb{P}_{X_{ij}\sim\mathcal{N}_{\mathbb{Z}}(0,\sigma_i^2)}\left[\sum_{i=1}^k\frac{1}{\sigma_i^2}\sum_{j=1}^{n_{i}} X_{ij} > \epsilon - \sum_{i=1}^k\sum_{j=1}^{n_i}\frac{1}{2\sigma_i^2}\right]
    \\
    &\quad\quad- \ex^{\epsilon} \cdot \mathbb{P}_{X_{ij}\sim\mathcal{N}_{\mathbb{Z}}(0,\sigma_i^2)}\left[\sum_{i=1}^k\frac{1}{\sigma_i^2}\sum_{j=1}^{n_{i}} X_{ij} > \epsilon + \sum_{i=1}^k\sum_{j=1}^{n_i}\frac{1}{2\sigma_i^2}\right],
\end{split}
\end{align}
which is further simplified to 
\begin{align*}
    \delta(\epsilon) 
    &= \mathbb{P}_{X_{ij}\sim\mathcal{N}_{\mathbb{Z}}(0,\sigma_i^2)}\left[\sum_{i=1}^k a_i \sum_{j=1}^{n_{i}} X_{ij} > \epsilon \cdot \frac{n}{2 \rho}  - \sum_{i=1}^k \sum_{j=1}^{n_{i}} \frac{a_i}{2}\right]
    \\
    &\quad\quad- \ex^{\epsilon} \cdot \mathbb{P}_{X_{ij}\sim\mathcal{N}_{\mathbb{Z}}(0,\sigma_i^2)}\left[\sum_{i=1}^k a_i \sum_{j=1}^{n_{i}} X_{ij} > \epsilon \cdot \frac{n}{2 \rho} + \sum_{i=1}^k\sum_{j=1}^{n_{i}} \frac{a_i}{2}\right]
    \\
    &= \mathbb{P}_{X_{ij}\sim\mathcal{N}_{\mathbb{Z}}(0,\sigma_i^2)}\left[\sum_{i=1}^k a_i \sum_{j=1}^{n_{i}} X_{ij} > \epsilon \cdot \frac{n}{2 \rho}  - \frac{n}{2} \right]
    \\
    &\quad\quad- \ex^{\epsilon} \cdot \mathbb{P}_{X_{ij}\sim\mathcal{N}_{\mathbb{Z}}(0,\sigma_i^2)}\left[\sum_{i=1}^k a_i \sum_{j=1}^{n_{i}} X_{ij} > \epsilon \cdot \frac{n}{2 \rho} + \frac{n}{2}\right],
\end{align*}
where the last equality follows from the fact that $\sum_{i=1}^{k} \sum_{j=1}^{n_{i}} a_i = n$. 

{ Based on the above discussion, the main objective is to compute
$$
\mathbb{P}_{X_{ij}\sim\mathcal{N}_{\mathbb{Z}}(0,\sigma_i^2)}\left[\sum_{i=1}^k a_i \sum_{j=1}^{n_{i}} X_{ij} > t_{\epsilon}\right],
$$
where $t_{\epsilon} = \frac{n}{2} \left( \frac{\epsilon}{\rho} - 1 \right)$ or $\frac{n}{2} \left( \frac{\epsilon}{\rho} + 1 \right)$. This expression involves a sum over an $n$-dimensional lattice.
Direct computation is costly, which is why estimating the privacy profile remains an open challenge, as noted in \cite{kifer2022bayesian}.
We propose an efficient approximation to the privacy profile and outline it below.
}

\paragraph{Outline of our approximation of the privacy profile.}
Let $t_{\epsilon} = \frac{n}{2} \left( \frac{\epsilon}{\rho}  - 1\right)$ or $\frac{n}{2} \left( \frac{\epsilon}{\rho}  +1\right).$ To make our estimate of the privacy profile $\delta(\epsilon)$ easier to understand, we summarize our pipeline as follows.
\begin{align*}
    &\mathbb{P}_{X_{ij}\sim\mathcal{N}_{\mathbb{Z}}(0,\sigma_i^2)}\left[\sum_{i=1}^ka_i \sum_{j=1}^{n_{i}} X_{ij} > t_{\epsilon} \right]
    \overset{\text{Prop. \ref{prop:approx_0}}}{\approx} \nu \left( \sum_{i=1}^{k} a_i \Bar{X}_{i} \geq t_{\epsilon} \right)
    \\
    &\overset{\text{Prop. \ref{prop:approx_1}}}{\approx} \int_{0}^{\pi} F(t) dt
    \overset{\text{Fact \ref{fact:approx_2}}}{\approx} \int_{0}^{\frac{1}{100}} F(t) dt
    \\
    &\overset{\text{Fact \ref{fact:approx_3}}}{\approx} \sum_{k=1}^{\frac{10^{7}}{4}} \frac{2}{100 \times 45 \times 10^{7}} \times \left(7 \times F(x_{4k-4}) + 32F(x_{4k-3}) + 12 F(x_{4k-2}) + 32 F(x_{4k-1}) + 7 F(x_{4k}) \right),
\end{align*}
with some measure $\nu$ and function $F(t)$ defined later in Proposition \ref{prop:approx_0} and \ref{prop:approx_1}, respectively.
The errors in all the approximate equalities above can be bounded numerically, and we will demonstrate that these errors are small. Additionally, the final approximation comes from the Boole sum, which is also computable numerically. As a result, the privacy profile can be efficiently computed.

{\noindent\textbf{Remark.} The outline above shows that the privacy profile can be approximated by a summation over a function $F(t)$, which is derived using the Fourier transform.
Precisely, one has 
\begin{align*}
    &\mathbb{P}_{X_{ij}\sim\mathcal{N}_{\mathbb{Z}}(0,\sigma_i^2)}\left[\sum_{i=1}^ka_i \sum_{j=1}^{n_{i}} X_{ij} > t_{\epsilon} \right] \\
    &= \sum_{k=1}^{\frac{10^{7}}{4}} \frac{2}{100 \times 45 \times 10^{7}} \times \left(7 \times F(x_{4k-4}) + 32F(x_{4k-3}) + 12 F(x_{4k-2}) + 32 F(x_{4k-1}) + 7 F(x_{4k}) \right) + \mathcal{E},
\end{align*}
where $\mathcal{E}$ is an error term.
The approximation error $\mathcal{E}$ can be bounded using Proposition \ref{prop:approx_0}, Proposition \ref{prop:approx_1}, Fact \ref{fact:approx_2}, and Fact \ref{fact:approx_3}.
Our approach is hybrid, combining rigorous mathematical derivation of the error bound with numerical evaluation.
In the following, we use $t_\epsilon = \frac{n}{2}(\frac{\epsilon}{\rho} - 1)$ as an example, but the same method applies to $t_\epsilon = \frac{n}{2}(\frac{\epsilon}{\rho} + 1)$.
Although our numerical results are based on the Allocation 2022-08-25, our hybrid method generalizes to other allocation schemes with different numbers of compositions and noise levels, as the numerical error depends only on the noise parameters and the number of folds of composition.
}

\paragraph{Details of our approximation of the privacy profile.} 

\begin{proposition}
\label{prop:approx_0}
We have
    \begin{align*}
        &\Bigg| \mathbb{P}_{X_{ij}\sim\mathcal{N}_{\mathbb{Z}}(0,\sigma_i^2)}\left[\sum_{i=1}^ka_i \sum_{j=1}^{n_{i}} X_{ij} > t_{\epsilon} \right] -  \nu \left( \sum_{i=1}^{k} a_i \Bar{X}_{i} \geq t_{\epsilon} \right) \Bigg| < \mathcal{E}^{(0)}_{0} + \mathcal{E}^{(0)}_{1} + \mathcal{E}^{(0)}_{2},
    \end{align*}
    where $\mathcal{E}^{(0)}_{i}$, for $i \leq 3,$ are constants satisfying the following conditions:
    \begin{align*}
        &\mathcal{E}^{(0)}_{0} = k \times \ex^{-12^2/2},
        \\
        &\mathcal{E}^{(0)}_{1} = k \times \ex^{-12^2/2},
        \\
        &\mathcal{E}^{(0)}_{2} = \prod_{i=1}^{k} \left( 24 \sqrt{n_{i}} \sigma_{i} \right) \cdot \sum_{i=1}^k r_{n_i, \sigma_i} \cdot \max_{\{y_{j}\}_{j=1} ^{k} \in \ZZ^{k}} \Bigg| \left( \prod_{j=1}^{i-1} c_j \right) \left( \prod_{j=i+1}^{k} b_j \right) \Bigg|,
    \end{align*}
    with
    $$
    c_i = \mathbb{P}_{X_{ij}\sim\mathcal{N}_{\mathbb{Z}}(0,\sigma_i^2)} \left[ \sum_{j=1}^{n_{i}} X_{ij} = y_{i}  \right] \leq \frac{1}{\sqrt{n_{i} \sigma_{i}^{2}}} \phi\left(\frac{y_i}{\sqrt{n_{i} \sigma_{i}^{2}}} \right) + r_{n_{i}, \sigma_{i}} \leq \frac{1}{\sqrt{2 \pi n_{i} \sigma_{i}^{2}}} + r_{n_{i}, \sigma_{i}}, 
    $$
    and
    $$
    b_i = \frac{1}{\sqrt{n_{i} \sigma_{i}^{2}}} \phi\left(\frac{y_i}{\sqrt{n_{i} \sigma_{i}^{2}}} \right)\leq \frac{1}{\sqrt{2 \pi n_{i} \sigma_{i}^{2}}},
    $$
    and $\{\Bar{X}_{i}\}_{1 \leq i \leq k}$ is a sequence of independent discrete measurable functions on $\ZZ$, with measure (not a probability measure)
    \begin{align*}
        \nu \left(\Bar{X}_{i} = \xb \right) = \frac{1}{\sqrt{n_{i} \sigma_{i}^{2}}} \phi \left(\frac{x}{\sqrt{n_{i} \sigma_{i}^{2}}} \right),
    \end{align*}
    for $\xb \in \ZZ$.
\end{proposition}

\paragraph{Upper bound on \texorpdfstring{$\mathcal{E}^{(0)}_{i}$}{}.} According to Table \ref{table: iid error terms}, numerically, we have 
$$
\mathcal{E}^{(0)}_{2} = \prod_{i=1}^{k} \left( 24 \sqrt{n_{i}} \sigma_{i} \right) \cdot \sum_{i=1}^k r_{n_i, \sigma_i} \cdot \max_{\{y_{j}\}_{j=1} ^{k} \in \ZZ^{k}} \Bigg| \left( \prod_{j=1}^{i-1} c_j \right) \left( \prod_{j=i+1}^{k} b_j \right) \Bigg|\leq 3.4 \times 10^{-29}.
$$
Since $\ex^{(-12^2/2)}<5.4 \times 10^{-32}$, overall, we have 
$$
\mathcal{E}^{(0)}_{0}  +\mathcal{E}^{(0)}_{1} + \mathcal{E}^{(0)}_{2}<3.4 \times 10^{-29} + 7 \times 5.4 \times 10^{-32} + 7 \times 5.4 \times 10^{-32} < 3.41 \times 10^{-29}.
$$

\begin{proposition}\label{prop:approx_1}
\label{prop:error-inid-2}
The following approximation to $\nu$ holds. 
\begin{align} \label{eqn:approx_1}
\begin{split}
    \left| \nu \left( \sum_{i=1}^{k} a_i \Bar{X}_{i} \geq t_{\epsilon} \right) - \int_{0}^{\pi} F(t) dt \right| < \mathcal{E}^{(1)},
\end{split}
\end{align}
where 
\begin{align*}
    F(t) = \frac{1}{2 \pi} \left[\cos( \left\lceil t_{\epsilon} L \right\rceil \cdot t) + \cos(\left\lceil 6 t_{\epsilon}L \right\rceil \cdot t) + \frac{\cos(t/2)}{\sin(t/2)} \left(\sin(\left\lceil 6 t_{\epsilon}L \right\rceil \cdot t) - \sin(\left\lceil t_{\epsilon} L \right\rceil \cdot t) \right) \right] \prod_{i=1}^{k} f_{\Bar{X}_{i}}(a_i L t),
\end{align*}
$$
\mathcal{E}^{(1)} = \nu \left( \Bar{X}_{i} > \frac{6 \times t_{\epsilon}L}{k \cdot L \cdot a_i} \right), \hbox{ for } L = 10^3 \hbox{ and } a_i \hbox{ given in Table \ref{table:real-allocation}}.
$$
Moreover, the characteristic function $ f_{\Bar{X}_{i}}(a_i L t)$ is given by
\begin{align*}
    f_{\Bar{X}_{i}}(a_i L t) = \frac{1}{\sqrt{2 \pi n_{i} \sigma_{i}^{2}}} + 2 \sum_{u=1}^{\infty} \cos( u a_i L t) \cdot \frac{\ex^{ - \frac{u^2}{2} \cdot  \frac{1}{n_{i} \sigma_{i}^{2}}}}{\sqrt{2 \pi n_{i} \sigma_{i}^{2}}}.
\end{align*}
A similar result holds by replacing $t_{\epsilon}$ to $T_{\epsilon} = \epsilon + \sum_{i=1}^k\sum_{j=1}^{n_i}\frac{1}{2\sigma_i^2}$. 
\end{proposition}

\paragraph{Upper bound on \texorpdfstring{$\mathcal{E}^{(1)}$}{}.} Numerically, one can verify that $\mathcal{E}^{(1)}\leq 5.6\times 10^{-29}.$

Although \eqref{eqn:approx_1} is complicated, the following decomposition simplified the computation by further approximating \eqref{eqn:approx_1}. 
Precisely, the following fact indicates that we only need to consider the integral from $0$ to $\frac{1}{100}$. 
\begin{fact} \label{fact:approx_2}
The equation below shows that the integral of $F(t)$ over the interval $[0,\pi]$ is almost the same as the integral over $[0,1/100]$: 
\begin{align} \label{eqn:approx_2}
\begin{split}
    & \left| \int_{0}^{\pi} F(t) dt - \int_{0}^{\frac{1}{100}} F(t) dt \right| = \mathcal{E}^{(2)},
\end{split}
\end{align}
with $\mathcal{E}^{(2)} \leq 1.3 \times 10^{-30}$.
The remaining portion of the integral $\mathcal{E}^{(2)}$ beyond $\frac{1}{100}$ is negligible.
\end{fact}

Although \eqref{eqn:approx_2} simplified the integral, it is still complicated to numerically compute the integral. 
We has perform the numerical integral by using Mathematica, vpaintegration in Matlab and mpmath.quad in Python. 
Unfortunately, none of them give us the accurate answer when the error tolerance is $10^{-30}$. 
Therefore, we choose to manually compute the Boole's Sum of \eqref{eqn:approx_2}.

\begin{fact} \label{fact:approx_3}
Recall the definition of $F(t)$ from Proposition \ref{prop:approx_1}.
We numerically evaluate the integral of $F(t)$ over the interval $[0, 1/100]$ using Boole's rule, employing a partition of $N = 10^7 + 1$ points, denoted as $\{x_i\}_{i=1}^N$, where
\begin{align*}
    x_{i} = i \times h, \quad \hbox{with } h = \frac{1}{100(N-1)},
\end{align*}
for $i = 0, \cdots N-1$. 
Then, the following approximation holds
\begin{align*}
    \left| \int_{0}^{\frac{1}{100}} F(t) d t
    - \sum_{l=1}^{\frac{N-1}{4}} \frac{2 h}{45} \times \left(7 F(x_{4l-4}) + 32F(x_{4l-3}) + 12 F(x_{4l-2}) + 32 F(x_{4l-1}) + 7 F(x_{4l}) \right) \right| = \mathcal{E}^{(3)},
\end{align*}
with $\mathcal{E}^{(3)} \leq 2.54 \times 10^{-24}$.
\end{fact}

\paragraph{Total approximation error.}
Overall, we can approximate the first term of the privacy profile $\delta(\epsilon)$, $\mathbb{P}_{X_{ij}\sim\mathcal{N}_{\mathbb{Z}}(0,\sigma_i^2)}\left[\sum_{i=1}^ka_i \sum_{j=1}^{n_{i}} X_{ij} > t_{\epsilon} \right],$ by using the sum: 
    \begin{align*}
        \sum_{k=1}^{\frac{N-1}{4}} \frac{2 h}{45} \times \left(7 \times F(x_{4k-4}) + 32F(x_{4k-3}) + 12 F(x_{4k-2}) + 32 F(x_{4k-1}) + 7 F(x_{4k}) \right).
    \end{align*}
This can be efficiently computed with numerical methods. The total approximation error is bounded by $\mathcal{E}^{(0)}_0 + \mathcal{E}^{(0)}_1 + \mathcal{E}^{(0)}_2 + \mathcal{E}^{(1)} + \mathcal{E}^{(2)} + \mathcal{E}^{(3)} < 2.6 \times 10^{-24}.$

\paragraph{Computation time.}
\label{rmk:computation_time}
Computing the privacy budget $(\epsilon, \delta(\epsilon))$ within each geographical level as in Figure \ref{fig:improve_epsilon} takes less than 5 minute. 
However, calculating the overall privacy budget as in Section \ref{sec:inid-overall} across all eight levels requires more time. For each $\epsilon$, $\delta(\epsilon)$ can be computed within 9.5 hours using an AWS EC2 c5.metal instance with 96$\times$2GB virtual CPUs.

Accounting for privacy budgets using characteristic functions has been widely employed in previous literature, \cite{Koskela2020computing,gopi2021numerical, zhu2022optimal}. However, the computation time of our approach significantly outperforms previous methods.
To attain an error below $10^{-12}$, prior methods relied on the Riemann sum for numerical integral, resulting in a computational cost of at least $O(N)$ with $N = 3.4 \times 10^{16}$, which is computationally infeasible.
In contrast, our approach in Proposition \ref{prop:approx_1} and Fact \ref{fact:approx_3} leverages Boole's sum rather than the Riemann sum to calculate the Fourier transform, resulting in a significant improvement in computational efficiency. 
Additionally, \eqref{eqn:approx_2} enhances computation by splitting the integral into a main body and a remainder, with the remainder bounded by $1.3 \times 10^{-30}$. 
Consequently, only the main part of the integral needs to be computed. Overall, this results in a computational cost of $O(N)$ with $N = 10^7$.

\paragraph{Limitations in the computations.}
\label{sec:numerical_difficulty}
We would like to briefly discuss the primary limitation encountered in the numerical computations. Recall the privacy profile defined in \eqref{eqn:inid_profile}. The parameter $\delta$ used in the Privacy-loss Budget Allocation released on August 25, 2022, is set to $10^{-10}$, which imposes a requirement that the second probability in \eqref{eqn:inid_profile} be less than $10^{-10} / \ex^{21.97} \approx 2.8 \times 10^{-20}$. This value is smaller than the precision limit of Python’s floating-point arithmetic. Consequently, it is extremely difficult to compute this term numerically, even with high-precision libraries such as mpmath or scipy. As a result, in Section \ref{sec:enhanced_overall}, for any $\epsilon$, we present the following upper bound of the overall privacy budget $\delta(\epsilon)$: 
\begin{align*}
    \delta(\epsilon) < \mathbb{P}_{X_{ij}\sim\mathcal{N}_{\mathbb{Z}}(0,\sigma_i^2)}\left[\sum_{i=1}^k\frac{1}{\sigma_i^2}\sum_{j=1}^{n_{i}} X_{ij} > \epsilon - \sum_{i=1}^k\sum_{j=1}^{n_i}\frac{1}{2\sigma_i^2}\right].
\end{align*}
This limitation results in the overall privacy budget calculated in Section \ref{sec:enhanced_overall} showing less significant improvement compared to Figure \ref{fig:improve_epsilon}.

\subsection{Counting the overall trade-off function of Allocation 2022-08-25}
\label{sec:overall_trade_off}

\begin{figure*}[!htp]
    \centering
    \begin{subfigure}[b]{0.4\textwidth}
    \includegraphics[width=\textwidth]{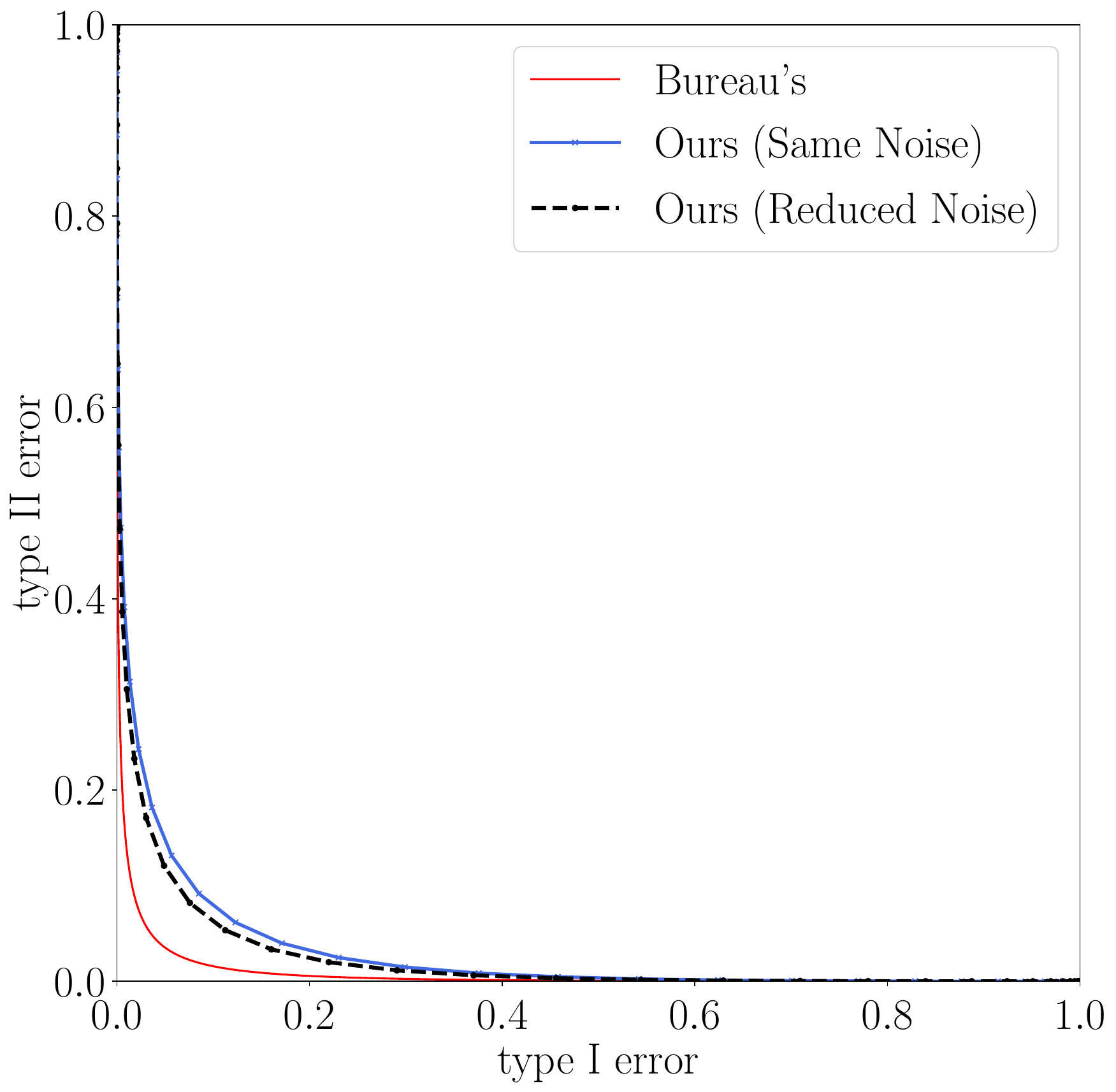}
    \end{subfigure}
  \caption{Trade-off functions for all geographical level of the 2020 U.S.\ Census, under the same noise level or after reducing the variance proxy by 8.59\% in our method.}
  \label{fig:tradeoff_alphabeta}
\end{figure*}

This section is to count the overall trade-off function among all 8 geographical levels that corresponds to the allocation adopted by the bureau (the row corresponding to Bureau's) in Table \ref{table:2020 allocation}.
{The results presented here are used to derive the trade-off functions shown in Figure \ref{fig:alpha_beta_geo_more} and Figure \ref{fig:tradeoff_alphabeta}.}
The technical details of all this section is similar to Section \ref{sec:inid-overall}.
The trade-off function is uniquely determined by the following parametric equation.
\begin{align*}
    \alpha(\zeta) =\ & \PP_{X_{i} \sim \cN_{\ZZ}(0, \sigma_{i}^2)} \left( \sum_{i=1}^k\frac{1}{\sigma_i^2}\sum_{j=1}^{n_{i}} X_{ij} > \zeta \right) + c \cdot \PP _{X_{i} \sim \cN_{\ZZ}(0, \sigma_{i}^2)} \left( \sum_{i=1}^k\frac{1}{\sigma_i^2}\sum_{j=1}^{n_{i}} X_{ij} = \zeta \right)
    \\
    \beta(\zeta) =\ & \PP_{X_{i} \sim \cN_{\ZZ}(0, \sigma_{i}^2)} \left( \sum_{i=1}^k\frac{1}{\sigma_i^2}\sum_{j=1}^{n_{i}} X_{ij} \leq \zeta - \sum_{i=1}^k\sum_{j=1}^{n_i} \frac{\mu}{\sigma_{i}^2} \right) - c \cdot \PP _{X_{i} \sim \cN_{\ZZ}(0, \sigma_{i}^2)} \left( \sum_{i=1}^k \frac{1}{\sigma_i^2}\sum_{j=1}^{n_{i}} X_{ij} = \zeta - \sum_{i=1}^k\sum_{j=1}^{n_i} \frac{\mu}{\sigma_{i}^2} \right)
\end{align*}
As $\sigma_{i}^{2} = n/2 a_i \rho$, we have
\begin{align*}
    \alpha(\zeta) =\ & \PP_{X_{i} \sim \cN_{\ZZ}(0, \sigma_{i}^2)} \left( \sum_{i=1}^k\frac{2 a_i \rho}{n}\sum_{j=1}^{n_{i}} X_{ij} > \zeta \right) + c \cdot \PP _{X_{i} \sim \cN_{\ZZ}(0, \sigma_{i}^2)} \left( \sum_{i=1}^k \frac{2 a_i \rho}{n} \sum_{j=1}^{n_{i}} X_{ij} = \zeta \right)
    \\
    =\ & \PP_{X_{i} \sim \cN_{\ZZ}(0, \sigma_{i}^2)} \left( \sum_{i=1}^k a_i \sum_{j=1}^{n_{i}} X_{ij} > \frac{n}{2 \rho} \cdot \zeta \right) + c \cdot \PP _{X_{i} \sim \cN_{\ZZ}(0, \sigma_{i}^2)} \left( \sum_{i=1}^k a_i \sum_{j=1}^{n_{i}} X_{ij} = \frac{n}{2 \rho} \cdot \zeta \right)
    \\
    \beta(\zeta) =\ & 
    \PP_{X_{i} \sim \cN_{\ZZ}(0, \sigma_{i}^2)} \left( \sum_{i=1}^k\frac{2 a_i \rho}{n}\sum_{j=1}^{n_{i}} X_{ij} \leq \zeta - \sum_{i=1}^k\sum_{j=1}^{n_i} \frac{2 a_i \rho}{n} \mu \right)
    \\
    &\qquad - c \cdot \PP _{X_{i} \sim \cN_{\ZZ}(0, \sigma_{i}^2)} \left( \sum_{i=1}^k \frac{2 a_i \rho}{n}\sum_{j=1}^{n_{i}} X_{ij} = \zeta - \sum_{i=1}^k\sum_{j=1}^{n_i} \frac{2 a_i \rho}{n} \mu \right)
    \\
    =\ & \PP_{X_{i} \sim \cN_{\ZZ}(0, \sigma_{i}^2)} \left( \sum_{i=1}^k a_i \sum_{j=1}^{n_{i}} X_{ij} \leq \frac{n}{2 \rho} \cdot \zeta - \sum_{i=1}^k a_i \sum_{j=1}^{n_{i}} \mu \right) 
    \\ 
    &\qquad - c \cdot \PP _{X_{i} \sim \cN_{\ZZ}(0, \sigma_{i}^2)} \left( \sum_{i=1}^k a_i \sum_{j=1}^{n_{i}} X_{ij} = \frac{n}{2 \rho} \cdot \zeta - \sum_{i=1}^k a_i \sum_{j=1}^{n_{i}} \mu \right)
\end{align*}
As $\mu = 1$, the following holds after reparametrization. 
\begin{align*}
    \alpha(\zeta) =\ & \PP_{X_{i} \sim \cN_{\ZZ}(0, \sigma_{i}^2)} \left( \sum_{i=1}^k a_i \sum_{j=1}^{n_{i}} X_{ij} > \zeta \right) + c \cdot \PP _{X_{i} \sim \cN_{\ZZ}(0, \sigma_{i}^2)} \left( \sum_{i=1}^k a_i \sum_{j=1}^{n_{i}} X_{ij} = \zeta \right)
    \\
    \beta(\zeta) =\ & \PP_{X_{i} \sim \cN_{\ZZ}(0, \sigma_{i}^2)} \left( \sum_{i=1}^k a_i \sum_{j=1}^{n_{i}} X_{ij} \leq \zeta - n \right) 
    - c \cdot \PP _{X_{i} \sim \cN_{\ZZ}(0, \sigma_{i}^2)} \left( \sum_{i=1}^k a_i \sum_{j=1}^{n_{i}} X_{ij} = \zeta - n \right).
\end{align*}
With $L$ defined in Proposition \ref{prop:error-inid-2}, we have
\begin{align*}
    \alpha(\zeta) =\ & \PP_{X_{i} \sim \cN_{\ZZ}(0, \sigma_{i}^2)} \left( \sum_{i=1}^k a_i L \sum_{j=1}^{n_{i}} X_{ij} > \zeta L \right) + c \cdot \PP _{X_{i} \sim \cN_{\ZZ}(0, \sigma_{i}^2)} \left( \sum_{i=1}^k a_i L \sum_{j=1}^{n_{i}} X_{ij} = \zeta L \right)
    \\
    \beta(\zeta) =\ & \PP_{X_{i} \sim \cN_{\ZZ}(0, \sigma_{i}^2)} \left( \sum_{i=1}^k a_i L \sum_{j=1}^{n_{i}} X_{ij} \leq \zeta L - n L \right) 
    - c \cdot \PP _{X_{i} \sim \cN_{\ZZ}(0, \sigma_{i}^2)} \left( \sum_{i=1}^k a_i L \sum_{j=1}^{n_{i}} X_{ij} = \zeta L - n L \right).
\end{align*}
For any $\zeta \notin \ZZ$, the following holds due to Proposition \ref{prop:approx_0} and \ref{prop:approx_1}.
\begin{align*}
    &\left| \alpha(\zeta) - \int_{0}^{\frac{1}{100}} F(t) dt \right| < 2.6 \times 10^{-24}
\end{align*}
where $F_{\alpha}(t)$ is defined as 
\begin{align*}
    F_{\alpha}(t) = \frac{1}{2 \pi} \left[\cos( \left\lceil \zeta L \right\rceil \cdot t) + \cos(\left\lceil U \right\rceil \cdot t) + \frac{\cos(t/2)}{\sin(t/2)} \left(\sin(\left\lceil U \right\rceil \cdot t) - \sin(\left\lceil \zeta L \right\rceil \cdot t) \right) \right] \prod_{i=1}^{k} f_{\Bar{X}_{i}}(a_i L t),
\end{align*}
with $U = 1.5 \times 10^{5}$. 
Moreover, we have
\begin{align*}
    \left| \beta(\zeta) - \int_{0}^{\frac{1}{100}} F_{\beta}(t) dt \right| < 2.4 \times 10^{-23},
\end{align*}
with 
\begin{align*}
    F_{\beta}(t)
    =\ & \frac{1}{2 \pi} \left[\cos( \left\lceil U \right\rceil \cdot t) + \cos(\left\lceil \zeta L - n L \right\rceil \cdot t) + \frac{\cos(t/2)}{\sin(t/2)} \left(\sin(\left\lceil \zeta L - n L \right\rceil \cdot t) + \sin(\left\lceil U \right\rceil \cdot t) \right) \right] \prod_{i=1}^{k} f_{\Bar{X}_{i}}(a_i L t).
\end{align*}

\section{Omitted details of Section \ref{sec:useful_facts}}
\label{proof:useful_facts}

\subsection{Proof of Lemma \ref{lemma:comparison}}
By the Poisson Summation Formula, we have
\begin{align*}
    \sum_{x \in \ZZ} \ex^{-\frac{(x - \mu)^2}{2 \sigma^2}} = \sqrt{2 \pi \sigma^2}\sum_{x \in \ZZ} \ex^{-2 \pi^2 \sigma^2 x^2} \ex^{-2 \pi i \mu x}.
\end{align*}
According to the Jacobi triple product, 
for $q = \ex^{-2 \pi^2 \sigma^2}$ and $z = \ex^{-2 \pi i \mu}$, 
the following equality holds. 
\begin{align*}
    \sum_{x \in \ZZ} \ex^{-2 \pi^2 \sigma^2 x^2} \ex^{-2 \pi i \mu x} = \prod_{m=0}^{\infty} (1 - q^{2m+2})(1 + z q^{2m+1})(1 + z^{-1} q^{2m+1}).
\end{align*}
Therefore, one has
\begin{align*}
    \frac{\sum_{x \in \ZZ} \ex^{-\frac{(x - \mu)^2}{2 \sigma^2}}}{\sum_{x \in \ZZ} \ex^{-\frac{(x - \nu)^2}{2 \sigma^2}}}
    =\ & \frac{\sum_{x \in \ZZ} \ex^{-2 \pi^2 \sigma^2 x^2} \ex^{-2 \pi i \mu x}}{\sum_{x \in \ZZ} \ex^{-2 \pi^2 \sigma^2 x^2} \ex^{-2 \pi i \nu x}}
    \\
    =\ & \prod_{m=0}^{\infty} \frac{(1 + \ex^{-2 \pi i \mu} q^{2m+1})(1 + \ex^{2 \pi i \mu} q^{2m+1})}{(1 + \ex^{-2 \pi i \nu} q^{2m+1})(1 + \ex^{2 \pi i \nu} q^{2m+1})}
    \\
    =\ & \prod_{m=0}^{\infty} \frac{1 + q^{4m + 2} + 2 \cos(2 \pi \mu) q^{2m+1}}{1 + q^{4m + 2} + 2 \cos(2 \pi \nu) q^{2m+1}}.
\end{align*}
Since $q > 0$ and $\cos(x)$ is an decreasing function in $[0,\pi]$, we have
\begin{align*}
    \frac{\sum_{x \in \ZZ} \ex^{-\frac{(x - \mu)^2}{2 \sigma^2}}}{\sum_{x \in \ZZ} \ex^{-\frac{(x - \nu)^2}{2 \sigma^2}}}
    = \prod_{m=0}^{\infty} \frac{1 + q^{4m + 2} + 2 \cos(2 \pi \mu) q^{2m+1}}{1 + q^{4m + 2} + 2 \cos(2 \pi \nu) q^{2m+1}} > 1. 
\end{align*}
This completes the proof of this lemma.

\subsection{Proof of Proposition \ref{prop:char_monotone}}
Recall \eqref{eqn:char_func}. It suffices to show that $\sum_{u=- \infty}^{\infty} \ex^{- \sigma^2 (t/B_{n} - 2 \pi u)^2/2}$ is non-increasing with respect to $t \in (0, \pi B_{n})$.
To see this, consider the following derivative: 
\begin{align*}
    &\frac{d}{d t} \sum_{u=- \infty}^{\infty} \ex^{- \sigma^2 (t/B_{n} - 2 \pi u)^2/2}
    \\
    =\ & \frac{d}{d t} \sum_{u=- \infty}^{\infty} \ex^{- (2 \pi \sigma)^2 (t/(2 \pi B_{n}) - u)^2/2}.
\end{align*}
Let $\mu$ and $\sigma^2$ in \eqref{eqn:dec_char_fn} be $t/(2 \pi B_{n})$ and $1/(2 \pi \sigma)^2$, respectively. Then, we have
\begin{align*}
    \frac{d}{d t} \sum_{u \in \ZZ} \ex^{-\frac{(2 \pi \sigma)^2 (u - t/(2 \pi B_{n}))^2}{2}} < 0,
\end{align*}
for any $0 < t/(2 \pi B_{n}) < 1/2$.

\section{Omitted details of Section \ref{sec:residual}}
\label{sec:appe-estimate-residual}
Recall that $B_n = \sqrt{n\sigma^2}.$
The main ingredient is to characterize the distribution of $S_n$ and bound the difference between the characteristic function of $S_n$ and that of $\frac{1}{B_n}\phi\left(\frac{i}{B_n} \right).$
As the residual term is estimated numerically and the numerically error depends on both $\sigma$ and $n$, for conciseness, we adopt the example $\sigma^2=5$ and $n=10$ (the smallest $n$ and $\sigma$ in real allocation files that implies the largest numerical error in our method).
The numerical estimate of the residual can be extended to any $\sigma$ and $n$.


\begin{fact}
[Estimate the residual of $N_{\ZZ}(0, 5)$ and $n=10$]
\label{fact:iid_approx}
    For any $x \in \ZZ$, we have
    \begin{align*}
        \sup_{x \in \ZZ} r_{n,\sigma} \left(\frac{x}{B_{n}} \right) = \sup_{x \in \ZZ} \left|\PP \left(S_{n} = \frac{x}{B_{n}} \right) - \frac{1}{B_{n}} \phi \left(\frac{x}{B_{n}} \right) \right| < 2.6 \times 10^{-37}.
    \end{align*}
\end{fact}

\subsection{Calculation of Fact \ref{fact:iid_approx}}

For any $y \in \ZZ/B_n$, recall the residual term 
\begin{align*}
    r_{n,\sigma}(y) &= \left| \mathbb{P}\left( S_n = y\right) - \frac{1}{B_n} \phi\left(y\right)\right|
    \\
    &=\frac{1}{2 \pi B_n} \left| \int_{-\pi B_n}^{\pi B_n} \ex^{-i t y} f_{S_n}(t) dt - \int^{\infty}_{\infty} \ex^{-t^2/2} \ex^{-i t y} dt \right|,   
\end{align*}
where $f_{S_n}$ is the characteristic function of $S_n$. Recall the closed-form representation of $f_{S_n}$ in \eqref{eqn:char_func}, i.e.,
\begin{align*}
\begin{split}
    f_{S_n}(t) = \EE \ex^{it S_n} 
    = \ex^{-t^2/2} \left( \frac{\theta_3\left( -i \sigma \pi t / \sqrt{n}, \ex^{-2 \sigma^2 \pi^2} \right)}{\theta_3\left(0,\ex^{-2\sigma^2\pi^2}\right)}\right)^{n}.
\end{split}
\end{align*}
Then, we have
\begin{align*}
    &\left|\PP \left(S_{n} = y \right) - \frac{1}{B_n} \phi \left(y \right) \right|\\
    \leq\ & \frac{1}{2 \pi B_n} \left|\int_{-\pi B_n}^{\pi B_n} \ex^{-i t y} f_{S_n}(t) dt - \int^{\infty}_{\infty} \ex^{-t^2/2} \ex^{-i t y} dt \right|\\
    \leq\ & \frac{1}{2 \pi B_n} \left|\int_{-\pi B_n}^{\pi B_n} \ex^{-i t y} f_{S_n}(t) dt - \int_{- \pi B_n}^{\pi B_n} \ex^{-t^2/2} \ex^{-i t y} dt \right| + \frac{1}{\pi B_n} \int_{\pi B_n}^{\infty} \ex^{-t^2/2} dt\\
    \leq\ & \frac{1}{2 \pi B_n} \int_{-\pi B_n}^{\pi B_n} \left| f_{S_n}(t) - \ex^{-t^2/2} \right| dt  + \frac{1}{\pi B_n} \int_{\pi B_n}^{\infty} \ex^{-t^2/2} dt.
\end{align*}
We decompose the upper bound into following parts: 
\begin{align*}
    &\frac{1}{2 \pi B_n} \int_{-\pi B_n}^{\pi B_n}  \left| f_{S_n}(t) - \ex^{-t^2/2} \right| dt  + \frac{1}{\pi B_n} \int_{\pi B_n}^{\infty} \ex^{-t^2/2} dt\\
    =\ &\sum_{i=1}^{\lfloor \pi B_n \rfloor} \frac{1}{\pi B_n} \int_{i}^{i+1} \left| f_{S_n}(t) - \ex^{-t^2/2} \right| dt + \frac{1}{\pi B_n} \int_{\lfloor \pi B_n \rfloor}^{\pi B_n} \left| f_{S_n}(t) - \ex^{-t^2/2} \right| dt + \frac{1}{\pi B_n} \int_{\pi B_n}^{\infty} \ex^{-t^2/2} dt\\
    =:\ &\Omega_{1} + \Omega_{2} + \Omega_{3}.
\end{align*}

\paragraph{Upper bound on \texorpdfstring{$\Omega_{1}$}{}.}
For $n=10$ and $\sigma^2 = 5,$ it holds $\Omega_{1} < 2.57 \times 10^{-37}$.

Consider $\Omega_{1}$ that corresponds to the case $t \in \left[0, \lfloor \pi B_n \rfloor \right]$. 
We observe that
\begin{align}
\label{eqn:char_trend}
    \frac{\partial}{\partial t} \left( \frac{\theta_3\left(-i \sigma \pi t/\sqrt{n}, \ex^{-2 \sigma^2 \pi^2}\right)}{\theta_3\left(0, \ex^{-2 \sigma^2 \pi^2} \right)} \right) 
    \left\{
    \begin{array}{rcl}
    < 0,     \qquad t < 0,\\
    = 0 ,    \qquad t = 0,\\
    > 0 ,    \qquad t > 0.
    \end{array}
    \right.
\end{align}
To see this, we note that
\begin{align*}
    \frac{\partial}{\partial t} \left( \frac{\theta_3\left(-i \sigma \pi t/\sqrt{n}, \ex^{-2 \sigma^2 \pi^2} \right)}{\theta_3\left(0, \ex^{-2 \sigma^2 \pi^2} \right)} \right) 
    =\ &  \frac{\partial}{\partial t} \left( \frac{\sum_{k = -\infty}^{\infty} \ex^{-2 \sigma^2 \pi^2 k^2} \ex^{2\pi \sigma k t/\sqrt{n}}}{\theta_3\left(0, \ex^{-2 \sigma^2 \pi^2} \right)} \right)
    \\
    =\ & \frac{\sum_{k = -\infty}^{\infty} 2\pi \sigma k/\sqrt{n} \cdot \ex^{-2 \sigma^2 \pi^2 k^2} \ex^{2\pi \sigma k t/\sqrt{n}}}{\theta_3\left(0, \ex^{-2 \sigma^2 \pi^2} \right)}
    \\
    =\ & \frac{\sum_{k = 1}^{\infty} 2\pi \sigma k/\sqrt{n} \cdot \ex^{-2 \sigma^2 \pi^2 k^2} \left( \ex^{2\pi \sigma k t/\sqrt{n}} - \ex^{-2\pi \sigma k t/\sqrt{n}} \right)}{\theta_3\left(0, \ex^{-2 \sigma^2 \pi^2} \right)},
\end{align*}
which obviously implies \eqref{eqn:char_trend}.
By \eqref{eqn:char_trend}, we  conclude that for any $t \in [j-1, j]$ and $1 \leq j \leq \lfloor \pi B_n \rfloor$, we have
\begin{align*}
    & \ex^{-t^2/2} \left|\left( \frac{\theta_3\left(-i \sigma \pi t / \sqrt{n}, \ex^{-2 \sigma^2 \pi^2} \right)}{\theta_3\left(0, \ex^{-2 \sigma^2 \pi^2} \right)} \right)^{n} - 1 \right|
    \\
    \leq\ & \ex^{-j^2/2} \left|\left( \frac{\theta_3\left(-i \sigma \pi (j+1) / \sqrt{n}, \ex^{-2 \sigma^2 \pi^2} \right)}{\theta_3\left(0, \ex^{-2 \sigma^2 \pi^2} \right)} \right)^{n} - 1 \right|. 
\end{align*}
Numerically, one can verify that, for any $1 \leq j \leq \lfloor \pi B_n \rfloor$, 
\begin{align*}
    \frac{1}{\pi B_n} \sum_{j=1}^{\lfloor \pi B_n \rfloor} \ex^{-(j-1)^2/2} \left|\left( \frac{\theta_3\left(-i \sigma \pi j / \sqrt{n}, \ex^{-2 \sigma^2 \pi^2} \right)}{\theta_3\left(0, \ex^{-2 \sigma^2 \pi^2} \right)} \right)^{n} - 1 \right| < 2.57 \times 10^{-37}.
\end{align*}
Therefore, $\Omega_1$ can be bounded as
\begin{align*}
    \Omega_1 = \sum_{i=1}^{\lfloor \pi B_n \rfloor} \frac{1}{\pi B_n} \int_{i}^{i+1} \left| f_{S_n}(t) - \ex^{-t^2/2} \right| dt < 2.57 \times 10^{-37}.
\end{align*}

\paragraph{Upper bound on \texorpdfstring{$\Omega_{2}$}{}.}
For $n=10$ and $\sigma^2 = 5,$ it holds 
$\Omega_{2} < 2.1 \times 10^{-106}.$

To bound $\Omega_2$, we decompose
\begin{align*}
    \Omega_2 =\ & \frac{1}{\pi B_n} \int_{\lfloor \pi B_n \rfloor}^{\pi B_n} \left| f_{S_n}(t) - \ex^{-t^2/2} \right| dt
    \\
    \leq\ & \frac{1}{\pi B_n} \int_{\lfloor \pi B_n \rfloor}^{\pi B_n} \left| f_{S_n}(t) \right| dt + \frac{1}{\pi B_n} \int_{\lfloor \pi B_n \rfloor}^{\pi B_n}  \ex^{-t^2/2} dt
    \\
    \leq\ & \Omega_4 + \Omega_5. 
\end{align*}
First, it is easy to see that
\begin{align*}
    \Omega_5 \leq \frac{\pi B_n - \lfloor \pi B_n \rfloor}{\pi B_n} \cdot \ex^{-\lfloor \pi B_n \rfloor^2/2} < 7.65 \times 10^{-110}. 
\end{align*}
Thus, it is enough to estimate  $\Omega_4$ numerically as follows.
Recall Proposition \ref{prop:char_monotone} that implies
$$
\max_{t \in [\lfloor \pi B_n \rfloor, \pi B_n]} f_{S_{n}}(t) = f_{S_{n}}(\lfloor \pi B_n \rfloor).
$$
Then, numerical results show that 
\begin{align*}
    \Omega_4 \leq \frac{\pi B_n - \lfloor \pi B_n \rfloor}{\pi B_n} \cdot f_{S_{n}}(\lfloor \pi B_n \rfloor) < 2.02 \times 10^{-106}.
\end{align*}

\paragraph{Upper bound on \texorpdfstring{$\Omega_{3}$}{}.}
For $n=10$ and $\sigma^2 = 5,$ it holds $\Omega_{3} < 1.45 \times 10^{-110}.$

For $x>0$, the Gaussian tail bound is given by
\begin{align} \label{eqn:gaussian tail bound}
    \int_{x}^{\infty} \ex^{-s^2/2} ds \leq \frac{1}{x} \ex^{-x^2/2}.
\end{align}
By \eqref{eqn:gaussian tail bound}, we have
\begin{align*}
    \Omega_{3} \leq \frac{1}{\pi B_n} \frac{\ex^{- (\pi B_n)^2/2}}{\pi B_n} < 1.45 \times 10^{-110}.
\end{align*}

\section{Omitted details of Section \ref{sec:inid-overall}}
\label{proof:inid-overall}

\subsection{Proof of Proposition \ref{prop:approx_0}}

Let $\Lambda_1$ and $\Lambda_2$ be the events defined as
\begin{align*}
    &\Lambda_1 := \bigcap_{i=1}^{k} \left\{\left| \sum_{j=1}^{n_{i}} X_{ij} \right| \leq 12 \cdot \sigma_{i} \sqrt{n_{i}} \right\}, 
    \\
    &\Lambda_2 := \bigcap_{i=1}^{k} \left\{\left| \Bar{X}_{i} \right| \leq 12 \cdot \sigma_{i} \sqrt{n_{i}} \right\}.
\end{align*}
By the triangle inequality, we have
\begin{align*}
    &\Bigg| \mathbb{P}_{X_{ij}\sim\mathcal{N}_{\mathbb{Z}}(0,\sigma_i^2)}\left[\sum_{i=1}^ka_i \sum_{j=1}^{n_{i}} X_{ij} > t_{\epsilon} \right] -  \nu \left( \sum_{i=1}^{k} a_i \Bar{X}_{i} \geq t_{\epsilon} \right) \Bigg|
    \\
    \leq\ &  \mathbb{P}_{X_{ij}\sim\mathcal{N}_{\mathbb{Z}}(0,\sigma_i^2)}\left[\sum_{i=1}^ka_i \sum_{j=1}^{n_{i}} X_{ij} > t_{\epsilon}, \Lambda_1^{c} \right] + \nu \left( \sum_{i=1}^{k} a_i \Bar{X}_{i} \geq t_{\epsilon}, \Lambda_2^{c} \right)
    \\
    &+  \Bigg| \mathbb{P}_{X_{ij}\sim\mathcal{N}_{\mathbb{Z}}(0,\sigma_i^2)}\left[\sum_{i=1}^ka_i \sum_{j=1}^{n_{i}} X_{ij} > t_{\epsilon}, \Lambda_1 \right] -  \nu \left( \sum_{i=1}^{k} a_i \Bar{X}_{i} \geq t_{\epsilon}, \Lambda_2 \right) \Bigg|.
\end{align*}
This further implies that 
\begin{align*}
    &\Bigg| \mathbb{P}_{X_{ij}\sim\mathcal{N}_{\mathbb{Z}}(0,\sigma_i^2)}\left[\sum_{i=1}^ka_i \sum_{j=1}^{n_{i}} X_{ij} > t_{\epsilon} \right] -  \nu \left( \sum_{i=1}^{k} a_i \Bar{X}_{i} \geq t_{\epsilon} \right) \Bigg|
    \\
    \leq\ &  \mathbb{P}_{X_{ij}\sim\mathcal{N}_{\mathbb{Z}}(0,\sigma_i^2)}\left(\Lambda_1^{c} \right) + \nu \left( \Lambda_2^{c} \right) 
    \\
    &+  \Bigg| \mathbb{P}_{X_{ij}\sim\mathcal{N}_{\mathbb{Z}}(0,\sigma_i^2)}\left[\sum_{i=1}^ka_i \sum_{j=1}^{n_{i}} X_{ij} > t_{\epsilon}, \Lambda_1 \right] -  \nu \left( \sum_{i=1}^{k} a_i \Bar{X}_{i} \geq t_{\epsilon}, \Lambda_2 \right) \Bigg|
    \\
    =\ & \Omega_6 + \Omega_7 + \Omega_8. 
\end{align*}

\paragraph{Upper bound on \texorpdfstring{$\Omega_6$}{}.}
We have
\begin{align*}
    \mathbb{P}_{X_{ij}\sim\mathcal{N}_{\mathbb{Z}}(0,\sigma_i^2)}\left[ \Lambda_1^c \right] \leq \sum_{i=1}^{k} \mathbb{P}_{X_{ij}\sim\mathcal{N}_{\mathbb{Z}}(0,\sigma_i^2)}\left[\left| \frac{1}{\sqrt{n_{i}}} \sum_{j=1}^{n_{i}} X_{ij} \right| > 12 \cdot \sigma_{i} \right].
\end{align*}
According to \eqref{eq:sub-Gaussian-sum}, $\sum_{j=1}^{n_{i}} X_{ij}$ is sub-Gaussian with variance proxy $\sqrt{n_i \sigma_i^2}.$
As a result, it holds
\begin{align}
\label{eq:inid-error-1-1}
\mathbb{P}_{X_{ij}\sim\mathcal{N}_{\mathbb{Z}}(0,\sigma_i^2)}\left[\left| \frac{1}{\sqrt{n_{i}}} \sum_{j=1}^{n_{i}} X_{ij} \right| > 12 \cdot \sigma_{i} \right]  \leq \ex^{- \frac{(12)^2 n \sigma_{i}^2}{2 n \sigma_{i}^2}} = \ex^{- \frac{12^2}{2}}.
\end{align}
Therefore, we have
\begin{align*}
\Omega_6\leq k \times \ex^{- \frac{12^2}{2}}=:\mathcal{E}^{(0)}_{0}.
\end{align*}

\paragraph{Upper bound on \texorpdfstring{$\Omega_7$}{}.}
Similar to the upper bound on $\Omega_6$, we have 
\begin{align*}
    \nu \left( \Lambda_2^{c} \right) 
    \leq\ & \sum_{i=1}^{k} \nu \left( \left| \Bar{X}_{i} \right| > 12 \cdot \sigma_{i} \sqrt{n_{i}} \right).
\end{align*}
Note that 
\begin{align*}
    \nu \left( \left| \Bar{X}_{i} \right| > 12 \cdot \sigma_{i} \sqrt{n_{i}} \right) 
    =\ & \sum_{ \{x \in \ZZ: x > 12 \cdot \sigma_{i} \sqrt{n_{i}}\} } \frac{1}{\sqrt{n_{i} \sigma_{i}^{2}}} \phi \left(\frac{x}{\sqrt{n_{i} \sigma_{i}^{2}}} \right)
    \\
    =\ & \sum_{ \{x \in \ZZ: x > 12 \cdot \sigma_{i} \sqrt{n_{i}}\} } \frac{1}{\sqrt{2 \pi n_{i} \sigma_{i}^{2}}} \ex^{-x^2/(2 n_i \sigma_i^2)}
    \\
    =\ & \frac{1}{\sqrt{2 \pi n_{i} \sigma_{i}^{2}}} \int_{\lfloor 12 \cdot \sigma_{i} \sqrt{n_{i}} \rfloor}^{\infty} \ex^{-x^2/(2 n_i \sigma_i^2)} dx
    \\
    \leq\ & \frac{1}{\sqrt{2 \pi n_{i} \sigma_{i}^{2}}} \ex^{-\frac{\lfloor 12 \cdot \sigma_{i} \sqrt{n_{i}} \rfloor^2}{2 n_i \sigma_i^2}} < \ex^{- \frac{12^2}{2}}.
\end{align*}
Therefore, we obtain
\begin{align*}
\Omega_7\leq k \times \ex^{- \frac{12^2}{2}} =: \mathcal{E}^{(0)}_{1}.
\end{align*}

\paragraph{Upper bound on \texorpdfstring{$\Omega_8$}{}.}
By the independence of $X_{ij}$ and $\Bar{X}_{i}$, we immediately have
\begin{align*}
    &\mathbb{P}_{X_{ij}\sim\mathcal{N}_{\mathbb{Z}}(0,\sigma_i^2)}\left[\sum_{i=1}^ka_i \sum_{j=1}^{n_{i}} X_{ij} > t_{\epsilon}, \Lambda_1 \right]
    \\
    =\ & \sum_{ \{y_{i}\}_{i=1}^{k} \in \ZZ^k} \one\left(\sum_{i=1}^k a_{i}  y_{i} > t_{\epsilon}, |y_{i}| \leq 12 \sqrt{n_{i}} \sigma_{i} \right) \cdot \prod_{i=1}^{k} \mathbb{P}_{X_{ij}\sim\mathcal{N}_{\mathbb{Z}}(0,\sigma_i^2)} \left[ \sum_{j=1}^{n_{i}} X_{ij} = y_{i}  \right],
\end{align*}
and 
\begin{align*}
    &\nu \left( \sum_{i=1}^{k} a_i \Bar{X}_{i} \geq t_{\epsilon}, \Lambda_2 \right) 
    \\
    =\ & \sum_{ \{y_{i}\}_{i=1}^{k} \in \ZZ^k} \one\left(\sum_{i=1}^k a_{i}  y_{i} > t_{\epsilon}, |y_{i}| \leq 12 \sqrt{n_{i}} \sigma_{i} \right) \cdot \prod_{i=1}^{k} \nu \left[ \Bar{X}_{i} = y_{i}  \right]
    \\
    =\ & \sum_{ \{y_{i}\}_{i=1}^{k} \in \ZZ^k} \one\left(\sum_{i=1}^k a_{i}  y_{i} > t_{\epsilon}, |y_{i}| \leq 12 \sqrt{n_{i}} \sigma_{i} \right) \cdot \prod_{i=1}^{k} \frac{1}{\sqrt{n_{i} \sigma_{i}^{2}}} \phi\left(\frac{y_i}{\sqrt{n_{i} \sigma_{i}^{2}}} \right).
\end{align*}
Recall the definition of $\Omega_8$.
It holds
\begin{align*}
    &\Bigg| \mathbb{P}_{X_{ij}\sim\mathcal{N}_{\mathbb{Z}}(0,\sigma_i^2)}\left[\sum_{i=1}^ka_i \sum_{j=1}^{n_{i}} X_{ij} > t_{\epsilon}, \Lambda_1 \right] -  \nu \left( \sum_{i=1}^{k} a_i \Bar{X}_{i} \geq t_{\epsilon}, \Lambda_2 \right) \Bigg|
    \\
    \leq & \sum_{\{y_{i}\}_{i=1} ^{k} \in \ZZ^{k}} \one\left(\sum_{i=1}^k a_{i}  y_{i} > t_{\epsilon}, |y_{i}| \leq 12 \sqrt{n_{i}} \sigma_{i} \right) \cdot \left| \prod_{i=1}^{k}  \mathbb{P}_{X_{ij}\sim\mathcal{N}_{\mathbb{Z}}(0,\sigma_i^2)} \left[ \sum_{j=1}^{n_{i}} X_{ij} = y_{i}  \right] - \prod_{i=1}^{k} \frac{1}{\sqrt{n_{i} \sigma_{i}^{2}}} \phi\left(\frac{y_i}{\sqrt{n_{i} \sigma_{i}^{2}}} \right) \right|
    \\
   < &  \sum_{\{y_{i}\}_{i=1} ^{k} \in \ZZ^{k}} \one\left(- 12 \sqrt{n_{i}} \sigma_{i} \leq y_{i} \leq 12 \sqrt{n_{i}} \sigma_{i} \right) \cdot \left| \prod_{i=1}^{k} \mathbb{P}_{X_{ij}\sim\mathcal{N}_{\mathbb{Z}}(0,\sigma_i^2)} \left[ \sum_{j=1}^{n_{i}} X_{ij} = y_{i}  \right] - \prod_{i=1}^{k} \frac{1}{\sqrt{n_{i} \sigma_{i}^{2}}} \phi\left(\frac{y_i}{\sqrt{n_{i} \sigma_{i}^{2}}} \right) \right|
    \\
    \leq &  \sum_{\{y_{i}\}_{i=1} ^{k} \in \ZZ^{k}} \one\left(- 12 \sqrt{n_{i}} \sigma_{i} \leq y_{i} \leq 12 \sqrt{n_{i}} \sigma_{i} \right)\\
    &\qquad \cdot \max_{\{y_{i}\}_{i=1} ^{k} \in \ZZ^{k}}\left| \prod_{i=1}^{k} \mathbb{P}_{X_{ij}\sim\mathcal{N}_{\mathbb{Z}}(0,\sigma_i^2)} \left[ \sum_{j=1}^{n_{i}} X_{ij} = y_{i}  \right] - \prod_{i=1}^{k} \frac{1}{\sqrt{n_{i} \sigma_{i}^{2}}} \phi\left(\frac{y_i}{\sqrt{n_{i} \sigma_{i}^{2}}} \right) \right|.
\end{align*}
Note that for two sequences $\{c_i\}_{i=1}^k$ and $\{b_i\}_{i=1}^k$, we have
\begin{align*}
    \Bigg| \prod_{i=1}^k c_i - \prod_{i=1}^k b_i \Bigg| =\ & \Bigg| \sum_{i=1}^k \left[ \left( \prod_{j=1}^{i-1} c_j \right) (c_i - b_i) \left( \prod_{j=i+1}^{k} b_j \right) 
    \right]\Bigg| \\
    \leq\ & \sum_{i=1}^k |c_i - b_i| \cdot \Bigg| \left( \prod_{j=1}^{i-1} c_j \right) \left( \prod_{j=i+1}^{k} b_j \right) \Bigg|.
\end{align*}
Therefore, let 
\begin{align*}
    &c_i = \mathbb{P}_{X_{ij}\sim\mathcal{N}_{\mathbb{Z}}(0,\sigma_i^2)} \left[ \sum_{j=1}^{n_{i}} X_{ij} = y_{i}  \right] \leq \frac{1}{\sqrt{n_{i} \sigma_{i}^{2}}} \phi\left(\frac{y_i}{\sqrt{n_{i} \sigma_{i}^{2}}} \right) + r_{n_{i}, \sigma_{i}} \leq \frac{1}{\sqrt{2 \pi n_{i} \sigma_{i}^{2}}} + r_{n_{i}, \sigma_{i}}, 
    \\
    &b_i = \frac{1}{\sqrt{n_{i} \sigma_{i}^{2}}} \phi\left(\frac{y_i}{\sqrt{n_{i} \sigma_{i}^{2}}} \right)\leq \frac{1}{\sqrt{2 \pi n_{i} \sigma_{i}^{2}}}, 
\end{align*}
and we have
\begin{align}
\label{eq:inid-error-1-2}
    &\Bigg| \mathbb{P}_{X_{ij}\sim\mathcal{N}_{\mathbb{Z}}(0,\sigma_i^2)}\left[\sum_{i=1}^k a_i \sum_{j=1}^{n_{i}} X_{ij} > t_{\epsilon},  \Lambda_1 \right]\nonumber
    \\
    & \qquad \qquad \qquad - \sum_{\{y_{i}\}_{i=1} ^{k} \in \ZZ^{k}} \one\left(\sum_{i=1}^k a_i  y_{i} > t_{\epsilon}, |y_{i}| \leq 12 \sqrt{n_{i}} \sigma_{i} \right) \cdot \prod_{i=1}^{k} \frac{1}{\sqrt{n_{i} \sigma_{i}^{2}}} \phi\left(\frac{y_i}{\sqrt{n_{i} \sigma_{i}^{2}}} \right) \Bigg|\nonumber
    \\
    \leq\ & \prod_{i=1}^{k} \left( 24 \sqrt{n_{i}} \sigma_{i} \right) \cdot \sum_{i=1}^k r_{n_i, \sigma_i} \cdot \max_{\{y_{j}\}_{j=1} ^{k} \in \ZZ^{k}} \Bigg| \left( \prod_{j=1}^{i-1} c_j \right) \left( \prod_{j=i+1}^{k} b_j \right) \Bigg|.
\end{align} 
Overall, we obtain 
\begin{align*}
    \Omega_9\leq  \prod_{i=1}^{k} \left( 24 \sqrt{n_{i}} \sigma_{i} \right) \cdot \sum_{i=1}^k r_{n_i, \sigma_i} \cdot \max_{\{y_{j}\}_{j=1} ^{k} \in \ZZ^{k}} \Bigg| \left( \prod_{j=1}^{i-1} c_j \right) \left( \prod_{j=i+1}^{k} b_j \right) \Bigg|=:\mathcal{E}^{(0)}_{2} .
\end{align*}
This completes the proof of Proposition \ref{prop:approx_0}.

\subsection{Proof of Proposition \ref{prop:error-inid-2}}
Let $L = 10^{3}$. For all $a_i L \in \ZZ$ with $a_i$ given in Table \ref{table:real-allocation}, we have
\begin{align*}
    \nu \left( \sum_{i=1}^{k} a_i \Bar{X}_{i} \geq t_{\epsilon} \right) 
    = \sum_{m \geq t_{\epsilon}L} \nu \left( \sum_{i=1}^{k} a_i L  \Bar{X}_{i} = m \right) = \sum_{6 \times t_{\epsilon}L \geq m \geq t_{\epsilon}L} \nu \left( \sum_{i=1}^{k} a_i L  \Bar{X}_{i} = m \right) + \cE^{(1)}, 
\end{align*}
where
\begin{align*}
    \cE^{(1)} = \nu \left( \sum_{i=1}^{k} a_i L \Bar{X}_{i} > 6 \times t_{\epsilon}L \right) \leq \sum_{i=1}^{k} \nu \left( \Bar{X}_{i} > \frac{6 \times t_{\epsilon}L}{k \cdot L \cdot a_i} \right).
\end{align*}
By discrete Fourier transform (Exercise 3.3.2 (iii) in \cite{durrett2019probability}), we have
\begin{align} \label{eqn:F_integrant}
\begin{split}
    & \sum_{6 t_{\epsilon}L \geq m \geq t_{\epsilon}L} \nu \left( \sum_{i=1}^{k} a_i L \Bar{X}_{i} = m \right)
    \\
    =\ & \sum_{6 t_{\epsilon}L \geq m \geq t_{\epsilon}L} \frac{1}{2 \pi} \int_{- \pi}^{\pi} \ex^{- i t m} \prod_{i=1}^{k} f_{a_i L \Bar{X}_{i}}(t) dt
    = \frac{1}{\pi} \int_{0}^{\pi} \sum_{6 t_{\epsilon}L \geq m \geq t_{\epsilon} L} \cos (t m) \prod_{i=1}^{k} f_{a_i L \Bar{X}_{i}}(t) dt
    \\
    =\ & \frac{1}{2 \pi} \int_{0}^{\pi} \left[\cos( \left\lceil t_{\epsilon} L \right\rceil \cdot t) + \cos(\left\lceil 6 t_{\epsilon}L \right\rceil \cdot t) + \frac{\cos(t/2)}{\sin(t/2)} \left(\sin(\left\lceil 6 t_{\epsilon}L \right\rceil \cdot t) - \sin(\left\lceil t_{\epsilon} L \right\rceil \cdot t) \right) \right] \prod_{i=1}^{k} f_{\Bar{X}_{i}}(a_i L t) dt,
\end{split}
\end{align}
where $f_{a_i L \Bar{X}_{i}}(t)$ and $f_{\Bar{X}_{i}}(t)$ are the characteristic functions of $a_i L \cdot \Bar{X}_{i}$ and $\Bar{X}_{i},$ correspondingly.
The closed-form representation of the $ f_{a_i L \Bar{X}_{i}}$ is given by
\begin{align*}
    f_{a_i L \Bar{X}_{i}}(t) = f_{\Bar{X}_{i}}(a_i L t) = \ & \sum_{u=-\infty}^{\infty} \ex^{i u a_i L t} \frac{\ex^{ - \frac{u^2}{2} \cdot  \frac{1}{n_{i} \sigma_{i}^{2}}}}{\sqrt{2 \pi n_{i} \sigma_{i}^{2}}}
    \\
    =\ & \frac{1}{\sqrt{2 \pi n_{i} \sigma_{i}^{2}}} + 2 \sum_{u=1}^{\infty} \cos( u a_i L t) \cdot \frac{\ex^{ - \frac{u^2}{2} \cdot  \frac{1}{n_{i} \sigma_{i}^{2}}}}{\sqrt{2 \pi n_{i} \sigma_{i}^{2}}}.
\end{align*}
Similar results hold if we replace $t_{\epsilon}$ with $T_{\epsilon} = \frac{n}{2} \left( \frac{\epsilon}{\rho}  + 1 \right)$.

\subsection{Calculation of Fact \ref{fact:approx_2}}

First, we have
\begin{align*}
    &\left| \frac{1}{2} \left[\cos( \left\lceil t_{\epsilon} L \right\rceil \cdot t) + \cos(\left\lceil 6 t_{\epsilon}L \right\rceil \cdot t) + \frac{\cos(t/2)}{\sin(t/2)} \left(\sin(\left\lceil 6 t_{\epsilon}L \right\rceil \cdot t) - \sin(\left\lceil t_{\epsilon} L \right\rceil \cdot t) \right) \right] \right|
    \\
    =\ & \left| \sum_{6 \times t_{\epsilon}L \geq m \geq t_{\epsilon} L} \cos (t m) \right|
    \leq 5 t_{\epsilon}L < 1.3 \times 10^{5}. 
\end{align*}
Let $c$ be a constant such that $|f_{\Bar{X}_{i}}(a_i L t)| = |\ex^{i \Bar{X}_{i} a_i L t}| \leq c$. Then, it holds
    \begin{align*}
        \left|\prod_{i=1}^{k} f_{\Bar{X}_{i}}(a_i L t)\right| = \left| \prod_{i=1}^{k} f_{a_i L \Bar{X}_{i}}(t) \right| \leq c^{k-1} \cdot \min \left\{|f_{\Bar{X}_{1}}(a_1 L t)|, \cdots, |f_{\Bar{X}_{k}}(a_k L t)| \right\}.
    \end{align*}
 Numerically, one can verify that $c < 1 + 1.0 \times 10^{-50}$. 
    \textnormal{Output of Characteristic Function Evaluation} in \href{https://github.com/BuxinSu/Census_Privacy.git}{GitHub} records the numerical value of $\{f_{\Bar{X}_{1}}(a_1 L t), \cdots, f_{\Bar{X}_{k}}(a_k L t)\}$ for all $t\in \Lambda$ with $\Lambda$ given by 
    \begin{align*}
        \Lambda = \left\{\frac{j}{200} \times \frac{\pi}{a_i L}: 0 \leq j \leq 200 \times a_i L, 1 \leq i \leq k \right\}. 
    \end{align*}
    Recall Proposition \ref{prop:char_monotone} that $f_{\Bar{X}_{i}}(a_i L t)$ is a ${2 \pi}/{(a_i L)}$-periodic function and monotone within any $\left[ k \times \frac{\pi}{a_i L}, (k+1) \times \frac{\pi}{a_i L} \right], k \in \ZZ$. 
    Inspired by the Sieve Method from number theory, we make the following observations in the order of decreasing $\sigma_{i}^2$: 
    \begin{align*}
        &\left|\prod_{i=1}^{k} f_{\Bar{X}_{i}}(a_i L t)\right| < c^{k-1} \cdot f_{\Bar{X}_{7}}(a_7 L t) < 10^{-35}, \quad \text{for}\ t \in [0.062831853, 2.031563249] \cup [2.157226955, \pi].
        \\
        &\left|\prod_{i=1}^{k} f_{\Bar{X}_{i}}(a_i L t)\right| < c^{k-1} \cdot f_{\Bar{X}_{1}}(a_1 L t) < 10^{-35}, \quad \text{for}\ t \in [0.024347343, 0.062831853] \cup [2.031039651, 2.158274153].
        \\
        &\left|\prod_{i=1}^{k} f_{\Bar{X}_{i}}(a_i L t)\right| < c^{k-1} \cdot f_{\Bar{X}_{3}}(a_3 L t) < 10^{-35}, \quad \text{for}\ t \in [0.011827172, 0.024578343].
        \\
        &\left|\prod_{i=1}^{k} f_{\Bar{X}_{i}}(a_i L t)\right| < c^{k-1} \cdot f_{\Bar{X}_{2}}(a_2 L t) < 10^{-35}, \quad \text{for}\ t \in [0.006592758, 0.011866965].
    \end{align*}
    Therefore, we conclude for all $t \in [0.006592758, \pi] \subset [1/100, \pi]$, it holds
    \begin{align*}
        \left|\prod_{i=1}^{k} f_{\Bar{X}_{i}}(a_i L t)\right| < 10^{-35},
    \end{align*}
    which further implies that
    \begin{align*}
        \mathcal{E}^{(2)} < \frac{1}{\pi} \times \pi \times 10^{-35} \times 1.3 \times 10^{5} < 1.3 \times 10^{-30}.
    \end{align*}

\subsection{Calculation of Fact \ref{fact:approx_3}}
Recall \eqref{eqn:F_integrant} that,
\begin{align*}
    F(t) = \frac{1}{\pi} \sum_{6 t_{\epsilon}L \geq m \geq t_{\epsilon} L} \cos (t m) \EE \ex^{i t X} = \frac{1}{\pi} \sum_{6 t_{\epsilon}L \geq m \geq t_{\epsilon} L} \EE \cos (t m) \cos(t X),
\end{align*}
where $X = a_1 L \bar{X}_1 + \cdots  + a_k L \bar{X}_k$. 

\paragraph{Moments of \texorpdfstring{$X$}{}.}
We compute the moments of $X$ using the triangle inequality
\begin{align*}
    \|X\|_{L_p}^{p} \leq \left( \|a_1 L \bar{X}_1\|_{L_p} + \cdots + \| a_k L \bar{X}_k\|_{L_p} \right)^{p}. 
\end{align*}
Moreover, numerically, it holds 
$\EE |X| \leq 7.0 \times 10^{3}$, 
$\EE |X|^2 \leq 7.6 \times 10^{7}$,
$\EE |X|^3 \leq 1.1 \times 10^{12}$,
$\EE |X|^4 \leq 1.8 \times 10^{16}$,
$\EE |X|^5 \leq 3.3 \times 10^{20}$,
$\EE |X|^6 \leq 6.6 \times 10^{24}$.

\paragraph{Upper Bound on \texorpdfstring{$6$-th order differentiation of $F(t)$}{}.}
The $6$-th order differentiation of $F(t)$ is bounded as follows. 
\begin{align*}
    \sup_{t} \left| F^{(6)}(t) \right| \leq\ & \frac{1}{\pi} \cdot \sum_{6 t_{\epsilon}L \geq m \geq t_{\epsilon} L} \left| \frac{d^6}{d t^6} \EE \cos (t m) \cos(t X) \right| 
    \\
    =\ & \sum_{6 t_{\epsilon} L \geq m \geq t_{\epsilon} L} \EE \bigg|  2 m X \left(3 m^4+10 m^2 X^2+3 X^4\right) \sin (t  m) \sin(t  X)
    \\
    &\qquad -\left(m^6+15 m^4X^2+15 m^2 X^4+X^6\right) \cos(t  m) \cos (t  X) \bigg|
    \\
    \leq\ & \frac{1}{\pi} \cdot \sum_{6 t_{\epsilon}L \geq m \geq t_{\epsilon} L} m^6 + 6 m^5 \EE|X| + 15 m^4 \EE X^2 + 20 m^3 \EE |X|^3 + 15 m^2 \EE X^4 + 6 m \EE |X|^5 + \EE |X|^6
    \\
    \leq\ & \frac{1}{\pi} \cdot \sum_{6 t_{\epsilon}L \geq m \geq t_{\epsilon} L} m^6 + 6 m^5 (7.0 \times 10^{3}) + 15 m^4 (7.6 \times 10^{7})
    \\
    &\quad + 20 m^3 (1.1 \times 10^{12}) + 15 m^2 (1.8 \times 10^{16}) + 6 m (3.3 \times 10^{20}) + (6.6 \times 10^{24})
    \\
    <\ & \frac{1}{\pi} \times 3.54 \times 10^{35} = 1.2 \times 10^{35}.
\end{align*}

\paragraph{Boole's sum and error bound.}
\label{rmk:Boole_error_bound}
    Consider the integral $\int_{a}^{b} F(x) dx.$ Let $\{x_i\}_{i=1}^N$ be a partition of $[a,b]$ with
    \begin{align*}
        x_{i} := a + i \times h, \quad h = \frac{b-a}{N-1},
    \end{align*}
    for $i = 0, \cdots N-1$. 
    Consider the following discretization of the integration.
    \begin{align*}
        &\sum_{l=1}^{\frac{N-1}{ 4}} \frac{2 h}{45} \times \left(7 F(x_{4l-4}) + 32F(x_{4l-3}) + 12 F(x_{4l-2}) + 32 F(x_{4l-1}) + 7 F(x_{4l}) \right)
        \\
        =\ & \frac{2 h}{45} \times \left[ 7 \left( F(a) + F(b) \right) + \sum_{k=1}^{(N-1)/4} \left( 32F(x_{4l-3}) + 12 F(x_{4l-2}) + 32 F(x_{4l-1}) \right) + 14 \times \sum_{k=1}^{m-1} F(x_{4l}) \right].
    \end{align*}
    Equation 3.2 in \cite{Sablonniere2010error} indicates that the error is bounded as 
    \begin{align*}
        &\left|\int_{a}^{b} F(t) dt - \sum_{k=1}^{\frac{N-1}{4}} \frac{2 h}{45} \times \left(7 F(x_{4l-4}) + 32F(x_{4l-3}) + 12 F(x_{4l-2}) + 32 F(x_{4l-1}) + 7 F(x_{4l}) \right) \right|
        \\
        \leq\ & \frac{2}{945} \cdot \max_{t \in [a,b]} f^{(6)}(t) \cdot h^6 \cdot (b-a).
    \end{align*}
    In our \href{https://github.com/BuxinSu/Census_2020_Privacy/blob/main/Numerical/Census_2022_08_eps_delta_division_eps.py}{numerical results}, we set $h = \frac{1}{100 \times 10^{7}}$, $a = 0$, and $b= \frac{1}{100}$. Therefore, the error bound is given by 
    \begin{align*}
        &\left|\int_{0}^{\frac{1}{100}} F(t) dt - \sum_{k=1}^{\frac{N-1}{4}} \frac{2 h}{45} \times \left(7 F(x_{4l-4}) + 32F(x_{4l-3}) + 12 F(x_{4l-2}) + 32 F(x_{4l-1}) + 7 F(x_{4l}) \right) \right|
        \\
        \leq\ & \frac{2}{945} \times 1.2 \times 10^{35} \times \frac{1}{100} \times \left( \frac{1}{100 \times 10^{7}} \right)^{6} 
        \\
        <\ & 2.54 \times 10^{-24}.
    \end{align*}

\section{Supplementary figures}
\label{sec:supp_fig}

\begin{figure*}[!htp]
    \centering
    \begin{subfigure}[b]{0.4\textwidth}
    \includegraphics[width=\textwidth]{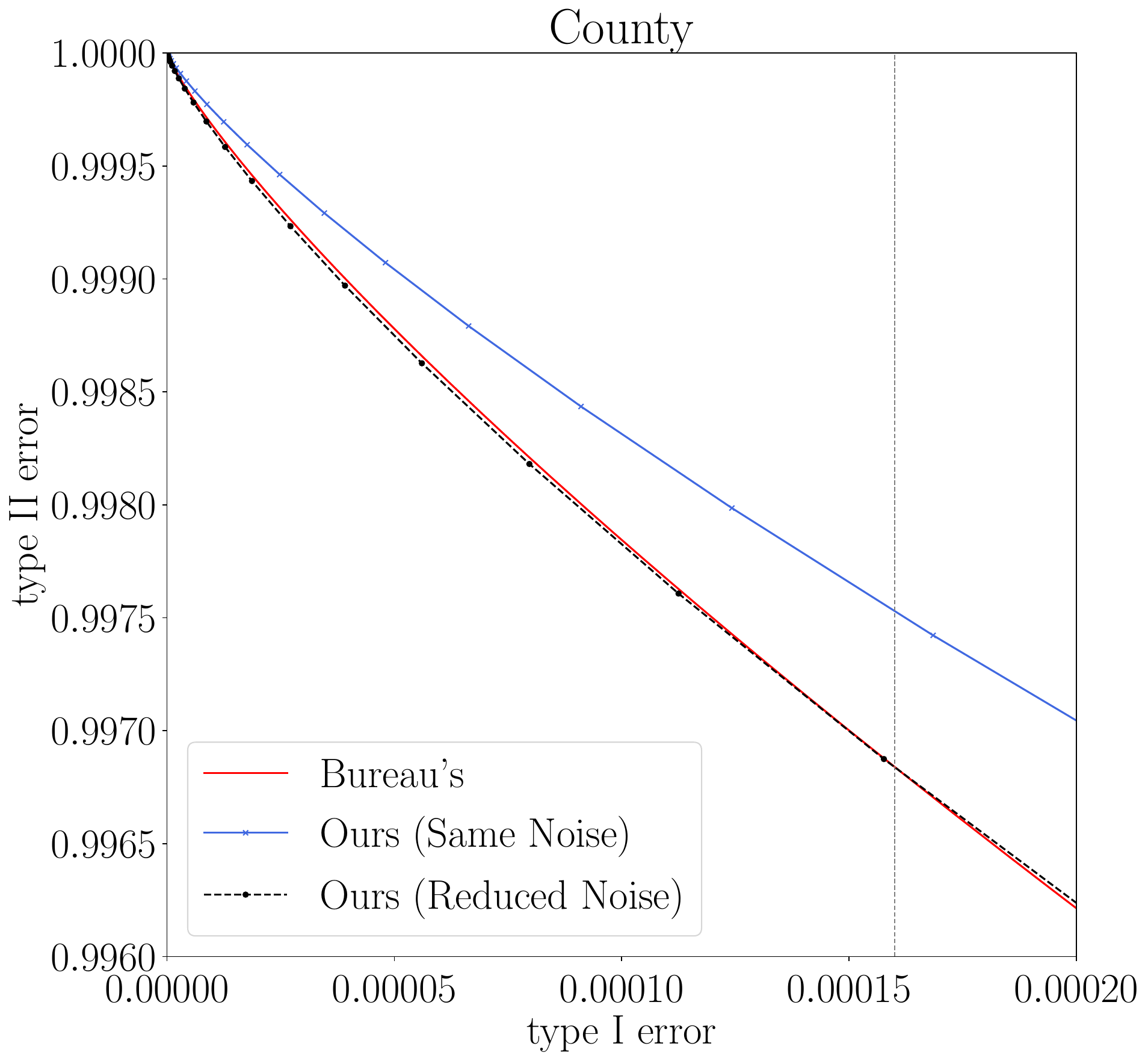}
    \end{subfigure}
  \caption{Zoomed-in trade-off functions for County of the 2020 U.S.\ Census, under the same noise level or after reducing the variance proxy by 8.59\% in our method.}
  \label{fig:tradeoff_alphabeta_zoom_in}
\end{figure*}

\begin{figure*}[!htp]
    \centering
    \begin{subfigure}[b]{0.23\textwidth}
        \includegraphics[width=\textwidth]{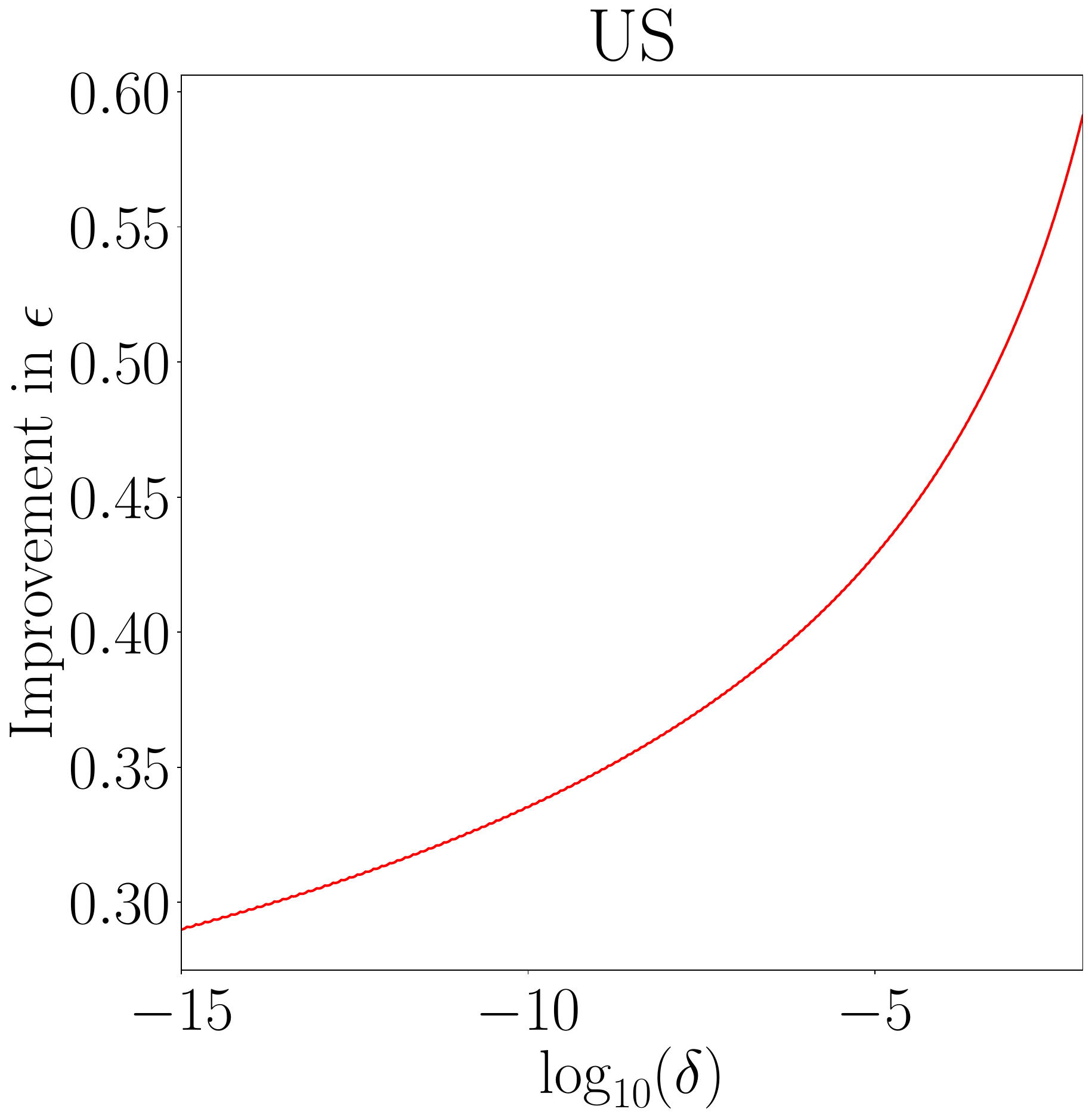}
    \end{subfigure}
    \begin{subfigure}[b]{0.23\textwidth}
        \includegraphics[width=\textwidth]{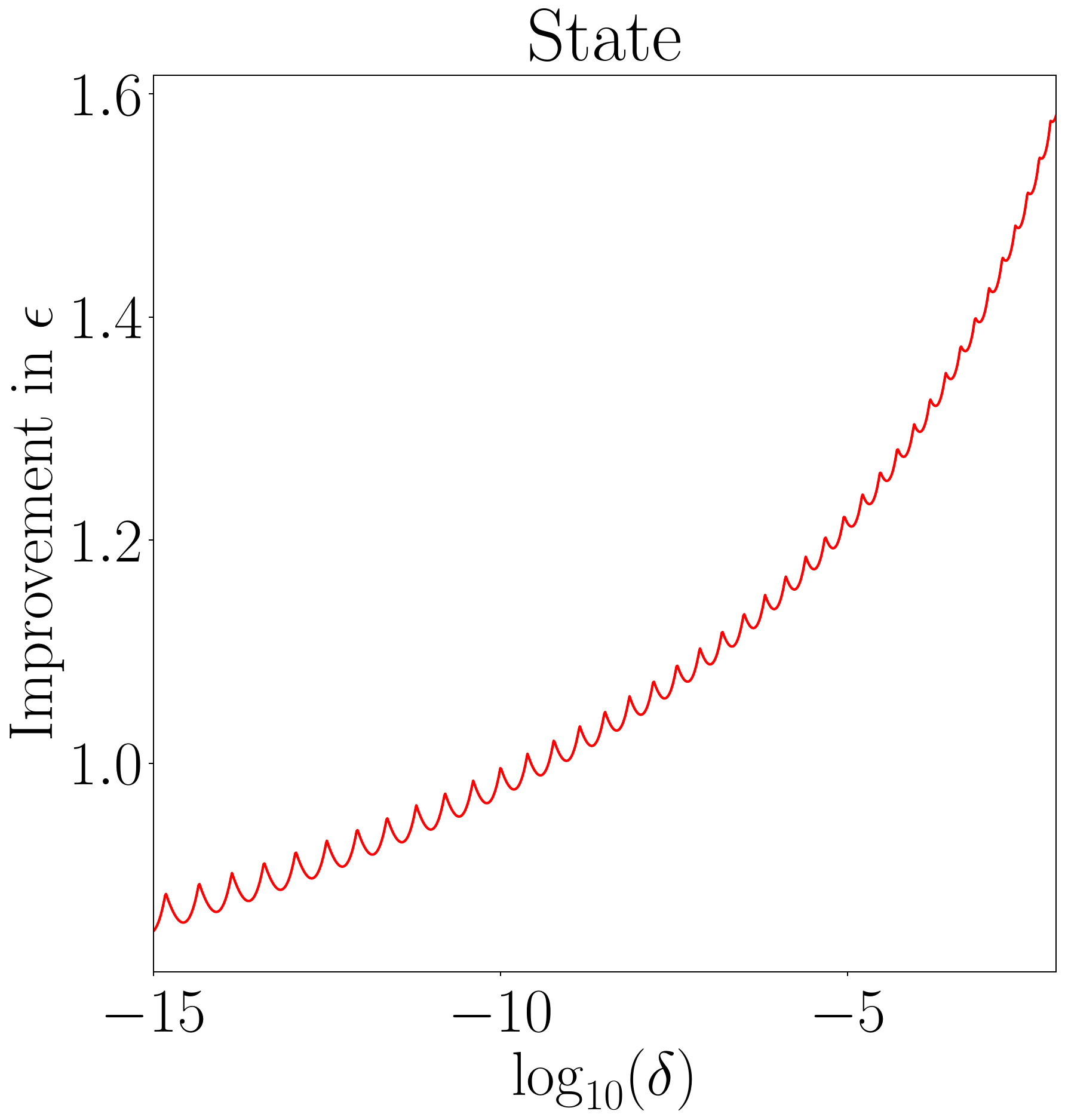}
    \end{subfigure}   
    \begin{subfigure}[b]{0.23\textwidth}
        \includegraphics[width=\textwidth]{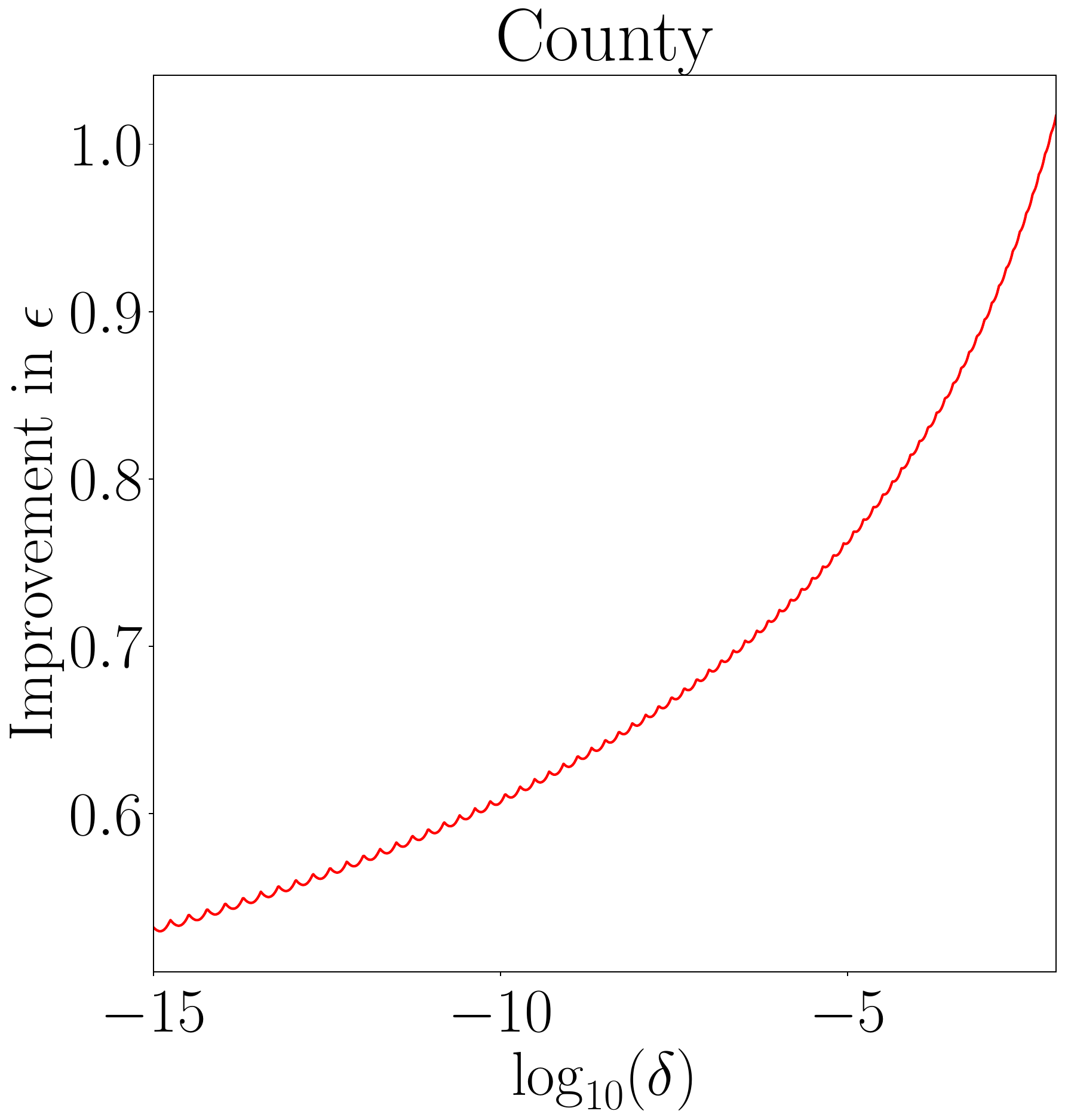}
    \end{subfigure}
    \begin{subfigure}[b]{0.23\textwidth}
        \includegraphics[width=\textwidth]{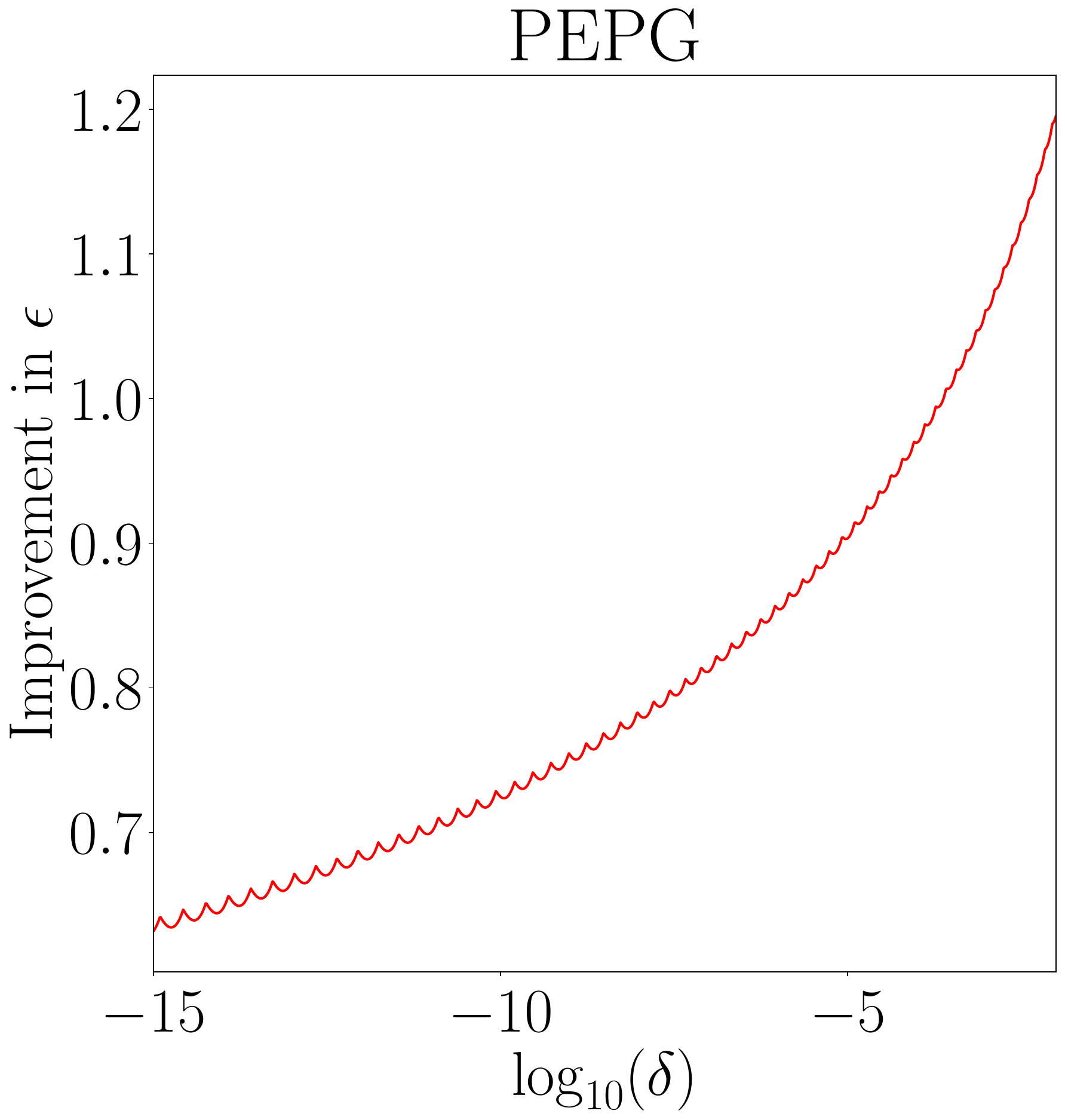}
    \end{subfigure}
    
    \hfill
    
    \begin{subfigure}[b]{0.23\textwidth}
        \includegraphics[width=\textwidth]{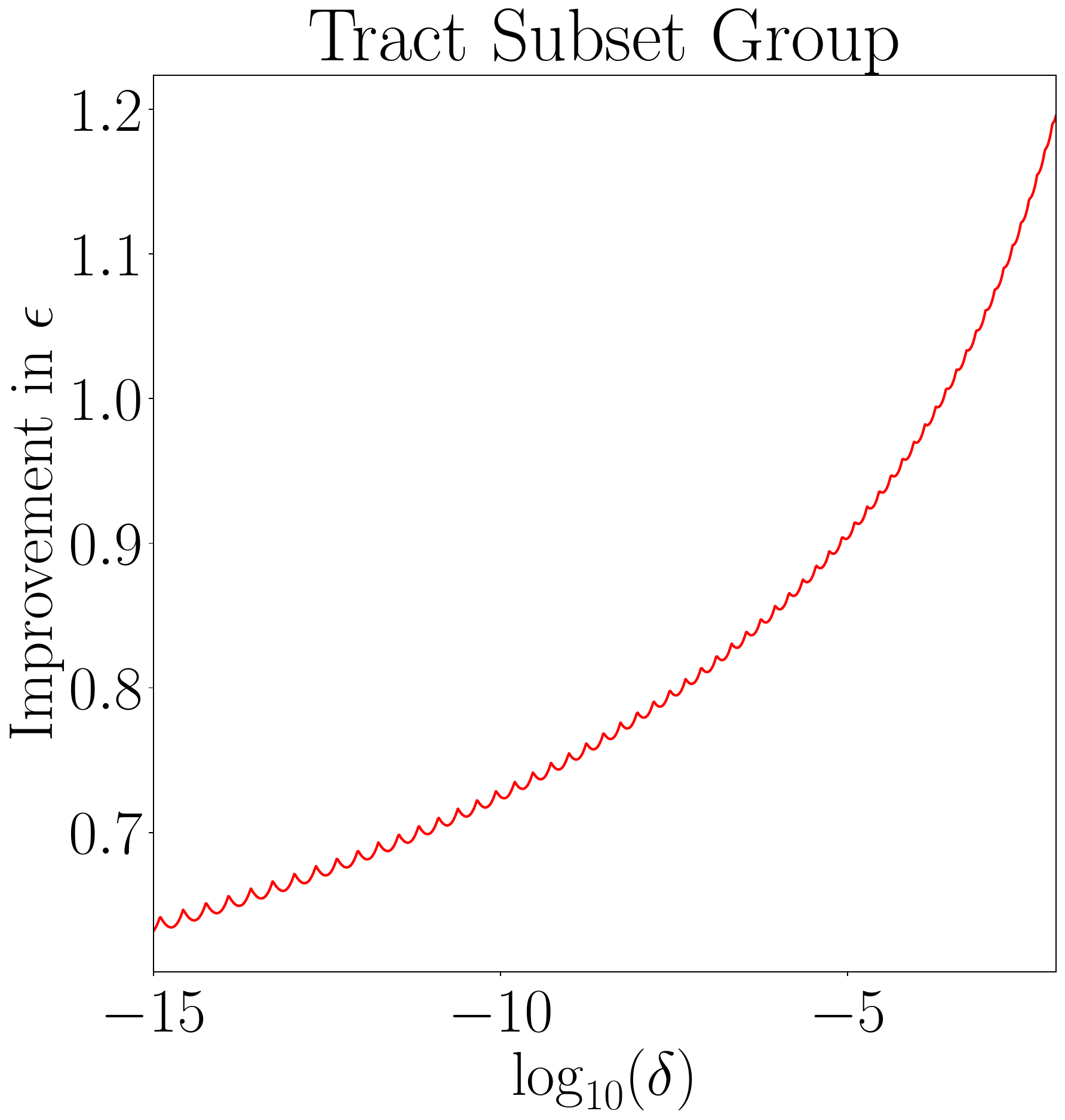}
    \end{subfigure}
    \begin{subfigure}[b]{0.23\textwidth}
        \includegraphics[width=\textwidth]{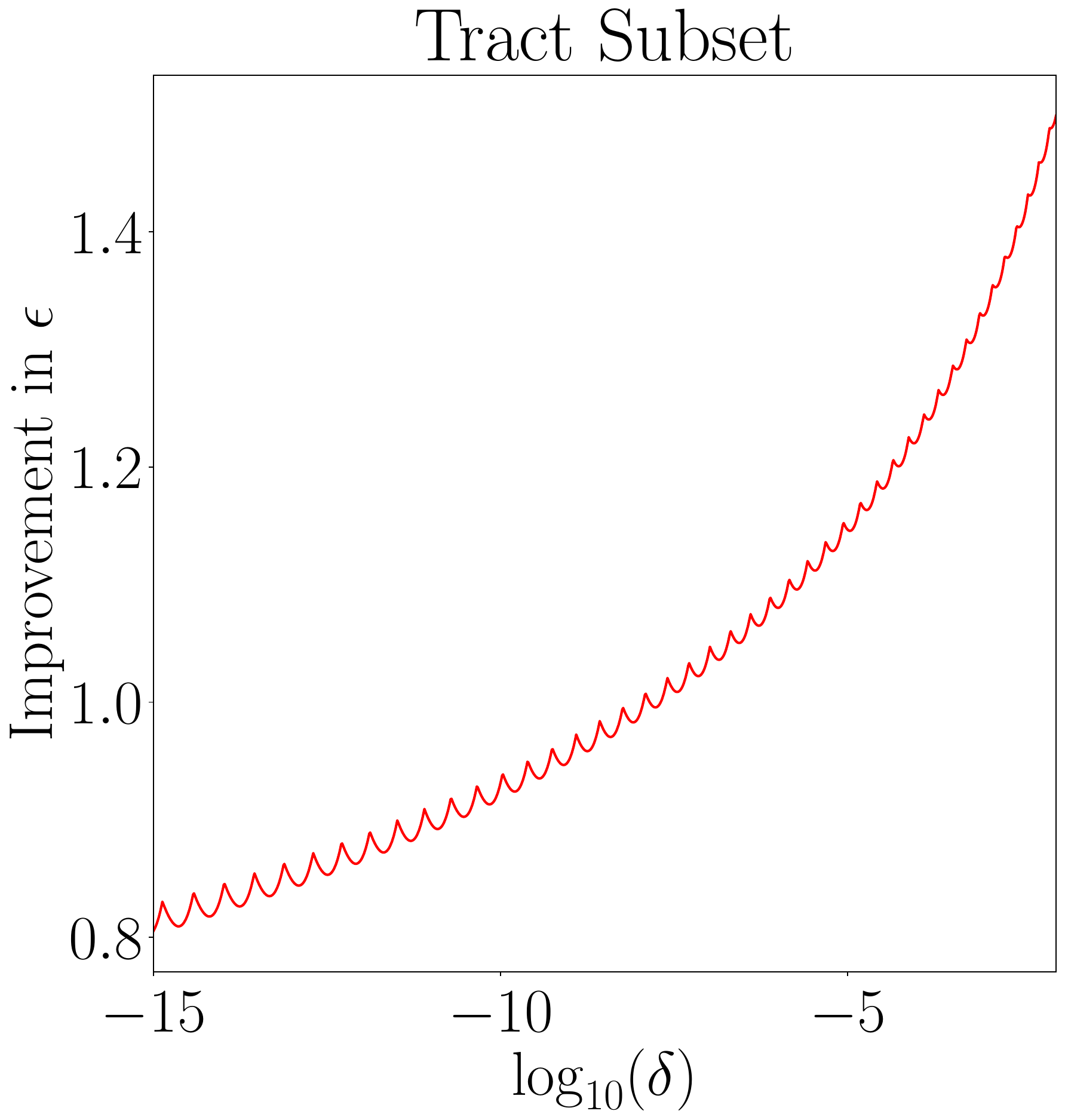}
    \end{subfigure}
    \begin{subfigure}[b]{0.23\textwidth}
        \includegraphics[width=\textwidth]{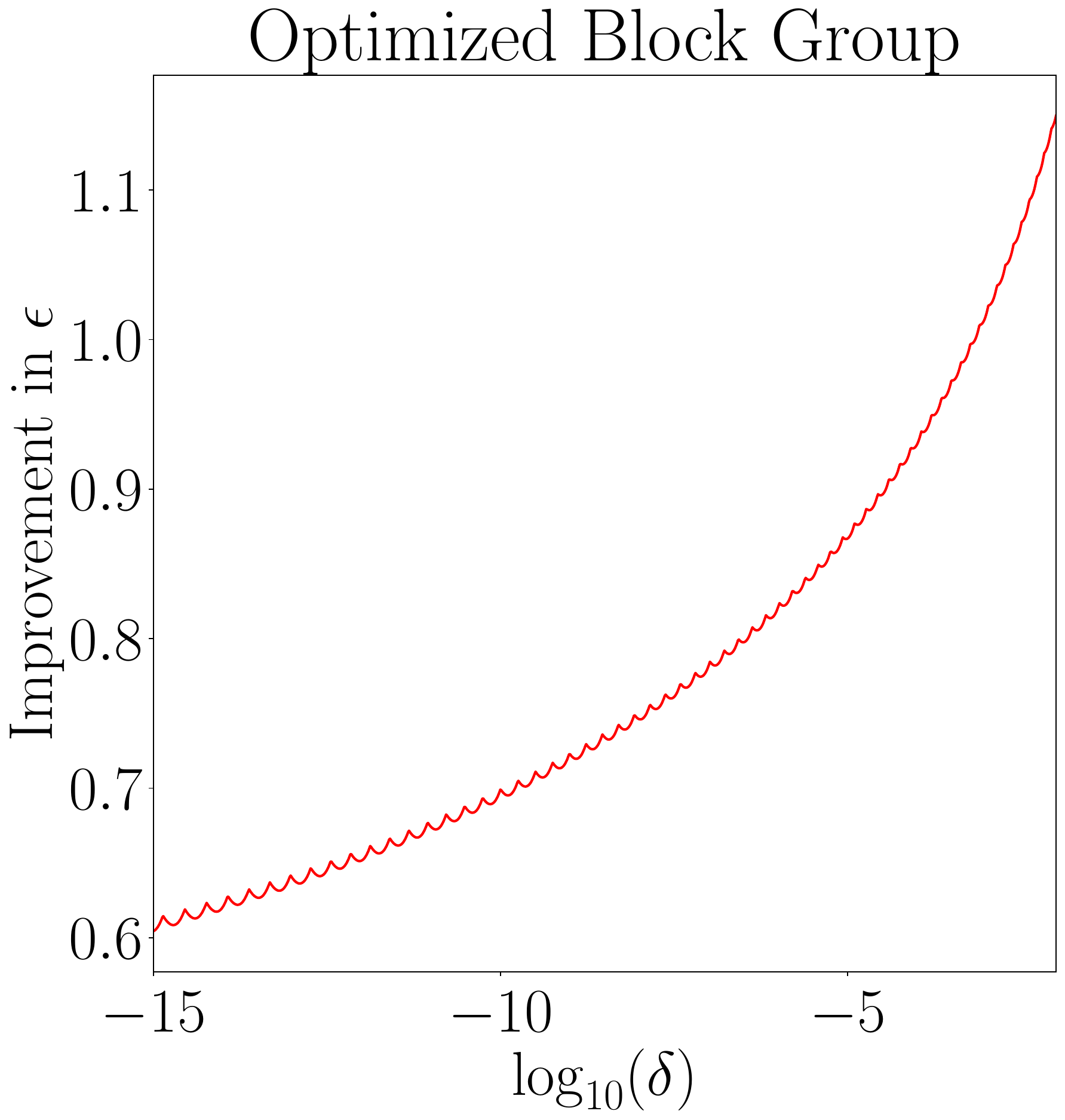}
    \end{subfigure}
    \begin{subfigure}[b]{0.23\textwidth}
        \includegraphics[width=\textwidth]{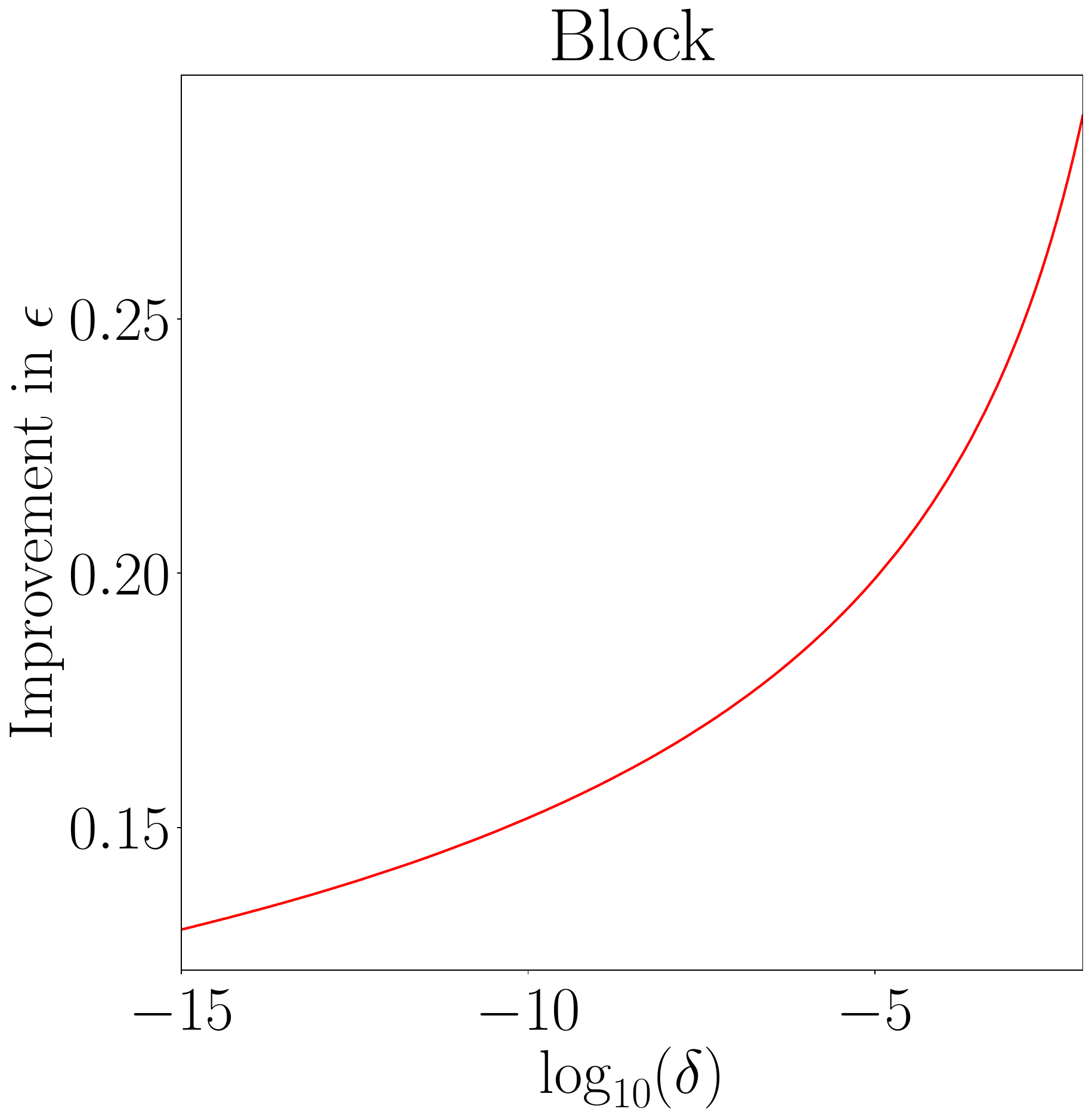}
    \end{subfigure}
 \caption{Improvement in $\epsilon$ (x-axis) for each geographical level of the 2020 U.S.\ Census, using $f$-DP based accounting method under the same setting as Figure~\ref{fig:improve_epsilon}. Our method achieves improved privacy analysis for any value of $\delta$ and gets better as $\delta$ become larger. }
    \label{fig:eps_delta_geo_gap}
\end{figure*}

\begin{figure*}[!htp]
    \centering
    \begin{subfigure}[b]{0.23\textwidth}
        \includegraphics[width=\textwidth]{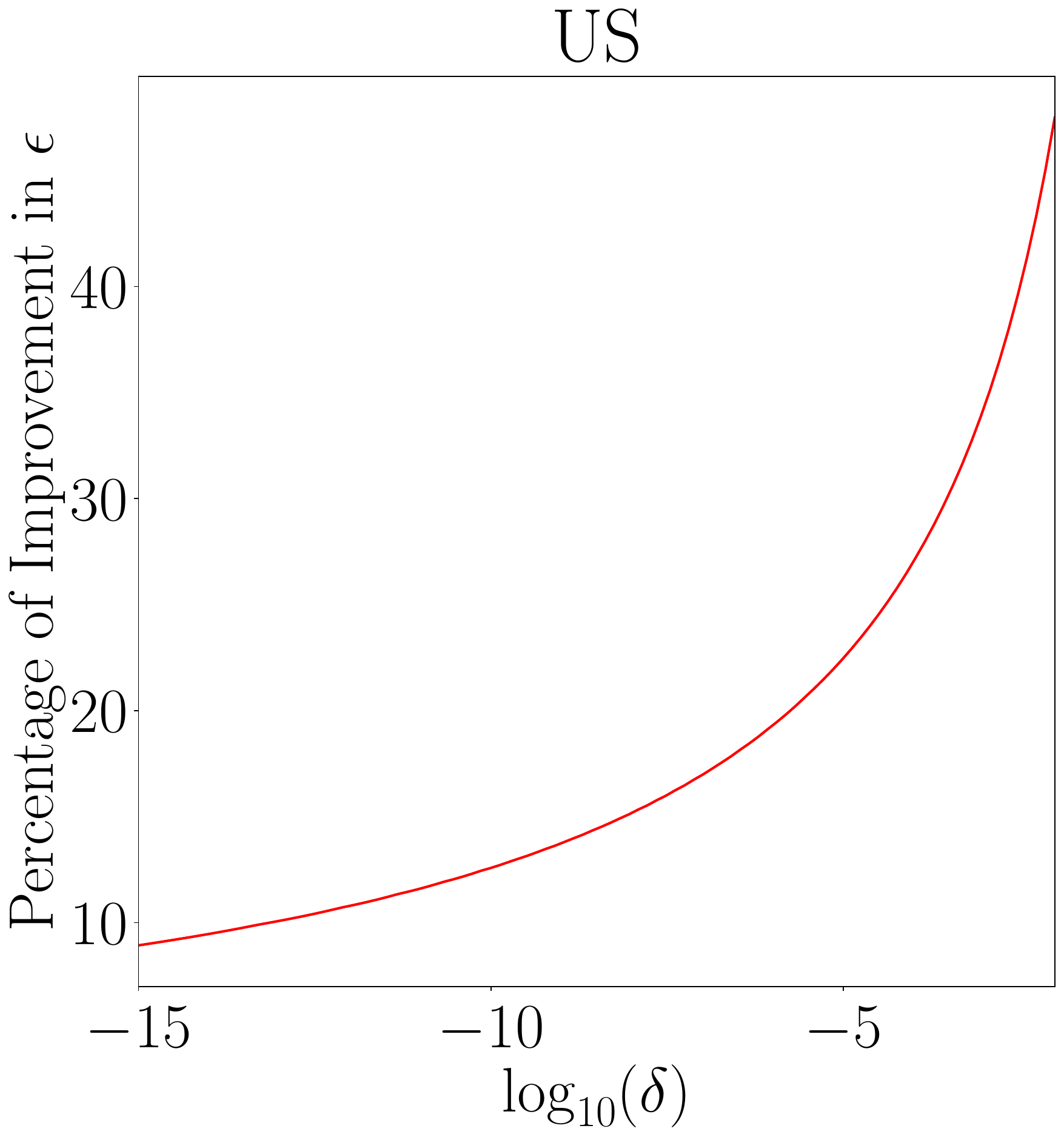}
    \end{subfigure}
    \begin{subfigure}[b]{0.23\textwidth}
        \includegraphics[width=\textwidth]{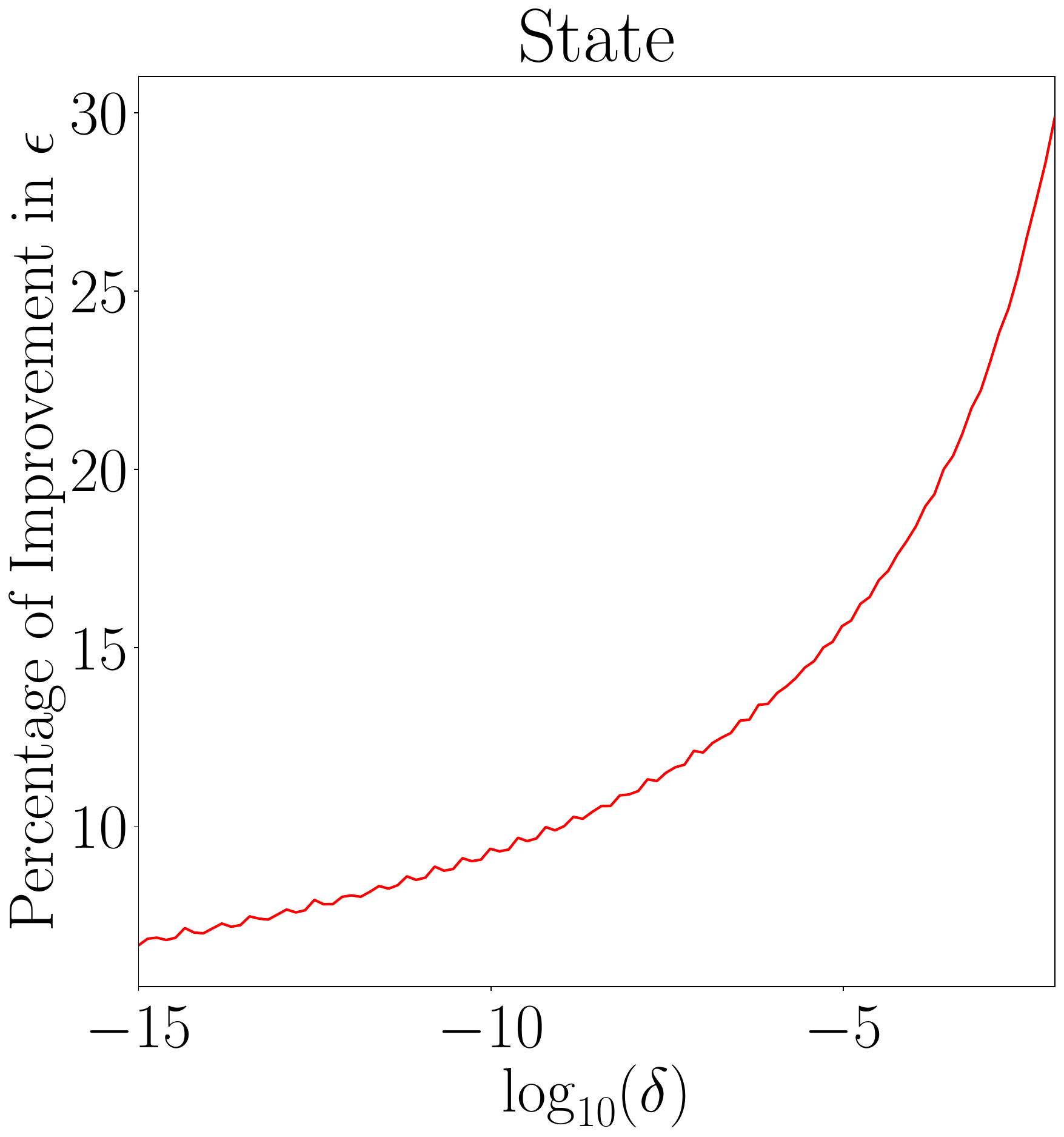}
    \end{subfigure}
    \begin{subfigure}[b]{0.23\textwidth}
        \includegraphics[width=\textwidth]{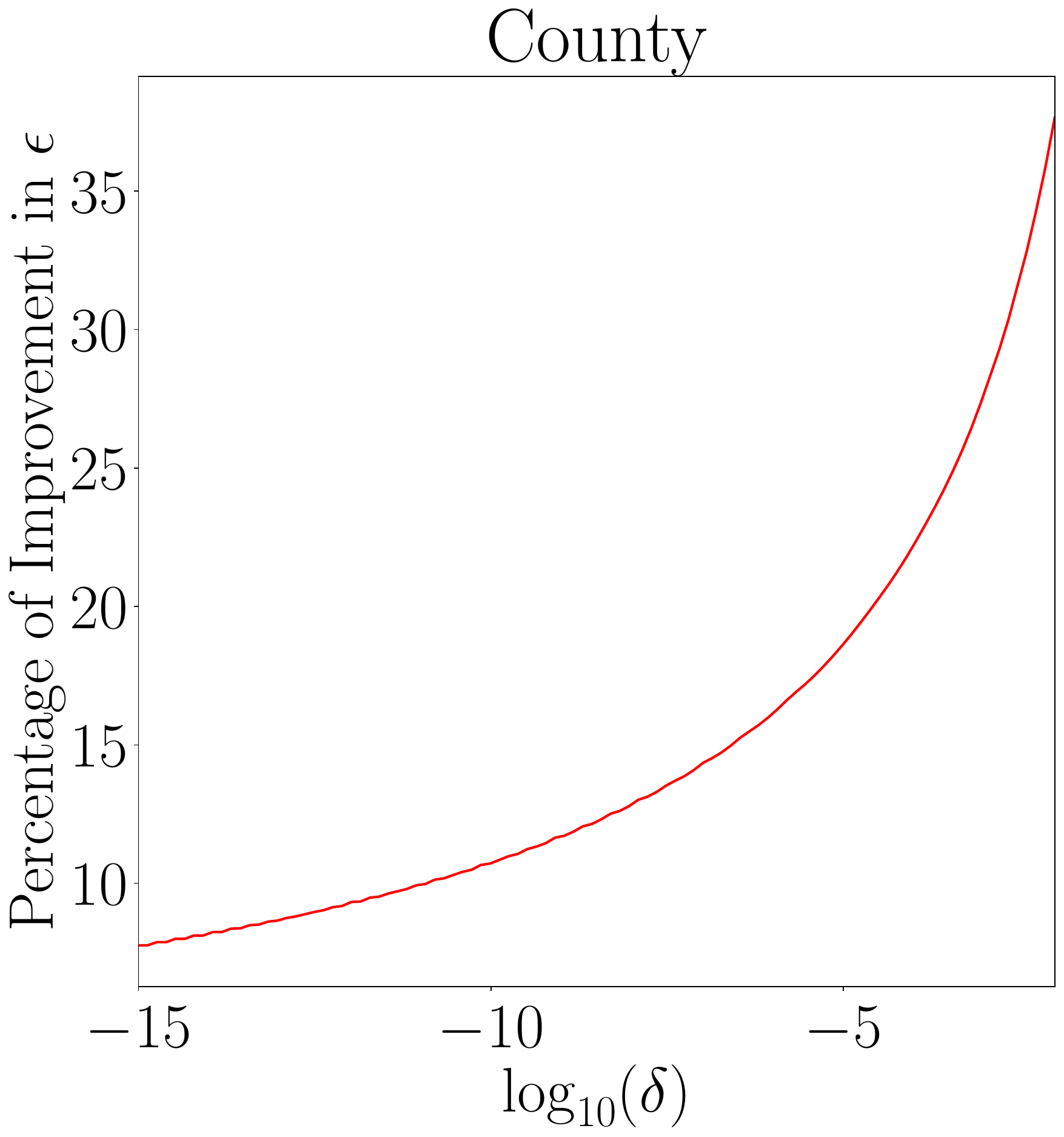}
    \end{subfigure}
    \begin{subfigure}[b]{0.23\textwidth}
        \includegraphics[width=\textwidth]{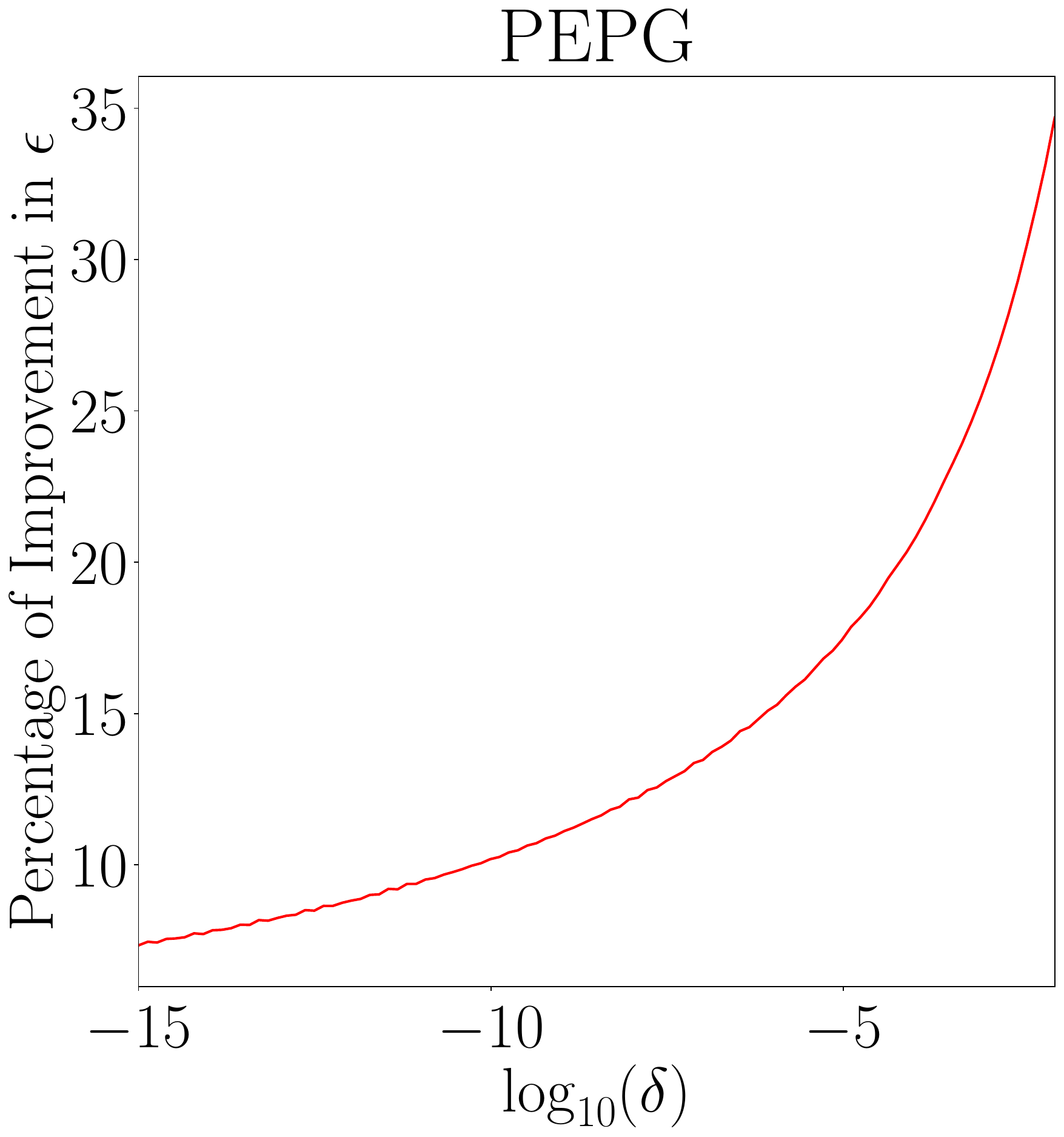}
    \end{subfigure}
    
    \hfill
    
    \begin{subfigure}[b]{0.23\textwidth}
        \includegraphics[width=\textwidth]{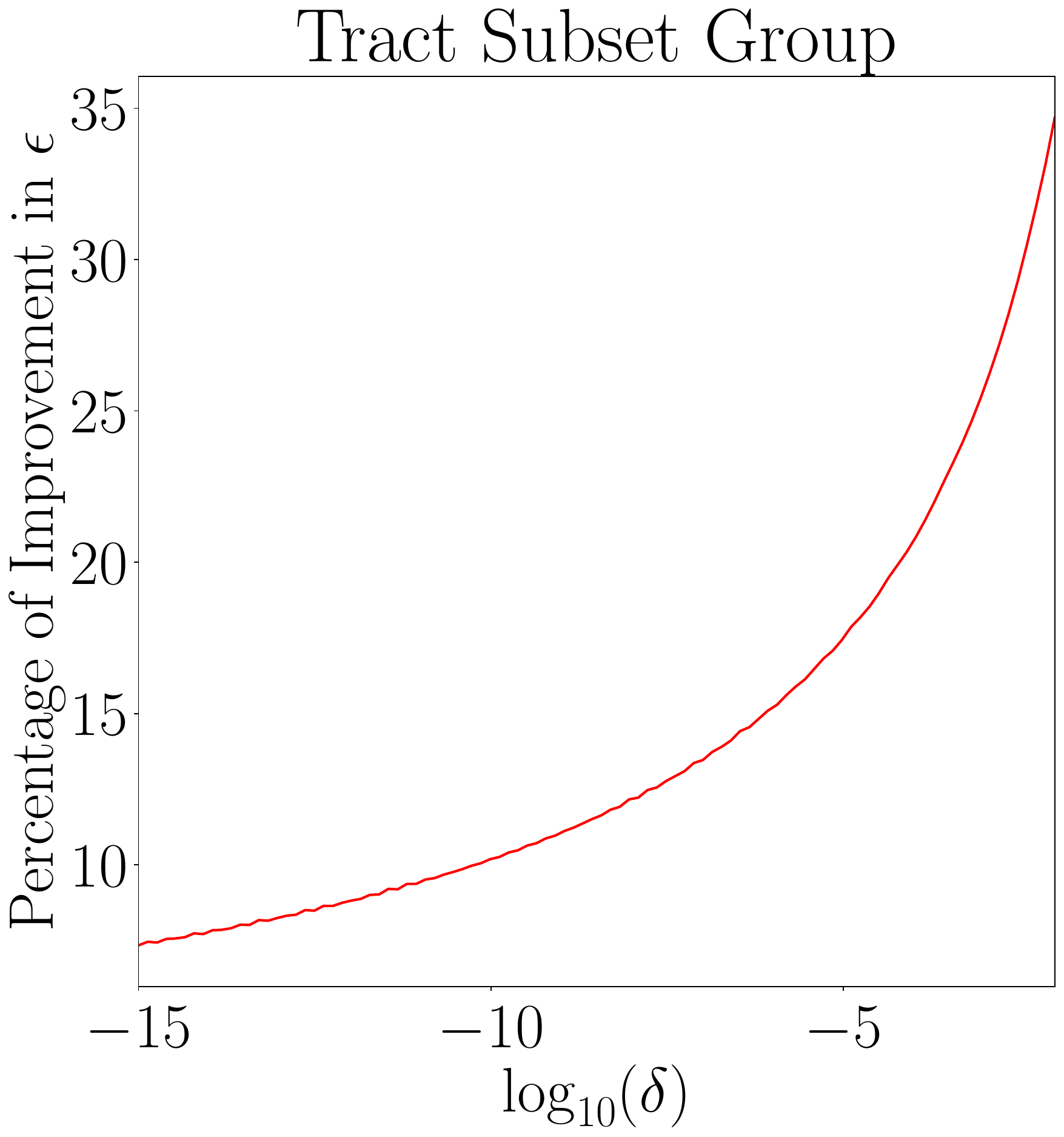}
    \end{subfigure}
    \begin{subfigure}[b]{0.23\textwidth}
        \includegraphics[width=\textwidth]{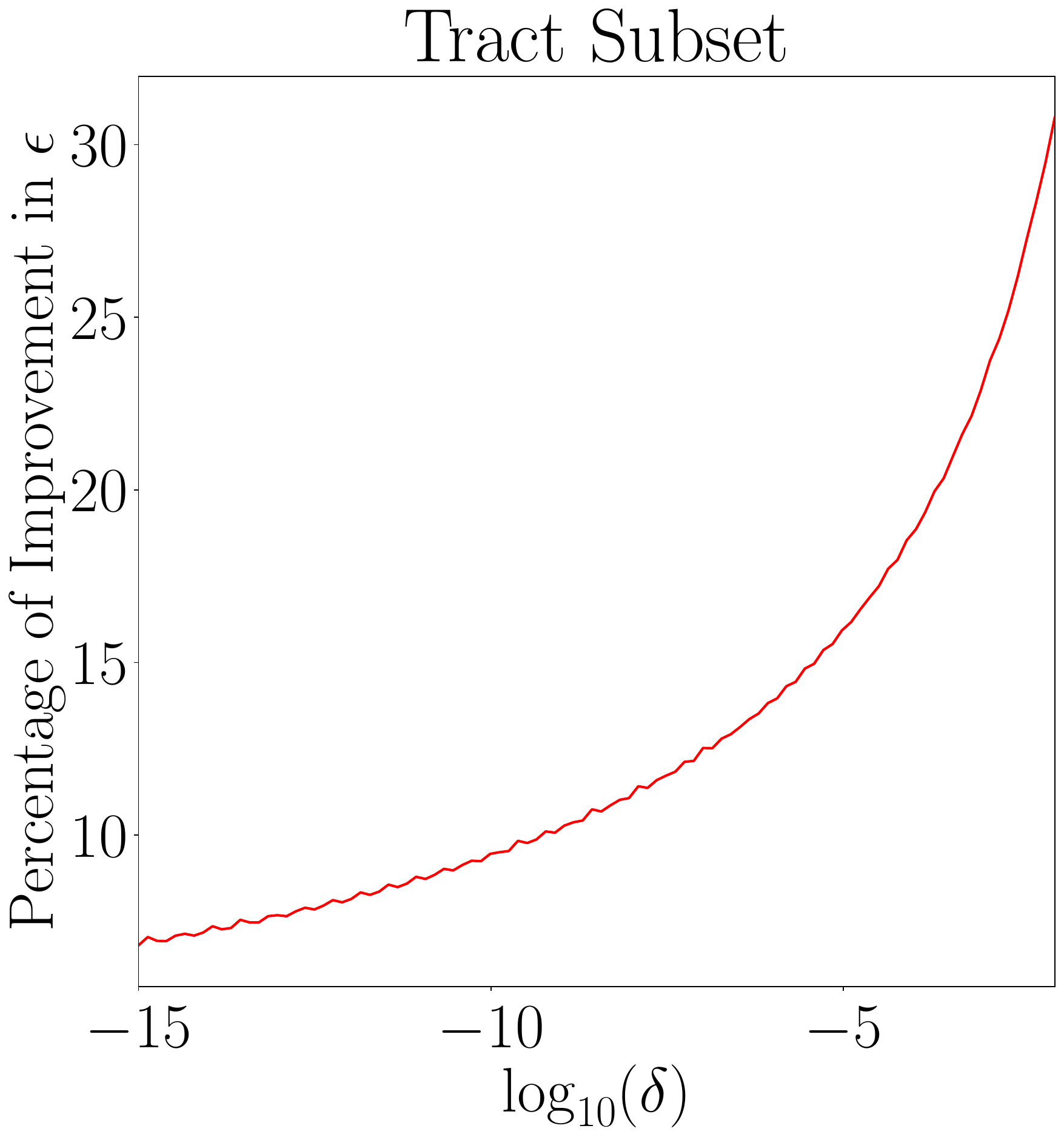}
    \end{subfigure}
    \begin{subfigure}[b]{0.23\textwidth}
        \includegraphics[width=\textwidth]{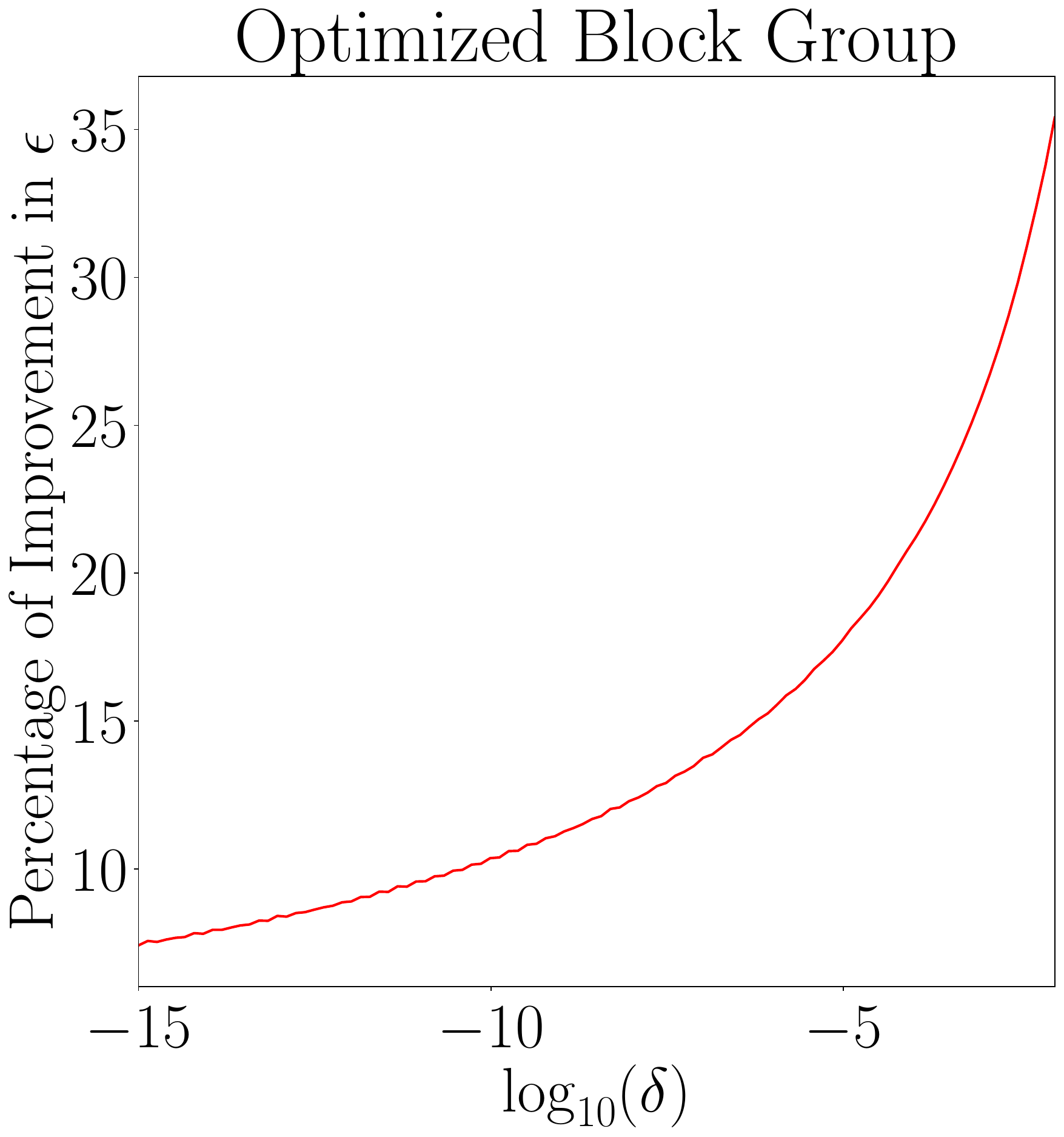}
    \end{subfigure}
    \begin{subfigure}[b]{0.23\textwidth}
        \includegraphics[width=\textwidth]{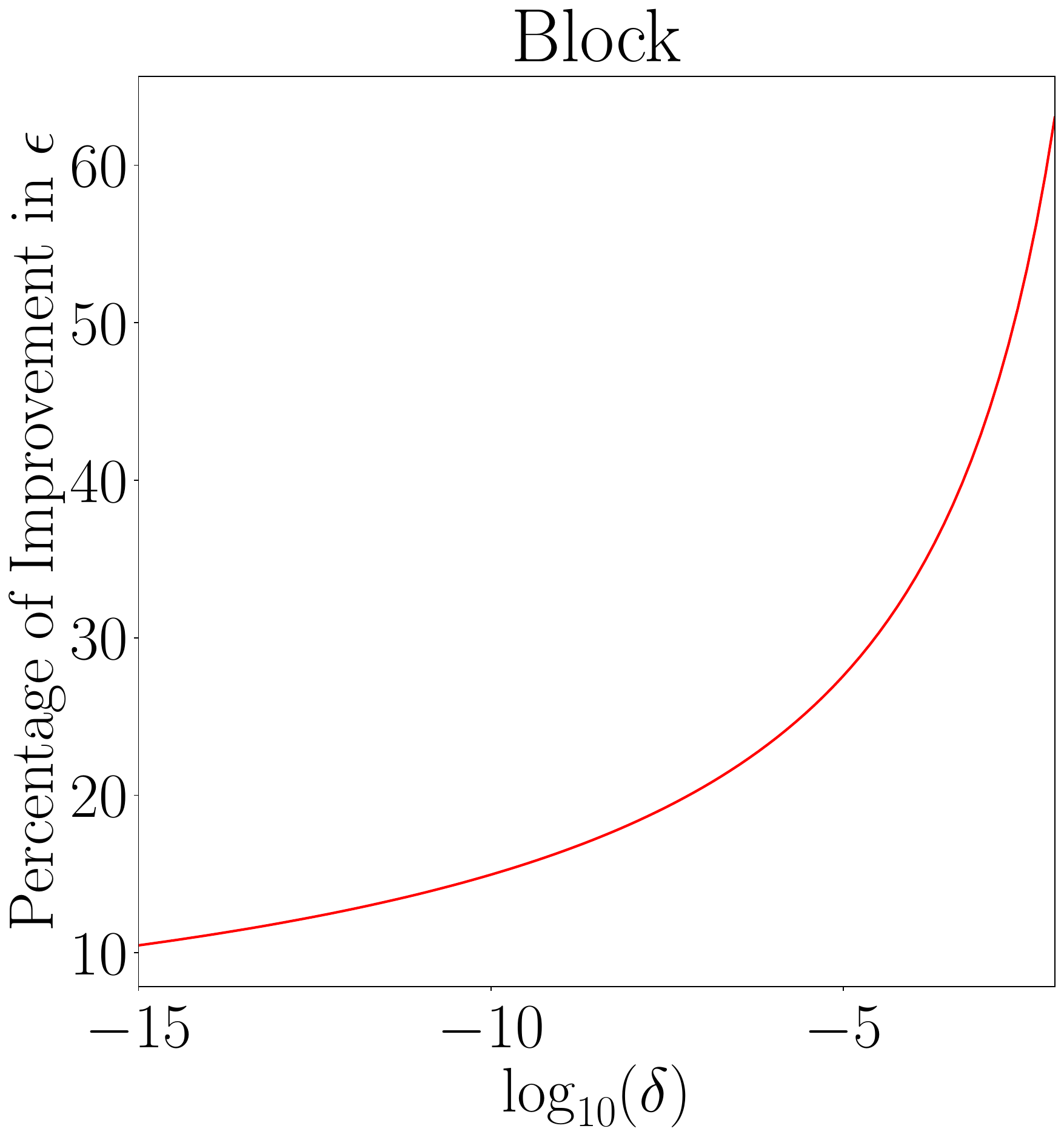}
    \end{subfigure}
    \caption{Percentage of improvement in $\epsilon$ by using $f$-DP based accounting method for each geographical level of the 2020 U.S.\ Census, under the same setting as Figure~\ref{fig:improve_epsilon}. Our method achieves improved privacy analysis for any value of $\delta$ and obtains better improvements as $\delta$ become larger.}
    \label{fig:eps_delta_geo_percentage}
\end{figure*}

\end{document}